\documentstyle[pre,aps,preprint,epsfig]{revtex}
\tightenlines
\begin{document} 
%\draft
\title{From Car Parking to Protein Adsorption: An Overview
of Sequential Adsorption Processes}
\author{J. Talbot$^1$, G. Tarjus$^2$, P. R. Van Tassel{$^3$}, and
P.   Viot{$^{2,4}$}}    \address{$^1$  Department  of    Chemistry  and
Biochemistry, Duquesne University, Pittsburgh, PA 15282-1530}
\address{$^2$  Laboratoire de Physique
Th{\'e}orique des Liquides,  Universit{\'e}  Pierre et Marie Curie,  4,  place
Jussieu 75252 Paris, Cedex 05 France}
\address{{$^3$}  Department of Chemical Engineering  and Material
Science, Wayne State University, 5050 Anthony Wayne Drive, MI 48202}
\address{{$^4$} Laboratoire de Physique
Th{\'e}orique, Bat. 211, Universit{\'e}  de Paris-Sud 91405 ORSAY Cedex France
}

\maketitle

\begin{abstract}
The  adsorption or  adhesion of  large particles (proteins,  colloids,
cells, \ldots) at  the liquid-solid interface plays  an  important role in
many  diverse  applications. Despite the  apparent   complexity of the
process, two features are particularly important: 1) the adsorption is
often  irreversible on experimental  time scales and 2) the adsorption
rate     is limited by  geometric blockage    from previously adsorbed
particles.  A coarse-grained description   that encompasses these  two
properties is provided by sequential  adsorption models whose simplest
example is the  random  sequential adsorption (RSA) process.  In  this
article, we review the theoretical  formalism and tools that allow 
the systematic  study  of  kinetic and  structural   aspects of  these
sequential adsorption models. We also show how the reference RSA model
may be generalized to  account for a  variety of experimental features
including   particle  anisotropy,   polydispersity,   bulk  diffusive
transport,  gravitational effects,  surface-induced conformational and
orientational change,   desorption, and multilayer  formation.  In all
cases,  the significant theoretical  results  are presented and  their
accuracy (compared to computer simulation) and applicability (compared
to experiment) are discussed.
\end{abstract}

\section{Introduction}\label{sec:Introduction}
The  adsorption  of   large particles (proteins,   viruses,  bacteria,
colloids,  macromolecules) at the  liquid-solid interface has received
considerable theoretical    and   experimental  attention  in   recent
years. Among the many issues pertaining to this problem, we address in
this  review  the consequences  on the  adsorption   kinetics and  the
adsorbed  layer  structure  of  two key  features:  i) the  absence of
reversibility and ii) the deceleration  of  adsorption due to  surface
exclusion            from          previously                 adsorbed
particles\cite{F80b,OL86,AH86,N86,R93,ASZB94}. We choose as a starting
point a  {\it  mesoscopic} approach,   where atomic-level  detail   is
coarse-grained and incorporated as a set  of effective parameters such
as rate constants, particle  shapes and sizes, effective interparticle
interaction parameters, etc\ldots The two  essential features listed above
are    well   accounted    for     in  {\it    sequential   adsorption
models}\cite{BP91,E93}; these will  be   the  central topic   of  this
review.

The outline of our review is as follows. In
section~\ref{sec:RSA:-reference-model}, we introduce the simplest of 
sequential adsorption  processes, the random sequential adsorption
model (RSA), and its one-dimensional version, the car parking problem.
Since these processes result in structures that are generally far from 
equilibrium, the well-developed methods of liquid-state statistical mechanics 
are not directly
applicable. We present, in section~\ref{sec:Theor-form-tools}, the
formalism and principal theoretical tools that we have developed for
describing the kinetics of  the adsorption processes and the structure
of the adsorbed  layers. The next  sections  are devoted  to  
extensions of these methods to processes involving additional physical
features. In section~\ref{sec:Basic-extensions-RSA}, we consider the 
adsorption of nonspherical particles and of mixtures
and in section~\ref{sec:Bulk-transport-issu}, we address bulk transport 
issues such  as  particle  diffusion and the influence of the
gravitational field. In section~\ref{sec:Surf-events:conf}, we treat
surface-induced events, like conformational and orientational changes 
of the  particles and in section~\ref{sec:Desorption-processes}, we 
introduce partial reversibility produced by desorption. Finally, we 
discuss multilayer formation in
section~\ref{sec:Multilayer-formation}.

\section{RSA: a reference model}\label{sec:RSA:-reference-model}
The random sequential  adsorption  model, the prototype for sequential 
addition processes, is  a  stochastic process  in
which ``hard'' particles are added   sequentially to a $D$-dimensional volume
at random positions with the  condition that no trial particle can
overlap previously inserted ones.
\subsection{The car parking problem}\label{sec:car-parking-problem}
The one-dimensional version  of  the model, known  as  the car parking
problem, was   first  introduced by  A.  R{\'e}nyi\cite{R63},  a Hungarian
mathematician.  Consider  an infinite  line, assumed
empty at $t=0$. Hard rods  of length $\sigma  $  are dropped randomly  and
sequentially at rate  $k_a$ onto the line, and   are adsorbed only  if
they do not overlap previously adsorbed rods. Otherwise, they are
rejected.  If $\rho(t)$ denotes the  number density  of particles on the
line at  time  $t$, the  kinetics of  this process is  governed by the
equation
\begin{equation}\label{eq:1}
\frac{\partial\rho(t)}{\partial t}=k_a\Phi(t),
\end{equation}
where $\Phi (t)$, the insertion probability at time $t$, is the fraction
of  the  substrate that is   available for  the    insertion of a  new
particle.  To calculate this quantity,  it is convenient to  introduce
the gap  distribution  function $G(h,t)$, which   is  defined so  that
$G(h,t)dh$ represents  the density of  voids of length between $h$ and
$h+dh$ at time $t$.   For a given void   of length $h$, the  available
length  for  inserting a new particle   is $h-\sigma $, and  therefore the
available  line function $\Phi  (t)$ is merely  the  sum over the number
density of available intervals, i.e.  $ G(h,t)$:

\begin{equation}\label{eq:2}
\Phi (t)=\int_\sigma^\infty dh (h-\sigma)G(h,t).
\end{equation}
Since each interval corresponds to one particle, the number density of
particles $\rho (t)$ can be expressed as
\begin{equation}\label{eq:3}
\rho (t)=\int_0^\infty dh G(h,t),
\end{equation} 
whereas the uncovered line is related to $G(h,t)$ by
\begin{equation}\label{eq:4}
1- \rho (t)\sigma =\int_0^\infty dh hG(h,t).
\end{equation}
These two equations, Eqs.~(\ref{eq:3}-~\ref{eq:4}), represents two sum 
rules for the gap distribution function.
During the process, the gap distribution function $G(h,t)$ evolves as
\begin{equation}\label{eq:5}
\frac{\partial{G(h,t)}}{\partial (k_a   t)}=   -H(h-\sigma)(h-\sigma )  G(h,t)   + 2
\int_{h+\sigma }^{\infty} dh' G(h',t),
\end{equation}
where $H(x)$ is   the  unit step   function.   The first term of   the
right-hand side of Eq.~(\ref{eq:5})  (destruction term) corresponds to
the insertion of  a particle within the  gap  of length $h$ (for  $h\geq
\sigma$), whereas  the second term  (creation   term) corresponds  to  the
insertion of particle in  a gap of  length $h'>h+\sigma $.  The factor $2$
is due to the two possibilities of creating a length $h$ from a larger
interval $h'$.  It is worth noting that the time evolution of $G(h,t)$
is entirely determined  by intervals larger  than $h$.  We now  have a
closed   set of   equations,  which results   from  the  fact that the
adsorption of a particle in one  gap has no  effect on other gaps. The
above equations can be solved by introducing the ansatz\cite{GHH74}
\begin{equation}\label{eq:6}
G(h,t)=\frac{F(k_a\sigma t)}{\sigma^2}\exp(-k_a(h-\sigma) t),
\end{equation}
which leads to
\begin{equation}\label{eq:7}
F(t)=t^2\exp\left(-2\int_0^t du \frac{1-e^{-u}}{u}\right).
\end{equation}
Inserting Eqs.~(\ref{eq:6}) and (\ref{eq:7}) , one
obtains $G(h,t)$ for $h>\sigma $ and integrating  Eq.~(\ref{eq:5}) with the
solution of $G(h,t)$ for $h>\sigma $ finally gives  $G(h,t)$ for $0<h<\sigma $,
\begin{equation}\label{eq:8}
G(h,t)=\frac{2}{\sigma^2}\int_0^{k_a\sigma t} du\exp(-uh/\sigma )\frac{F(u)}{u}.
\end{equation} 
The three equations,  (\ref{eq:1}), (\ref{eq:3})  or (\ref{eq:4}), all
lead  to the same result for the number density $\rho(t)$,
\begin{equation}\label{eq:9}
\rho(t)=\frac{1}{\sigma}\int_0^{k_a\sigma t}du\exp\left(-2\int_0^u dv \frac{1-e^{-v}}{v}\right),
\end{equation}
which was first derived by R{\'e}nyi\cite{R63}.

A first nontrivial property of the model is that the process reaches a
``jamming limit''  (when $ t\to\infty$) at which the  density saturates at a
value  $  \rho_\infty\sigma=0.7476 \ldots$; this value  is  significantly lower  than the
closed-packed   density ($\rho_\infty\sigma=1$) that is expected   when  the  surface is
allowed to restructure between adsorption steps.  Moreover, it is easy
to show that the jamming
limit depends upon the initial configuration,  here an empty line.  In
contrast, the final state of an equilibrium system is determined solely
by the chemical potential and has no memory of the initial state.  The
long-time kinetics, during which  a small number of  adsorption events
occurs, can be obtained from Eq.~(\ref{eq:9}) and are seen to display
a power-law behavior.
\begin{equation}\label{eq:10}
\rho_\infty\sigma -\rho(t)\sigma \simeq\left(\frac{e^{-2\gamma }}{ k_a\sigma^2}\right)\frac{1}{t}
\end{equation}
where   $\gamma$ is the Euler  constant. 

The structure of configurations generated by this irreversible process
has   several  noticeable  properties.   At  saturation,  the gap
distribution function shows a logarithmic  divergence at contact, $h\to
0$,
\begin{equation}\label{eq:11}
G(h,\infty)\simeq-e^{-2\gamma } \ln(h/\sigma ).
\end{equation}
The correlations between pairs of particles are extremely weak at long
distances. The pair correlation function behaves at long distances as
\begin{equation}
g(r)-1\propto\frac{1}{\Gamma(r/\sigma)}\left(\frac{2}{\ln\ r/\sigma}\right)^{r/\sigma}\ ,
\
\end{equation}
where   $\Gamma   (x)$  is    the    Gamma function,  i.e.,     it  decays
super-exponentially in contrast  with  the characteristic  exponential
decay of equilibrium systems\cite{BBV94}.

\subsection{Two-dimensional sequential adsorption}\label{sec:Two-dimensional-sequ}
The RSA model  can be studied in  two dimensions, which represents the
physical dimension for adsorption  problems.   Since the  RSA  process
generates random disordered configurations of ``hard particles'', {\it
a statistical geometric approach} is useful for describing the system.
For a given time  $t$,  or equivalently a  given  density $\rho$,  it is
convenient to  introduce the  set  of $n-$particle  density  functions
$\{\rho^{(n)}({\bf r}_1,{\bf r}_2,\cdots ,{\bf r}_n,t)\}$, where ${\bf r}_1,{\bf
r}_2,\cdots ,{\bf r}_n$ denote the   positions of the $n$ particles,  that
characterize completely the configuration of particles.

The time evolution of the density is still  given by Eq.~(\ref{eq:1}), where
$\Phi$  denotes the available fraction of  the surface for the insertion
of a   new particle. When  the particles  are  hard disks,  $\Phi$ has a
simple geometrical  interpretation:  It represents the  probability of
finding a circular  cavity of  diameter  equal to twice the   particle
diameter,  that is centered  at ${\bf r}_1$,   and is free of particle
centers\cite{ST89a,ST89b,TST91a}

\begin{equation}\label{eq:12}
\Phi({\bf  r}_1^*;\rho)=  \sum_{s=0}^{\infty  }\frac{1}{s!}\int  \cdots\int d{\bf r}_2\cdots
d{\bf   r}_{(s+1)}f_{12}\cdots     f_{1(s+1)}\rho^{(s)}({\bf    r}_2,\cdots,{\bf
r}_{(s+1)};\rho),
\end{equation}
where $*$  denotes a cavity (as  defined  above) and $f_{ij}\equiv f(|{\bf
r}_i-{\bf r}_j|)$ is  the  Mayer  function for  non  overlapping  hard
particles:
\begin{eqnarray}\label{eq:13}
f_{ij}& = &  0,\;\; |{\bf r}_i-{\bf r}_j| >  \sigma, \\ f_{ij}& = & -1,\;\;
|{\bf r}_i-{\bf r}_j| < \sigma,
\end{eqnarray}
where $\sigma$ is the diameter of the hard spherical particles. (Note that
because of the   macroscopic   uniformity  of the   system,   $\Phi({\bf
r}_1^*;\rho)$ is independent of the position ${\bf r}_1$, i.e., $\Phi({\bf
r}_1^*;\rho)\equiv \Phi(\rho)$.)  The terms of the sum in the right-hand side of
Eq.~(\ref{eq:12})  can be interpreted geometrically  : the first term,
which   is equal  to  one,  corresponds to  the certain  acceptance of
particles added   to  an empty surface.    The second  (negative) term
subtracts the fraction of  the surface occupied  by exclusion disks in
which no new particle center can be placed.  The third (positive) term
corrects  for  the fact that  two exclusion  disks may overlap, etc...
For hard-core interactions, the sum is  always finite since the number
of non-overlapping particles located  at   a distance less  than   the
particle diameter from  a given point  is  finite.  For instance,  for
hard disks  in  two dimensions, only  six first  terms have a  nonzero
contribution to Eq.~(\ref{eq:12}).

Higher order     density   functions  can  be systematically
defined.  For instance, $\Phi^{(2)}({\bf  r}_1^*,{\bf r}_2;\rho)$ represents the
density of finding one (unspecified) particle at the point ${\bf r}_2$
and a cavity of diameter $2\sigma$  free of particle  centers at the point
${\bf   r}_1$.   More  generally,   one can   introduce $\Phi^{(n)}({\bf
r}_1^*,{\bf r}_2,...{\bf   r}_n;\rho)$ as  the  probability  density  of
finding $n-1$ (unspecified)  particles  at the points ${\bf  r}_2,{\bf
r}_3,...,{\bf r}_n$ and  a cavity of diameter  $2\sigma$ free of  particle
centers at the point ${\bf r}_1$,
\begin{eqnarray}\label{eq:14}
\Phi^{(n)}({\bf    r}_1^*,{\bf   r}_2,\cdots    {\bf      r}_n;\rho)&=&\prod_{j=2}^n(1+f_{1j})\nonumber\\  &&\left( \sum_{s=0}^{\infty }\frac{1}{s!}\int  \cdots\int
d{\bf       r}_{n+1}\cdots              d{\bf       r}_{(n+s)}f_{1(n+1)}\cdots
f_{1(n+s)}\rho^{(n+s-1)}({\bf r}_2,\cdots,{\bf r}_{(n+s)};\rho)\right).
\end{eqnarray}

For a  RSA process,  the insertion of a new particle  amounts  to
finding a cavity  of  diameter $2\sigma$  for  a given set   of previously
adsorbed  particles.    The  kinetic equation for  the
density, Eq.~(\ref{eq:1}), can be  generalized to higher order density
functions,
\begin{equation}\label{eq:15} 
\frac{\partial \rho _{n}({\bf r}_1,{\bf r}_2,...,{\bf r}_n;\rho(t) )}{\partial(k_a t)}=
 \sum_{i=1}^n\Phi^{(n)}({\bf r}_1,.,{\bf  r}_{i-1},{\bf      r}_i^*,  {\bf
 r}_{i+1}, ..{\bf r}_n;\rho ),
\end{equation} 
or, by using Eq.~(\ref{eq:1}),
\begin{equation}\label{eq:16}
\frac{\partial \rho _{n}({\bf r}_1,{\bf r}_2,...,{\bf r}_n;\rho )}{\partial \rho}= 
\frac{1}{\Phi({\bf r}_1;\rho )}\sum_{i=1}^n\Phi^{(n)}({\bf r}_1,.,{\bf r}_{i-1},{\bf r}_i^*, {\bf r}_{i+1}, ..{\bf r}_n;\rho ). 
\end{equation}
The exact  and infinite hierarchy of equations   cannot be solved 
(nor is it possible in the equilibrium  case), but approximate
solutions and computer simulations (both to be detailed below) provide 
a good description of the process.

The main features  observed in the one-dimensional car parking problem 
persist in higher dimensions: existence of a jamming limit at which
the density of adsorbed particles saturates (in $2D$, $\theta_\infty =\rho_\infty
\frac{\pi\sigma^2}{4}\simeq 0.547$\cite{F80a,HFJ86}), slow kinetics when
approaching jamming (in $2D$,   $\theta_\infty -\theta(t) \sim t^{-1/2}$
\cite{F80a}), logarithmic divergence at contact of the pair
correlations at saturation,  configurations of adsorbed particles that
differ    from those characteristic   of  an  equilibrated system. The
qualitative   and quantitative relevance    of two-dimensional RSA for
describing the adsorption of macromolecules at liquid-solid interfaces
has   been       demonstrated   in     several      experiments     on
proteins\cite{F80b,R93b}                and                  colloidal
particles\cite{OL86,ASZB94,AZSW90,AB91,WSSZV93}.

\section{Theoretical  formalism and tools }\label{sec:Theor-form-tools}

\subsection{Statistical mechanics of sequential adsorption systems}\label{sec:Stat-mech-sequ}
Consider a system of $N$ particles in a $D$-dimensional volume $V$ (in
practice, adsorption involves  $D=2$). For simplicity,  we assume that
the particles are  identical and interact  through a pairwise-additive
spherically-symmetric     potential.   Generalization to     mixtures,
nonspherical   particles and   non  pairwise-additive potentials   is,
conceptually at least, straightforward (see below).   If the system of
particles  is    at equilibrium  at   a  given  temperature   $T$, the
probability density,   $\rho^{(N)}_{eq}(\{ {\bf r}_1,{\bf   r}_2,\cdots {\bf
r}_N\})$ associated with finding the $N$ particles in a configuration
with one (unspecified) particle whose center is an infinitesimal
volume element around position ${\bf r}_1$,\ldots,one (unspecified) particle whose center is an infinitesimal
volume element around position ${\bf r}_N$ is given by the Gibbs
distribution for the canonical ensemble, 
\begin{equation}\label{eq:17}
\rho^{(N)}_{eq}(\{ {\bf        r}_1,{\bf        r}_2,\cdots        {\bf
r}_N\})=\frac{e^{-\beta\sum_{1\leq  i   <j\leq  N  }v(|{\bf  r}_i-{\bf
r}_j|)}}{Z_N(V,\beta) },
\end{equation}
where $\beta=1/k_BT$, $k_B$ is the Boltzmann constant, $v(|{\bf r}_i-{\bf
r}_j|)$ is the pair potential, and  $Z_N(V,\beta)$ is the configurational
integral defined by
\begin{equation}\label{eq:18}
Z_N(V,\beta)=\int\ldots\int d{\bf  r}_1d{\bf r}_2\cdots  d{\bf r}_N   e^{-\beta\sum_{1\leq i
<j\leq N }v(|{\bf r}_i-{\bf r}_j|)}
\end{equation}
All the powerful tools  of  equilibrium statistical mechanics  and the
connection to  thermodynamics derive   from  the  Gibbs  distribution.
Consider now a situation in which the  configuration of particles does
not  correspond   to  thermal equilibrium,    but  is  generated  by a
sequential  adsorption process.   More  specifically,  we  consider  a
``cooperative sequential  adsorption'' (CSA) that is a simple generalization
of the random sequential adsorption  process introduced above. In this
process,   the rate  of   adsorption  of a   new  particle in  a given
configuration of pre-adsorbed particles is proportional to a Boltzmann
factor  involving the interaction energy  between the  new particle at
the  chosen  position  and all   preadsorbed   particles.  If  $\{{\bf
r}_1,{\bf   r}_2,\cdots  {\bf  r}_n\}$ denotes  the   position  of the $n$
preadsorbed  particles,  adsorption of  a  particle  at position ${\bf
r}_{n+1}$ is given by
\begin{equation}\label{eq:19}
k({\bf      r}_{n+1}|\{{\bf       r}_1,{\bf   r}_2,\cdots     {\bf    r}_n
\})=k_a\exp\left[-\beta \sum_{j=1}^nv(|{\bf r}_{n+1}-{\bf r}_j|)\right],
\end{equation}
where  $k_a$ is the rate   of adsorption of a   particle  in an  empty
$D$-dimensional volume.   What is the probability  of finding  a given
configuration   with  a total   number  of  $N$  particles  at $\{{\bf
r}_1,{\bf r}_2,\cdots {\bf r}_N \}$ which is generated by this cooperative
sequential adsorption, irrespective of the time required?  It is clear
that the  intrinsic  irreversibility,  resulting from   the fact  that   once a
particle is adsorbed it can no longer move  or desorb, creates a strong
memory effect. As a  result, one must  take into account the  order in
which  the  positions $ {\bf  r}_1,{\bf r}_2,\cdots  {\bf  r}_N$ have been
filled   by   particle centers and   consider    the $N!$ sequences of
insertion that lead to  a given configuration  of $N$ particles at $\{
{\bf r}_1,{\bf   r}_2,\cdots  {\bf r}_N  \} $.    With  the rate  given by
Eq.~(\ref{eq:19}) and the definition $e({\bf r}_{n+1}|\{{\bf r}_1,{\bf
r}_2,\cdots {\bf   r}_n \})=k({\bf r}_{n+1}|\{{\bf  r}_1,{\bf r}_2,\cdots {\bf
r}_n \})/k_a$, one obtains the following probability density

\begin{equation}\label{eq:20}
\rho^{(N)}_{CSA}(\{{\bf       r}_1,{\bf            r}_2,\cdots          {\bf
r}_N)\}=\frac{e^{-\beta\sum_{1\leq i <j\leq  N }v(|{\bf  r}_i-{\bf r}_j|)}}{N!}
\sum_{\sigma\in  S_N}\prod_{i=1}^{N}\left[  \int d{\bf  r}_{\sigma (i)}  e({\bf r}_{\sigma
(i)}|{\bf r}_{\sigma(1)},\cdots {\bf r}_{\sigma(i-1)} \})\right]^{-1},
\end{equation}
where the sum $\sigma\in S_N$ is over all permutations of $\{ 1,2,\cdots ,N\} $
and  where we have used   that $\prod_{i=1}^{N}e(\{ {\bf r}_{\sigma (i)}|{\bf
r}_{\sigma(1)},{\bf  r}_{\sigma(2)},\cdots {\bf   r}_{\sigma(i-1)} \})=\exp\left(  -\beta
\sum_{1\leq i <j\leq n }v(|{\bf r}_i-{\bf r}_j|)\right)$. (One easily checks
that  the normalization  condition $\int \cdots  \int   d{\bf r}_1d{\bf r}_2\cdots
d{\bf r}_N \rho^{(N)}_{CSA}(\{{\bf r}_1,{\bf r}_2,\cdots {\bf r}_N)\})=1$ is
satisfied).   Eq.~(\ref{eq:20}) is  the CSA  counterpart  of the Gibbs
distribution  for  equilibrium   systems,   Eq.~(\ref{eq:17}).   In  a
thermally   equilibrated system,  Eq.~(\ref{eq:17})  implies  that all
configurations having the same  energy are equiprobable.  On the other
hand, in  a sequential adsorption process there  is a bias that favors
some configurations over others  among all  those  that have the  same
energy:   this results from the fact   that  in Eq.~(\ref{eq:20}), the
factor $\sum_{\sigma\in  S_N}\prod_{i=1}^{N}\left[ \int d{\bf  r}_{\sigma (i)}  e({\bf
r}_{\sigma  (i)}|{\bf   r}_{\sigma(1)},\cdots  {\bf r}_{\sigma(i-1)}  \})\right]^{-1}$
depends  on the positions   ${\bf r}_1,{\bf  r}_2,\cdots  {\bf r}_N$  (for
$N>2$).  This feature has been already noted by Widom\cite{W66} in the
example of three hard spheres added  to a large volume.  The seemingly
formal    difference  between  the    equilibrium Gibbs  distribution,
Eq.~(\ref{eq:17}),  and the CSA  distribution, Eq.~(\ref{eq:20}),  has
important consequences.   Some of the dramatic  ones are  that the CSA
systems  are characterized by the  existence of a  jamming limit where
the density  of adsorbed particles   (for particles possessing a  hard
core) reaches a  saturation value that is  significantly less than the
optimum filling (as in RSA, cf section~\ref{sec:RSA:-reference-model})
and the absence  of thermodynamic phase transitions\footnote{Geometric
phase transitions such a percolation phenomena are still possible, see
section~\ref{sec:Ball-transport-gener}.}.  This latter feature results
from the infinite memory of  the adsorption process and is illustrated
in   Fig~\ref{fig:1}.   In the    figure,   we  compare two    typical
configurations   of  hard    nonspherical  two  dimensional  particles
(disco-rectangles , i.e. a planar spherocylinder, with an aspect ratio
of $15$)  on a  plane at the  same surface  density (or coverage), one
being an equilibrium configuration, the  other being generated by RSA.
At the chosen coverage of $0.445$, the equilibrium configuration is in
a nematic phase  with  long range orientational order  (the transition
from the disordered to nematic occurs for $\theta\simeq 0.4 $), whereas the RSA
configuration is at saturation in a disordered state.

The existence of a well-defined probability density for configurations
of      $N$ particles   produced   by   sequential    adsorption in  a
$D$-dimensional   volume $V$ at   a   given    temperature  $T$,  as    in
Eq.~(\ref{eq:20}),  makes   a   ``statistical   mechanical''  approach
possible, but the complicated  form of the probability density imposes
a nontrivial generalization of the equilibrium formalism.

The  kinetics  of a sequential  adsorption   process, as  well as  the
characteristics of  the  configurations of adsorbed  particles, can be
obtained  from  a knowledge  of   the grand ensemble   analog of the
``canonical'' probability  density introduced above, $ \rho^{(N)}(\{{\bf
r}_1,{\bf r}_2,\cdots  {\bf r}_N\};t)$, which is  associated with finding a
configuration  of  $N$  adsorbed  particles   positioned  at   $\{{\bf
r}_1,{\bf r}_2,\cdots {\bf r}_N\}$ at time $t$. However, the expression of
this  quantity is quite  intricate, and   it  is convenient  to derive
directly kinetic equations describing     the time evolution   of  the
infinite set of  $n$-particle densities, $ \rho^{(n)}(\{{\bf  r}_1,{\bf
r}_2,\cdots  {\bf  r}_n)\};t)$. ($  \rho^{(n)}(\{{\bf r}_1,{\bf r}_2,\cdots {\bf
r}_n  \};t)$   is associated with  the   probability of simultaneously
finding  an  (unspecified) particle at  ${\bf  r}_1$, another at ${\bf
r}_2$,\ldots, and  one at ${\bf    r}_n$,  irrespective of the   remaining
particles and is  thus obtained from  the $  \rho^{(N)}(\{{\bf r}_1,{\bf
r}_2,\cdots {\bf r}_n\};t)$'s for $N \geq  n$ by integrating over the positions
of $N-n$ remaining  particles.    This derivation  has been done    in
section~\ref{sec:Two-dimensional-sequ} for  the  RSA of hard particles
by means of  statistical   geometric  arguments.  The  result  can  be
generalized  for  all  sorts   of  cooperative  sequential  adsorption
processes  (as well    as  processes involving   desorption  and other
mechanisms, see  the next section).   Consider  a  cooperative  sequential
adsorption similar to that described  above with an adsorption rate that
generalizes Eq.~(\ref{eq:19}), namely
\begin{eqnarray}\label{eq:21}
k({\bf n+1}|\{{\bf 1},{\bf  2},\cdots {\bf n} \})&=&k_ae({\bf n+1} |\{{\bf
1},{\bf 2},\cdots {\bf n} \})\nonumber\\ &=&k_a\exp[-\beta W({\bf n+1}|\{{\bf
1},{\bf 2},\cdots {\bf n} \})]
\end{eqnarray}
where ${\bf j}$ is a  short-hand notation for center-of-mass   position
${\bf r}_j$ and  orientation  {\boldmath$\theta$\unboldmath  }$_j$  and  $
W({\bf  n+1}|\{{\bf  1},{\bf 2},\cdots  {\bf   n} \})$ is  the interaction
energy between  an incoming particle at  position  ${\bf r}_{n+1}$
with orientation {\boldmath$\theta$\unboldmath  }$_{n+1}$ and $n$
preadsorbed particles characterized by  $\{{\bf 1},{\bf 2},\cdots  {\bf n}
\}$.  There is no need here to assume pairwise additivity or spherical
symmetry for the interaction potential.

The time  evolution    of the    $1-$particle   density,  which,    for
macroscopically homogeneous (i.e.  both uniform and isotropic) systems
considered here, is equal to the average density of adsorbed particles
$\rho$, is governed by the following kinetic equation:
\begin{equation}\label{eq:22}
\frac{\partial\rho (t)}{\partial (k_a t)}=\Phi ({\bf 1}^*;t),
\end{equation} 
where $\Phi   ({\bf 1}^*;t)$  is the  probability    of inserting a  new
particle  at the position ${\bf r}_1$  with an orientation {\boldmath $
\theta$\unboldmath}$_1$.   (Note  that because of  the  homogeneity of the
system, one can  average  as well  over  ${\bf r}_1$ and {\boldmath  $
\theta$\unboldmath}$_1$).    Introducing   the  Ursell   (or   ``cluster'')
functions  associated with the  $e({\bf n+1}|\{{\bf 1},{\bf 2},\cdots {\bf
n} \})$'s,
\begin{equation}\label{eq:23}
f_{{\bf n+1}|\{{\bf  1},{\bf  2},\cdots  {\bf n}  \}}=\sum_{m=0}^n(-1)^{n-m}
\sum_{\{ i_1,i_2,\ldots,i_m\} \subset \{1,2,\ldots n \}} e({\bf n+1}|\{{\bf i}_1,{\bf
i}_2,\cdots {\bf i}_m \}),
\end{equation}
where the second  sum in the right-hand  side of the above equation is
over all  distinct   $m$-tuplets  chosen  from $\{1,2,\ldots,  n   \}$, and
recalling that for  sequential adsorption processes there  is a one-to-one
mapping   between  reduced time, $k_a t$,   and  density, $\rho$,  one generalizes
Eq.~(\ref{eq:12}) as follows:
\begin{equation}\label{eq:24}
\Phi({\bf  1}^*;\rho)= \sum_{s=0}^{\infty    }\frac{1}{s!}\int  \cdots\int d{\bf    2}\cdots
d{\bf(s+1)}f_{{\bf 1}|\{{\bf   2},\cdots{\bf(s+1)}\}     } \rho^{(s)}(\{{\bf
2},\cdots {\bf (s+1)} \};\rho),
\end{equation}
where the  integral  involves  integration over   both  center-of-mass
positions  and orientations. For  a pairwise additive potential, it is
easy to   show that the    Ursell function is just   a   product  of
conventional  Mayer functions    $f_{ij}=e^{-\beta v_{ij}}-1$, $f_{{\bf
1}|\{{\bf   2},\cdots{\bf(n+1)}   \}}=\prod_{j=1}^nf_{1(j+1)j} $, and that Eq.~(\ref{eq:24})
reduces  to  Eq.~(\ref{eq:12})   for   simple RSA.   For  higher-order
particle densities, generalizing Eq. (\ref{eq:16})  leads in a similar
way to
\begin{equation}\label{eq:25}
\frac{\partial \rho^{(n)}( {\bf 1},{\bf 2},\cdots{\bf n};\rho )}{\partial 
\rho    }=\frac{1}{\Phi({\bf  1}^*;\rho)} \sum_{i=1}^n\Phi^{(n)}({\bf
1},.,{\bf i-1},{\bf  i}^*, {\bf i+1}, ..{\bf n};\rho ),
\end{equation} 
with
\begin{equation}\label{eq:26}
\Phi^{(n)}({\bf  1}^*,{\bf 2},...{\bf n};\rho )=\sum_{s=0}^\infty \frac{1}{s!}\int
\ldots\int d{\bf (n+1)}\ldots d{\bf (n+s)}f^{\{ {\bf 2},...{\bf n}\} }_{({\bf 1}|\{
{\bf (n+1)}\cdots{\bf (n+s)}\})} \rho^{(s)}({\bf 2},...{\bf (n+s)};\rho),
\end{equation}
where    the  $f^{\{  {\bf    2},\cdots{\bf  n}\}     }_{{\bf 1}|\{   {\bf
(n+1)}\cdots{\bf(n+s)}\} }$'s are generalized Ursell functions defined by
\begin{equation}\label{eq:27}
f^{\{ {\bf 2},\cdots{\bf    n}\} }_{{\bf 1}|\{  {\bf  (n+1)}\cdots{\bf(n+s)}\}
}=\sum_{m=0}^s  (-1)^{s-m} \sum_{ \{  i_1,\ldots   i_m\} \subset \{  n+1,\ldots,n+s\} }
e({\bf 1}|\{ {\bf 2},\cdots{\bf n},{\bf i}_1,\ldots,{\bf i}_m\} )
\end{equation}

Again,    for pairwise additive   potentials,   the generalized Ursell
functions    reduce  to  products     of   Mayer   functions.      (In
section~\ref{sec:Ball-transport-gener}, we   shall  consider a process
with non-pairwise additive potentials.)

The above hierarchy of equations for the $n$-particle densities is the
counterpart for cooperative  sequential  adsorption  processes  of the
Kirkwood-Salsburg hierarchy for equilibrium systems. The latter reads
\begin{eqnarray}\label{eq:28}
\frac{ \rho^{(n)}( {\bf 1},{\bf 2},\cdots{\bf n};\rho )}{ \rho  }
&=&\frac{1}{\Phi({\bf    1}^*;\rho)}   \sum_{i=1}^n\Phi^{(n)}({\bf
1},.,{\bf i-1},  {\bf     i}^*,     {\bf   i+1},    ...,{\bf      n};\rho  ),\nonumber\\
&=&\frac{n\Phi^{(n)}({\bf 1}^*,{\bf 2},\ldots,{\bf n};\rho )}{\Phi({\bf 1}^*;\rho)}
\end{eqnarray}

where the   $\Phi^{(n)}$'s  are given   by the  same  expression  as  in
Eqs.~(\ref{eq:24})     and   (\ref{eq:26}).     The   comparison    of
Eqs.~(\ref{eq:25}) and  (\ref{eq:28}) illustrates again the difference
between CSA and equilibrium systems.

The       Kirkwood-Salsburg     hierarchy     for     CSA   processes,
Eqs.~(\ref{eq:22})-~(\ref{eq:26}), is a    good starting  point    for
generating diagrammatic expansions\cite{TST91a}.   These can be used to
formulate  various   approximate   descriptions,  such  as low-density
expansions for the adsorption probabilities and integral equations for
the pair correlations, that  parallel    those used in     equilibrium
statistical   mechanics of fluids.  In  the  case of pairwise additive
potentials, diagrammatic  expansions are  also conveniently derived by
using    the     ``replica  trick''  developed    for     spin   glass
models\cite{MPV87}. The trick relates the probability distribution for
CSA and RSA configurations of $N$ particles  to the Gibbs distribution
of  a fictitious equilibrium  system with $n_2$  replicas of particle
$2$,  $n_3$ replicas of particle  $3$,\ldots, $n_N$ replicas of particle
$N$, in the limit $n_2\to 0$,$n_3\to 0$,$n_N\to 0$.\cite{G92}

\subsection{Approximation scheme}\label{sec:Approximation-scheme}
\begin{figure}
(a)\resizebox{5cm}{!}{
\includegraphics{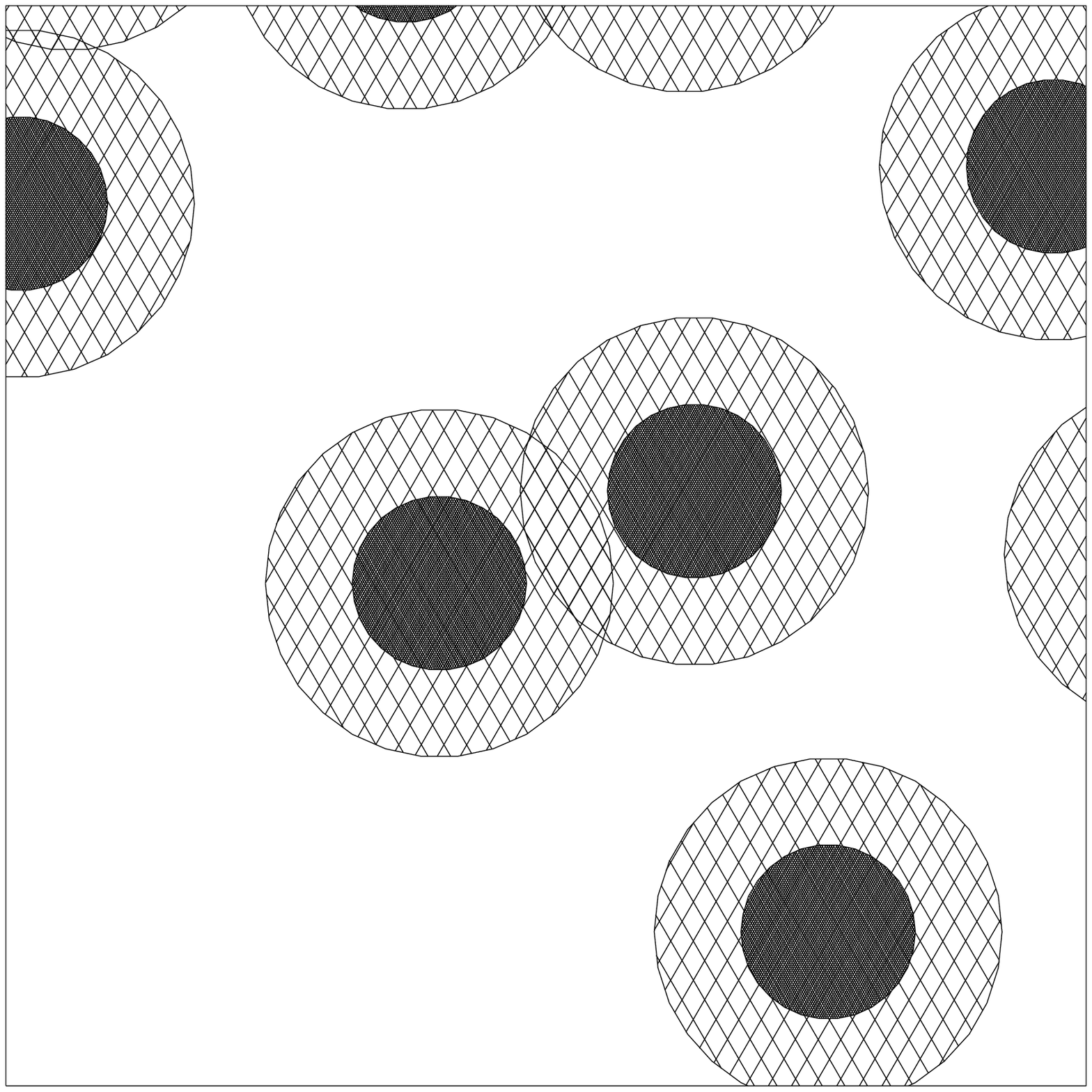}}(b)\resizebox{5cm}{!}{
\includegraphics{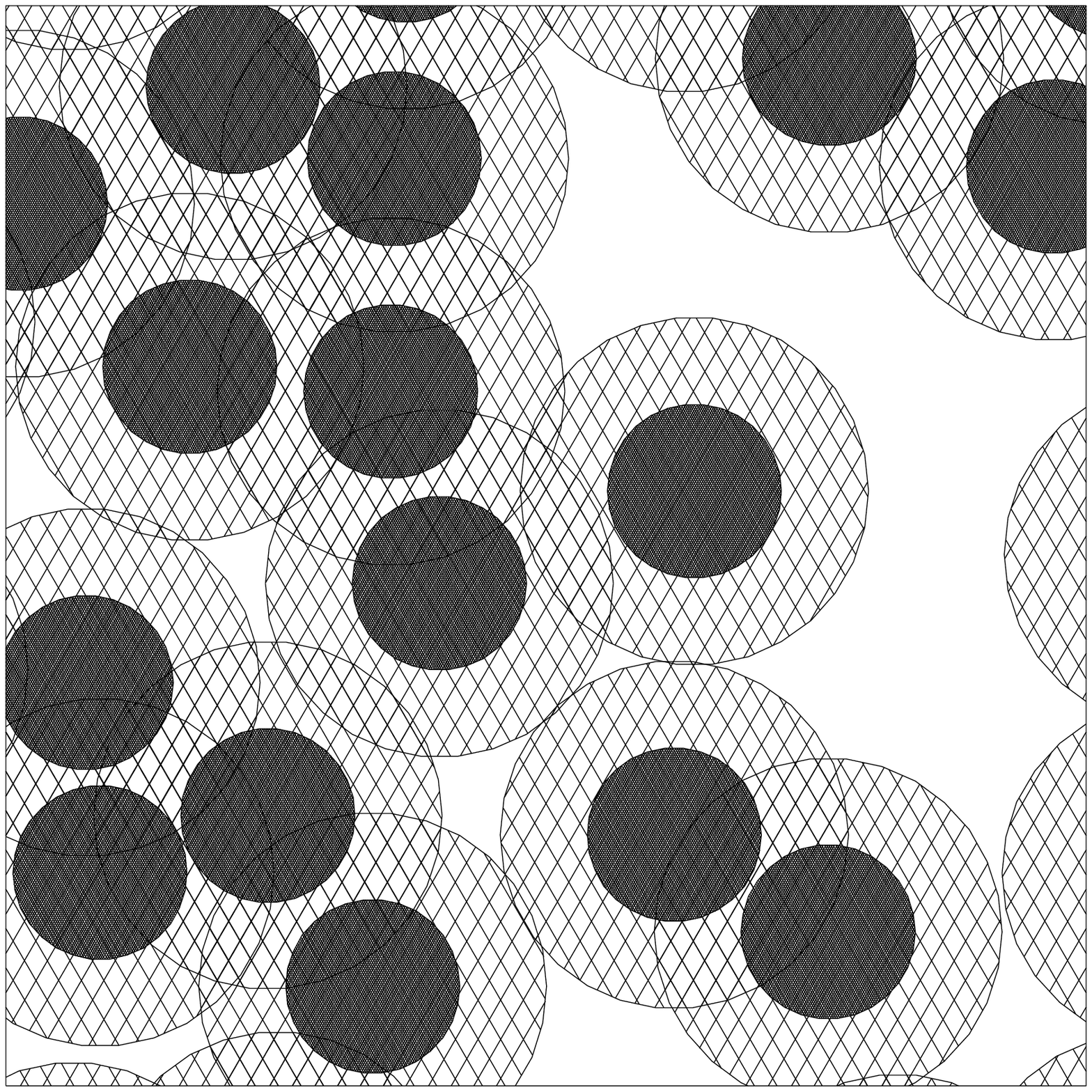}}(c)\resizebox{5cm}{!}{
\includegraphics{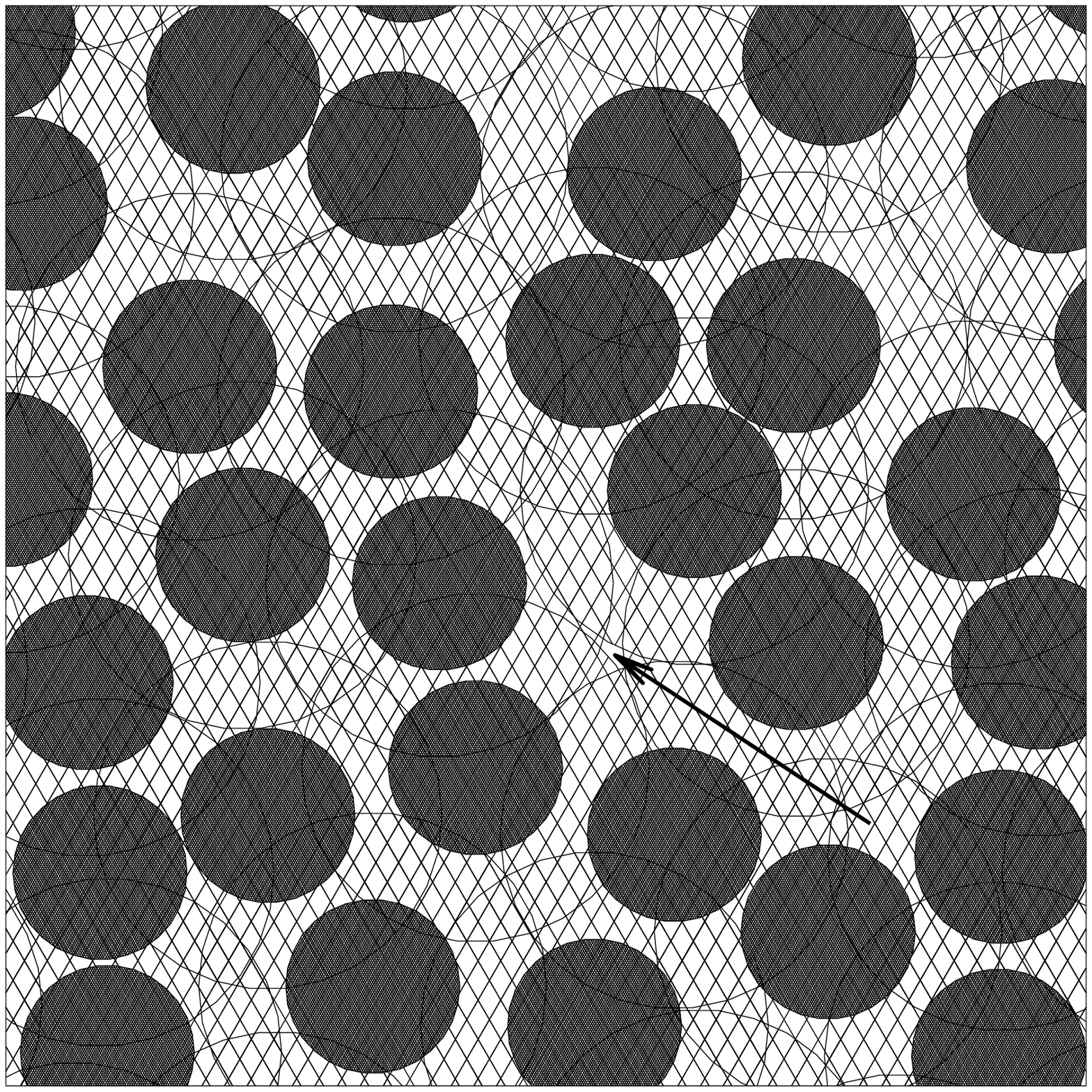}
}
\vspace*{0.5cm}
\caption[99]{ (a) Typical  RSA configurations of hard
disks for  three values of  the coverage (a) $\theta=0.1$,(b) $\theta=0.3$,(c)
$\theta=0.5$. The black disks are the adsorbed hard disks and the hatched region
represents each disk's exclusion area. The remaining white region
corresponds to the  surface that is available for adsorption of  a new
disk. In (c), close to the jamming limit, the remaining available surface
consists only of a small ``target'' indicated by the arrow.}
\label{fig:2}
\end{figure}

There is, unfortunately, no  exact solution to the sequential adsorption
problem when the  dimensionality $D$ of the  substrate is two or more,
(the one-dimensional case will  be discussed below).  To  describe the
kinetics of the   adsorption process  and   of the  structure of   the
adsorbed configuration,  one can use  computer simulation (see below).
In   addition, it is also fruitful   to develop approximate treatments
that allow one to  study a wide range of  control parameters and a variety
of  interactions, external  fields, and  mechanisms.  To do  so, it is
instructive to consider configurations of  hard disks generated by RSA
at  different    stages of  the  process.   This    is  illustrated in
Fig.~\ref{fig:2}a-c.  Also shown in white on the figures is the region
that is available  for the insertion of a  new  disk center and  whose
area gives the adsorption  probability (correspondingly, the region in
gray is that part of the surface which is excluded  to the center of a
new disk, over   and above the  region  covered by the  adsorbed disks
themselves       that      is       shown    in       black,        cf
section~\ref{sec:Two-dimensional-sequ}).   At    low surface coverage,
corresponding to short times, the available surface percolates through
the whole  space and    adsorption is  locally prohibited    by groups
involving only a small number of preadsorbed disks.  When the coverage
increases, the available  surface diminishes and no longer percolates.
Finally, close to the jamming limit (the saturation coverage is $\theta_\infty
\simeq  0.547$), the available surface  reduces to small, isolated regions
that  can accommodate only one   additional disk.  This suggests  that
different approximations must  be  used for describing the   different
stages of the process.

\begin{figure}
\begin{center}
\resizebox{10cm}{!}{\includegraphics{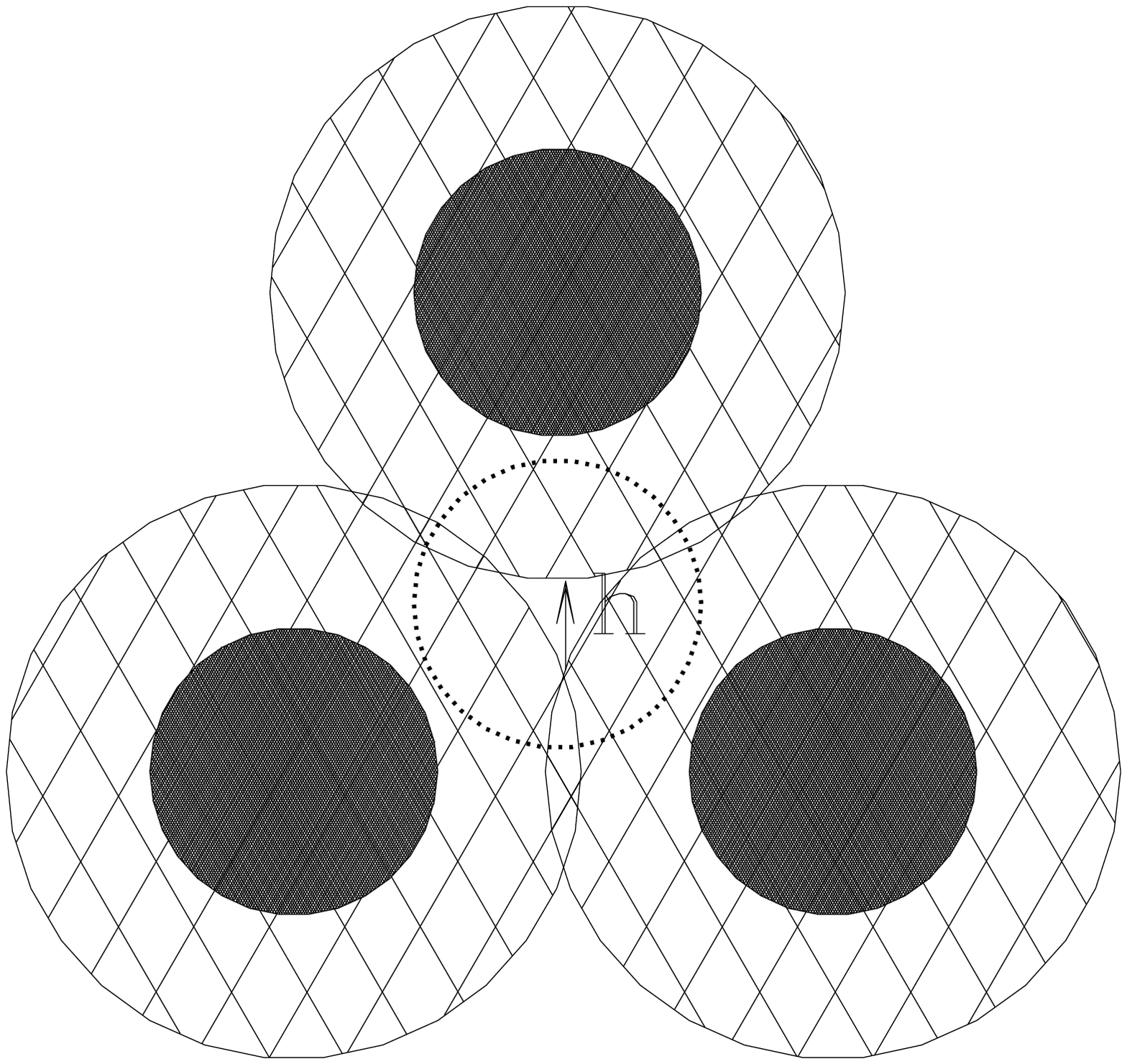}}
\caption[99]{Typical ``target'' in the asymptotic regime of the RSA of 
disks close to   saturation. (see  Fig.~\ref{fig:3}c).  Only a small
surface area characterized  by a  linear length $h$ is available
for adsorption of a new   disk (such a  disk  is shown by the  dotted
circle).} \label{fig:3}
\end{center}

\end{figure}

Consider first  the last-stage, long-time,  regime that corresponds to
the asymptotic approach to the jamming limit.  Note that the existence
of such a  jamming  limit is  a feature  of all sequential  adsorption
processes, be they  random   or cooperative, provided  the   adsorbing
particles  possess a  ``hard  core'' that results  in  an exclusion or
blocking effect.  When the surface coverage is sufficiently high, this
hard-core interaction  dominates   all additional  ``soft'' potentials
describing     for  instance           dispersion   and  electrostatic
interactions. (The case of  ``singular'' additional potentials such as
those encountered  for systems of  sticky disks and in the description
of            generalized      ballistic      deposition          (see
section~\ref{sec:Ball-transport-gener})   is   different).   In    the
asymptotic regime near  the jamming limit,  the surface available  for
adding new  particles is composed, as  shown above,  of small isolated
``targets'',  whose number exactly   provides the number of  particles
that  will still be   adsorbed  before  reaching   saturation.  Due to   the
irreversibility  and the infinite memory  of  the process,  the asymptotic
kinetics  is  simply  related to  the rate  of   disappearance and the
statistics of  the targets.  The   standard arguments for  the RSA  of
spherical  particles  have been put  forward   by Pomeau\cite{P80} and
Swendsen\cite{S81}.  Consider  a  typical target  at time  $t$ that is
characterized  by a small linear scale  $h$ so that the area available
for  the center of a  new  disk goes as   $h^2$ in $2$ dimensions (and
$h^D$ for general dimension $D$) (see Fig.~\ref{fig:3}). The rate of  disappearance of such a
target is  proportional to its  surface   area, hence  to $h^2$; as  a
consequence, the number density  of  targets characterized by a  linear
size $h$, $n(h,t)$ decays exponentially in time according to
\begin{equation}\label{eq:29}
n(h,t)\simeq n(h,t_c)e^{-k_aAh^D(t-t_c)},
\end{equation}
where $t_c$ is a   crossover time that    marks the beginning  of  the
asymptotic regime and  $A$ is a mean   shape factor (besides  $h$, the
geometric characteristics of  the targets are irrelevant).  Making the
reasonable assumption that $n(h,t)$ goes to a nonzero value $n(0,t_c)$
when $h\to 0$,  one  then obtains that  the  difference in  density  of
adsorbed particles between time $t$  and saturation ($t\to\infty$) is given
by
\begin{eqnarray}\label{eq:30}
\rho(\infty)-\rho(t)&=&\int_0^{h_c}dh\,n(h,t)\nonumber\\                     &\sim
&n(0,t_c)\int_0^{h_c}dh\,e^{-k_aAh^Dt}\sim t^{-1/D},
\end{eqnarray}  where
$h_c$ is an (unimportant) upper cutoff.  The approach to saturation is
thus very  slow in RSA  processes and is  described by a  power law in
time  (when  $D=1$,  one recovers  the    exact  $t^{-1}$ result,   cf
section~\ref{sec:car-parking-problem}).   The same  line of  reasoning
permits one  to show that  the pair correlation function at saturation
has a  logarithmic divergence when  two disks (or, more generally, two
$D-$dimensional spherical  particles) approach contact, a  result that
is     also    obtained     exactly     in    one    dimension     (cf
section~\ref{sec:car-parking-problem}).   The  arguments in   terms of
small,  isolated targets  can  be   generalized  to all  CSA and   RSA
processes involving  particles with a  hard core.  The key property is
the rate of  disappearance of targets  when their linear size $h$ goes
to zero.   If this rate  goes to a  nonzero limit, as  in  the case of
generalized ballistic deposition (which corresponds to singular sticky
potentials,  cf    section~\ref{sec:Ball-transport-gener}) and   of  a
mixture   involving point-like particles\cite{TT92},  the  approach to
saturation is fast  and is essentially exponential.   If the rate goes
to zero, generally as $h^{\cal  N}$ where ${\cal N}$  is the number of
degrees of freedom for the adsorbing species,  the approach is again a
power-law, $t^{-1/{\cal N}}$:  for instance, in two dimensions, ${\cal
N}=3$ for  non  spherical  (unoriented) particles\cite{TTS89,VT90}  as
well as for  a continuously polydisperse mixture of  disks\cite{TT91},
whereas ${\cal N}=D=2$ for a  monodisperse system of disks.  Note that
the above results are valid for  irreversible adsorption in which once
adsorbed, the  particles  stay fixed.   A quite  different behavior is
observed when  particles are allowed to desorb   from the surface (see
section~\ref{sec:Part-reversibility}).

In the low and intermediate density regime (Figs.~\ref{fig:2}a and
\ref{fig:2}b),   one expects that  the  approximation  employed in the
statistical mechanical  theory   of equilibrium   fluids can,  {\it mutatis
mutandis},   be useful as  well   for  describing systems of  particles
produced by sequential adsorption. In particular, this is the case for
low-density  expansions,  somewhat analogous  to  the  standard virial
expansion  for the imperfect  gas\cite{HM86}. As mentioned above, such
expansions    can be  generated   either  order  by  order  from the
Kirkwood-Salsburg-like hierarchy or   directly  from the  diagrammatic
series  when   they are  available     (e.g., for  pairwise   additive
potentials). For instance,  the adsorption probability $\Phi_{RSA}(\theta )$
for the RSA of hard disks onto a plane is given by
\begin{equation}\label{eq:31}
\Phi_{RSA}(\theta )=1-4\theta+\frac{6\sqrt{3}}{\pi}\theta^2+\left(
\frac{40\sqrt{3}}{3\pi}-\frac{176}{3\pi^2}\right)\theta^3+\ldots,
\end{equation}
where   $\sigma$ is the hard-disk   diameter and $\theta=(\pi\sigma^2/4)\rho    $ is the
surface coverage.  As noted before (section~\ref{sec:Stat-mech-sequ}),
this  expansion coincides with   its equilibrium counterpart at  order
$\theta^2$, but departs from it at the next order (in general, $\Phi_{RSA} <
\Phi_{EQ}$).  Similar formulas have  been derived for  the adsorption of
hard   convex   objects of  various  shapes,   for  mixtures,  and for
sequential  processes   involving  cooperative effects  and  competing
mechanisms (see below).  For  cases involving ``soft'' (i.e continuous
in the configurational  variables) interaction potentials in  addition
to hard cores, the computation of the coefficients is always numerical
since it requires   more than geometric considerations.   However, one
can  again   fruitfully borrow ideas  that have    proven efficient in
equilibrium statistical  mechanics\cite{HM86}:     weak     attractive
potentials can be  treated as perturbations and spherically-symmetric
soft repulsive potentials can be approximately replaced by a hard-core
potential with  a suitably chosen  effective diameter that  depends on
both density and temperature\cite{ASZB94,AW96}.

Even    for   hard-core   exclusion     interactions, calculating  the
coefficients  rapidly  becomes  intractable  as  the  order in density
increases. For hard disks, the fourth-order term has been obtained and
for  parallel   hard squares,   the  sixth-order  term,  but these are
exceptions, and in practice only coefficients up to $\rho^3$ (or $\theta^3$)
have been  calculated analytically.   This unfortunately restricts the
range of density over which the truncated expansion is accurate. It is
possible, however,   to enhance  the   convergence of  the low-density
expansion by  building interpolation   formulas that  incorporate  the
known asymptotic behavior.   For instance, in the  case of the RSA  of
hard disks, the  result  $\rho(\infty)- \rho(t)\sim t^{-1/2}$  converts, through
the  use of the  kinetic  equation $\partial\rho/\partial t=\Phi(\rho(t))$, to $\Phi(\rho)\sim
(\rho(\infty)-\rho)^3$ when $\rho \to \rho(\infty)$.   By employing standard techniques
such as Pad{\'e} approximants\cite{E89,DWJ91,BHM93}, e.g.,
\begin{equation}\label{eq:32}
\Phi(\rho)=\frac{(\rho(\infty)-\rho)^3}{\sum_{n=0}^p a_n\rho^n}
\end{equation}
where  the $(p+1)$ coefficients $a_n$ are  determined  by expanding in
coverage (or   density)  and matching   term  by  term to   the  exact
expansion, Eq.~(\ref{eq:31}),  one obtains an  improved description of
the kinetics over the whole process.   In particular, expressions such
as Eq.~(\ref{eq:32})  provide estimates   of the  saturation  coverage
$\theta(\infty)$.    The  procedure  works  very  well  for  the   RSA of hard
$D-$dimensional spheres  and  $D-$dimensional parallel hard  cubes: cf
Table I. For more  complicated shapes and more complicated cooperative
processes (see  below), the results are  fair, but not  as good.  This
trend  can usually  be rationalized  by invoking the  appearance of an
additional  regime in the  process  (see,  for  instance, the  case of
elongated nonspherical particles with random orientations discussed in
section~\ref{sec:RSA-anis-part}).

\begin{figure}
\begin{center}
\resizebox{16cm}{!}{\includegraphics{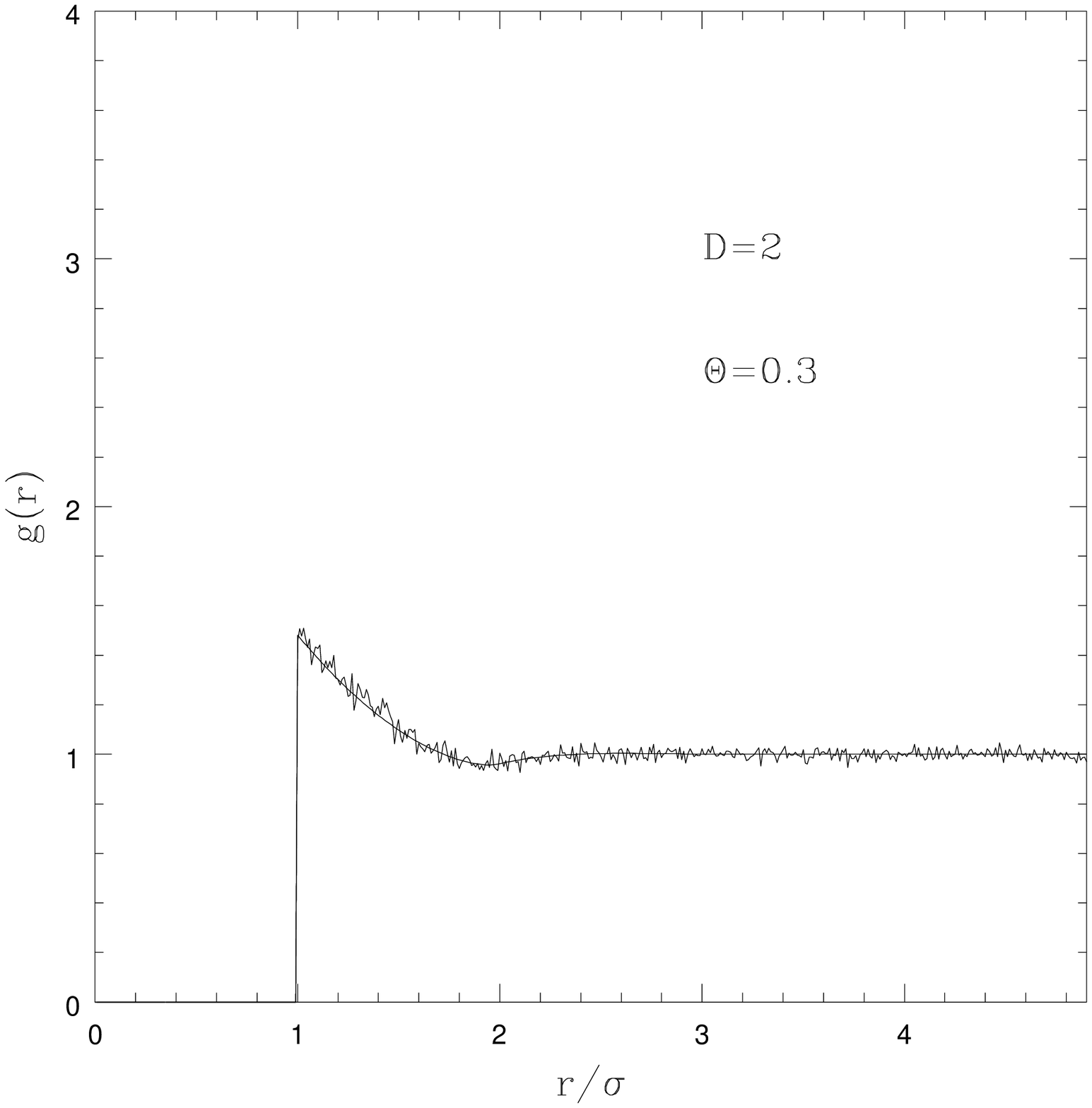}
\includegraphics{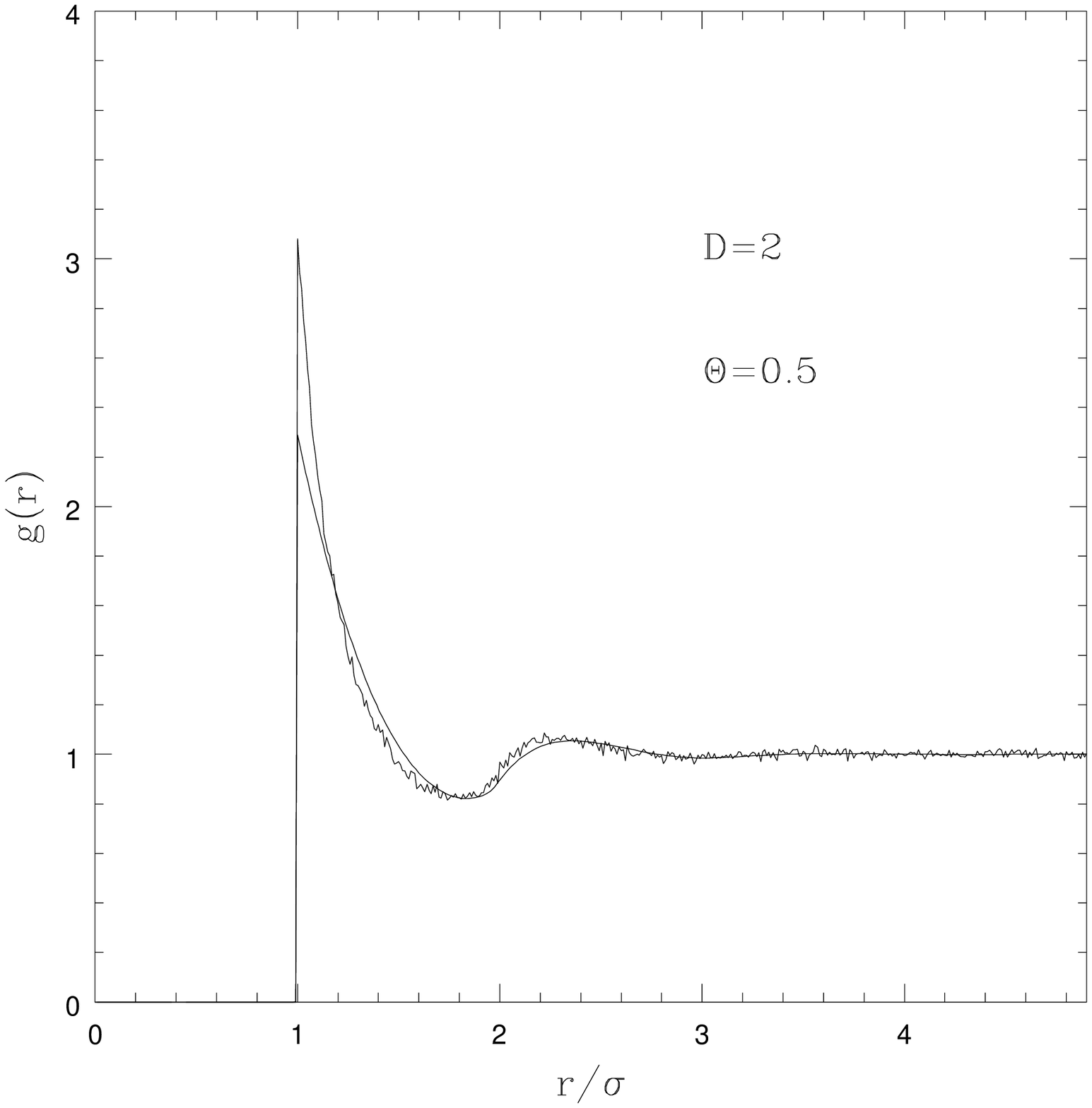}}
\caption[99]{Pair correlation function $g(r)$   for the  RSA
configurations of  hard  disks for  two   values of the  coverage:  (a)
$\theta=0.3$  (b) $\theta=0.5$.  The   continuous  line  is computed  from  the
Percus-Yevick-like  integral  equation for  RSA,   and  the wavy  line
represents the simulation result. } \label{fig:4}
\end{center}

\end{figure}

The approximation scheme just described focuses on the kinetics of the
process at the level of the density of  adsorbed particles (or surface
coverage).  However, valuable insight  on  the adsorption process   is
gained by    a proper description  of  the  spatial correlations among
adsorbed particles.  Actually, studying  the structure of the adsorbed
layer can facilitate the discrimination between different hypothesized
model         descriptions           of       real          adsorption
situations\cite{F80b,ASZB94,WSSZV93}.      In  this   regard,  it   is
interesting to  develop approximate treatments  for obtaining the pair
correlation function   in configurations   of particles  generated  by
sequential adsorption.  As  in equilibrium statistical  mechanics, the
most obvious  route  is  to derive  integral  equations  for  the pair
correlations, and this  can be done  by  making approximations  in the
diagrammatic   expansions,   as  obtained   for  instance   from  the
Kirkwood-Salsburg-like hierarchy presented above.   As an example, the
analog   of the  celebrated    Percus-Yevick  integral  equation   for
$D-$dimensional equilibrium   hard-sphere  fluids\cite{HM86}  has been
studied   for  the RSA  case\cite{BTVT95}.   The result   for the pair
correlation  function  for  an    adsorbed layer  of  hard  disks   is
illustrated in   Fig~\ref{fig:4}.     The  description  is    good  at
intermediate surface coverage,  but, as  expected, it deteriorates  in
the  asymptotic regime  near saturation;  the  approximation  fails to
reproduce     the     logarithmic    divergence    at  contact    (see
section~\ref{sec:Two-dimensional-sequ}),  but   it    captures  the
rapid,  super-exponential  decay  of  the   spatial correlations  with
distance      that  is     typical     of     sequential    adsorption
processes\cite{E93,BBV94}.

\subsection{Usefulness of one-dimensional  models}\label{sec:Usefulness-1-D}

The   vast  majority of  real    situations in  which large  particles
(proteins, cells, colloids,\ldots)  adsorb   at a  liquid/solid  interface
correspond to  a     two-dimensional substrate (that,   for   sake  of
simplicity, we consider  flat and uniform  in this whole review).  Why
therefore  should one   study  one-dimensional   models?   Equilibrium
statistical mechanics seems  to teach us  that one-dimensional systems
have  very little  relevance   to physical two-  and three-dimensional
fluids; indeed, no thermodynamic phase transitions are possible in one
dimension whereas   they  are a key  phenomenon  of  real fluids.  The
situation is, however,  different for sequential adsorption processes.
First, as already stressed, there is no thermodynamic phase transition
in  the adsorbed layer,  regardless of its dimensionality. Second, the
main features of the phenomenology of sequential adsorption processes,
such as the existence  of a jamming limit  and of a well-characterized
asymptotic approach toward saturation, the short (super-exponentially)
decay of  the spatial correlations  among adsorbed particles, and many
other properties that may depend upon the specific process under study
(see the following sections), are independent of space dimensionality.
One-dimensional    sequential  adsorption   models   already  have the
ingredients,  partial or   total  irreversibility with the  associated
memory  effect,  exclusion  or  blocking  phenomena,  that  make their
behavior nontrivial  and  qualitatively similar  to that  observed  in
higher   dimensions.  There are  of  course quantitative changes.  For
instance,  a sequential adsorption process  with  hard objects is less
and   less efficient as  the  dimension  increases.   (For the RSA  of
spherical particles,  $\theta(\infty)\simeq 0.75$ for  $D=1$,  $\theta(\infty)\simeq 0.55$ for
$D=2$, $\theta(\infty)\simeq 0.38$ for $D=3$, etc; it  is, however, interesting to
note   that,  as a rule  of   thumb,  the saturation   coverage in $D$
dimensions is rather well approximated by that in one dimension raised
to power $D$\cite{P60}).

An additional motivation for  studying one-dimensional models is that they
are  often amenable to  exact  analytical solutions, as illustrated by
the    case       of     the     car      parking      problem    (see
section~\ref{sec:car-parking-problem}).     The  underlying reason for
solvability is a {\it shielding property}\cite{E93}, that is generally found
only   in one   dimension.  The shielding property in one dimension means that
an empty interval of a minimum, finite  length separates the substrate
into  two disconnected regions such  that  adsorption in one region is
not   affected  by the configuration   of     particles in the   other
region\cite{E93}. As a result, a closed system involving only a finite
number of  kinetic  equations  can be   written  and,  often,  solved.
Further examples will be  given in sections \ref{sec:Bulk-transport-issu}
and
\ref{sec:Surf-events:conf}.

\subsection{Monte Carlo algorithms}\label{sec:Monte-Carlo-algor}

Computer   simulation provides  additional information  on  sequential
adsorption processes.   It was from  simulation results, for instance,
that   Feder first observed the  characteristic  power-law kinetics at
long   times\cite{F80a}.  The    basic procedure for    a  Monte Carlo
simulation of a sequential adsorption process with hard-core particles
is the following:  at each step,  one attempts to  place a particle in
the  simulation cell with  periodic  boundary conditions, at positions
sampled    from a uniform  probability  distribution.   Because of the
finite range, hard-core  interaction between particles, it is
advantageous to employ a grid   construction\cite{HFJ86}. The
computer time then depends  on the number of  particles instead of the
square of  the number.    Moreover, close  to the  jamming  limit  the
process is dominated  by very few  successful events of adsorption and
accelerating procedures have  been introduced.  Indeed, at long times,
the  available surface for  inserting    new particles is  very  small
compared to  the simulation cell.  If  the grid cell element is chosen
for accommodating only one particle, the  algorithm can be improved as
follows: first, an unoccupied grid  element is chosen  at random and a
second  random number determines  the trial position within the vacant
grid  element;  the  time is   then   incremented by  an   amount  $\Delta
t=N_{total}/N_{vacant}$ where $N_{total}$ is the  total number of grid
elements in  the simulation cell and $N_{vacant}$  is number of vacant
grid elements at   a given density\cite{VTV97,RTV94}.  (The  procedure
accelerates significantly the simulation, but it may occur that in the
last stage of the  process, the surface which is  the sum of the  grid
elements remaining vacant   exceeds largely the  available surface for
inserting new particles and many unsuccessful attempts still slow down
the simulation).  Event-driven  algorithms have  been used in  various
adsorption  models (RSA of nonspherical particles\cite{RTV94}, polymer
adsorption\cite{W98}).   Moreover,  in some cases, e.g.,  in one
dimension,   the available surface can    be determined exactly, and a
trial position can be always  chosen within the available surface such
that  no rejection (or wasted time)  occurs along the simulation.  The
procedure  leads  to a  correct   result  provided that   the time  is
incremented by the ratio of the system size  to the available surface.
It  can   also be used   in  the  simulation   of processes  involving
desorption (see Section \ref{sec:Parking-lot-model}).

\section{Adsorption of nonspherical particles and mixtures}\label{sec:Basic-extensions-RSA}
\subsection{RSA of anisotropic particles}
\label{sec:RSA-anis-part}
Biological  molecules have in  general a  nonspherical  shape and they
adsorb in an  irreversible way when the area  of the surface they have
in contact with the  substrate is maximum. Experimentally, Schaaf {\it
et. al}\cite{SDJS92} found that  the maximum coverage of the substrate
they   were  able  to obtain   in   the adsorption  of   fibrinogen (a
nonspherical protein with aspect   ratio of approximately  $7.5$)  was
around $40\%$,  i.e.  less than  the saturation  coverage predicted by
the RSA   of  hard disks ($\simeq   55  \%$) and observed  in  experiments
involving  fairly spherical globular proteins\cite{F80b,F80a,R93b}. (A
similar effect was  observed for adsorption of albumin\cite{MVBBS91}.)
A simple way to account for these  effects consists in considering the
irreversible adsorption  of anisotropic particles  onto a surface with
the RSA rules generalized so  that the positions and the  orientations
of   the    adsorbing particles   are  chosen   randomly  from uniform
distributions       (for      generalizations,       see       Section
\ref{sec:Surf-events:conf}.              We    and               other
authors\cite{TTS89,VT90,VZ89,VZ90,TVRT91,TV91,RTTV92,VTRT92,RTTV94}
have performed this study for  hard convex bodies of different shapes:
ellipses, squares,  disco-rectangles (i.e.  ``$2D$  spherocylinders'').
The relevant questions include the following:

\begin{enumerate}
\item What is the influence of the  aspect ratio on the saturation
coverage of the substrate?

\item What is the dependence on the particle shape of the kinetics
 (short and long times)?

\item What are the similarities and differences   between the  RSA generated
configurations and  equilibrium  configurations  at the   same surface
coverage?
\end{enumerate}

\begin{figure}
\begin{center}
\resizebox{10cm}{!}{\includegraphics{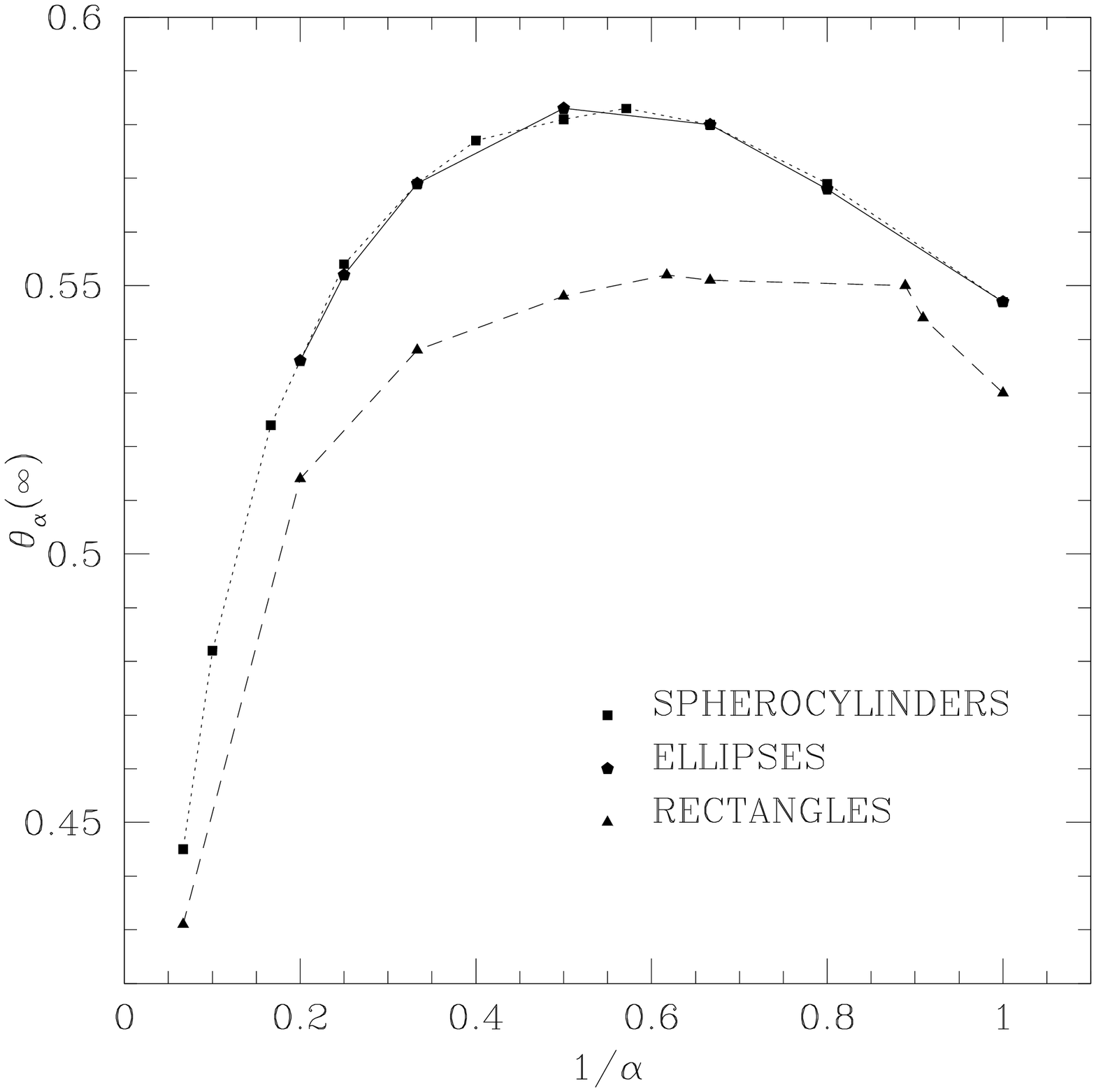}}
			    \end{center}

\caption[99]{RSA of rectangles, ellipses
and disco-rectangles onto a plane: saturation coverage, $\theta_{\alpha}(\infty)$,
as a function of the inverse of the aspect  ratio $\alpha$ (for $1\leq \alpha \leq
15$). The lines are drawn for visual guidance.  }
\label{fig:5}
\end{figure}

The   hard convex bodies   mentioned   above are characterized  by one
parameter which   is   the aspect  ratio,  $\alpha $:   for ellipses   and
disco-rectangles, $\alpha$ is the ratio of the long axis to the short axis
and for  rectangles is the  ratio  of the length to   the width.  Both
disco-rectangles and ellipses reduce  to a disk  when $\alpha \to 1$, while
rectangles tend to a square (i.e.   an anisotropic object) in the same
limit.  Fig.~\ref{fig:5} displays the  saturation coverage obtained by
computer simulation   of  two-dimensional RSA  as  a  function  of the
inverse  aspect ratio,  $1/\alpha   $,  for various shapes.    Three  main
conclusions can be drawn from the results: (i) for the three different
shapes, the maximum coverage is obtained for a value of $\alpha $ close to
2;  (ii) the maximum coverage  for ellipses  and disco-rectangles (for
$\alpha \simeq 2$) is larger than  that for disks;  and (iii), for all objects
considered here, the saturation coverage  goes to zero when the aspect
ratio $\alpha$ goes  to infinity, seemingly  with a power law  dependence.
It   is also interesting to note   that  for $\alpha\sim  7$-$8$ (typical of
fibrinogen,  see above), the     saturation coverage falls  below  the
hard-disk value and is close to $40\%$.

As far as  the kinetics  are  concerned,  computer  simulation also
showed that the approach  toward saturation is slower  for anisotropic
objects  than it is for disks\cite{TTS89}.   From a geometric analysis
of the available surface, Talbot {et  al.}\cite{TTS89} showed that the
long-time kinetics results  from adsorption of  particles in two types
of targets, the larger  ones being independent  of the orientation  of
the   trial   particle (nonselective targets)  and    the smaller ones
corresponding to the  insertion of a particle  with  a small  range of
possible orientations (selective  targets).  To minimize size  effects
while  still   retaining the effect   of  anisotropy,  Viot and Tarjus
\cite{VT90}  performed  extensive simulations   of the RSA  of  unoriented
squares,  and the results clearly  showed that the asymptotic approach
follows a power law  with an exponent equal  to $1/3$.  For  ellipses
and  disco-rectangles with  moderate and   large  aspect ratios,   the
long-time  kinetics  are described  by the   same  power law,  but the
crossover time beyond which the  asymptotic kinetics sets in increases
with $\alpha   $.  For weakly  elongated  objects, by introducing  a small
anisotropy  parameter, $\epsilon=\alpha-1$, it can  be  shown that the long-time
kinetics follows the equation\cite{VTRT92}
\begin{equation}\label{eq:33}
\rho(\infty)-\rho(t)\sim         \epsilon                \left(\frac{\int_0^z         dx
e^{-x^3}}{z^{1/3}}+Ce^{-z}{z^{1/2}}\right)
\end{equation}
where $z=\epsilon^2 t$ and $C$ is a positive constant. Eq.~(\ref{eq:33}) can
be simplified  in  the two  limits $z <   <  1$ and  $z >   > 1$. The
long-time    kinetics thus consists of two  consecutive
asymptotic regimes.  First,  for $1/\sqrt{\epsilon }> >   t> >1$, one has  a
$t^{-1/2}$  behavior and for  $t> >1/\sqrt{\epsilon  }$ this regime vanishes
and a $t^{-1/3}$ behavior emerges.  This last  regime disappears and the
contribution to the coverage goes  to zero when  the aspect ratio goes
to one. Simulations have been performed for weakly elongated objects and only
an effective      exponent    between $1/2$  and      $1/3$   has been
measured\cite{VTRT92}.  (To observe numerically  a clear  separation of
these two critical regimes,  simulations  of very slightly   elongated
particles should be performed.)

The low-coverage expansion of the available surface function $\Phi$
permits, as shown in section~\ref{sec:Stat-mech-sequ}, a systematic
description for low and, somewhat intermediate coverages. To third order, one
has
\begin{equation}
\Phi=1-B_2\theta+\left(2B_2^2-\frac{3}{2}B_3\right)\theta^2+[3B_2B_3-\frac{4}{3}
(B_2^3+B_4)-B'_4]\theta^3 
\end{equation}
where     $B_i$    denotes     the    $i$th     equilibrium     virial
coefficient\cite{HM86}  and   $B'_4$ is  the  specific RSA coefficient
which can be  expressed in terms of  Mayer diagrams\cite{RTTV92}.  All
coefficients  must be  calculated numerically in general\cite{TVRT91},
except the second  virial coefficient for  which an analytical formula
is known\cite{B75}.  The comparison with simulation results shows that
the third-order expansion provides a good  description as long as $\Phi$
is not too small.  It is worth noting that  the range of usefulness of
the  density  expansion  diminishes   when the  aspect  ratio  $\alpha   $
increases.  Even at  low  coverage, $\Phi$ decreases markedly  when $\alpha$
increases, which means that  adsorption  of very elongated   particles
rapidly becomes difficult.  An intermediate  regime appears before the
asymptotic regime, in  which the longest linear  size of the particles
is responsible for many rejections of trial  particles in the process.
As  displayed in Fig.~\ref{fig:1}, this regime  leads to the formation
of a local order whose (linear) domain size is given  by the length of
the particle long axis.  The feature  becomes more pronounced when the
aspect ratio becomes larger.  This   regime cannot be described by   a
low-density expansion  whose validity (or  quality) range is  of order
$1/ \alpha $.  (Notice that for equilibrium systems, a virial expansion is
also unable   to   describe  the  isotropic-nematic  transition.)   To
understand  the    emergence of this  third regime,    it is useful to
consider the adsorption    of infinitely  thin   needles whose  length
corresponds to  that  of the long axis   of  elongated particles.  For
needles, which do not have a proper area, no saturation is reached and
the     number  density of    needles  increases   algebraically  with
time\cite{TV91}.   By  studying the   one-dimensional problem,  two of
us\cite{TV91} showed that  the RSA of needles   can be mapped at  long
times    onto    a      solvable     two     variable    fragmentation
model\cite{TV91,BTV95} and that   the  density of   needles  increases
asymptotically as
\begin{equation}\label{eq:34}
\rho (t) \sim t^{\sqrt{2}-1}.
\end{equation}   
The same algebraic  law is expected  to  hold in two  dimension, as is
confirmed by computer simulations\cite{ZV91,Z91}.

Following           the       procedure          described          in
section~\ref{sec:Approximation-scheme},  approximation schemes can  be
constructed   by  combining   low-density expansions   and  asymptotic
regimes. For weakly and moderately  elongated particles  ( $\alpha \leq 5$),
a fit of  the  form  of Eq.~(\ref{eq:32})   gives results  which   are
indistinguishable  from the simulation data  during  most of the process.
However, the predicted saturation coverage is overestimated by the fitting
formula\cite{RTTV92}.

For very elongated particles,   a more efficient  approximation scheme
can be built.  From the discussion above, it  can be assumed  that the
filling  process is  mainly  composed  of  two   critical regimes:   a
needle-like  regime followed by the true  asymptotic regime. Matching the
two behaviors leads to the following result,
\begin{equation}\label{eq:35}
\theta_\infty(\alpha) \sim \alpha^{-1/(1+2\sqrt{2})},
\end{equation}
and the crossover time between the two regimes is 
\begin{equation}\label{eq:36}
t_c\sim \alpha^{3/(1+2\sqrt{2})}
\end{equation}
when $\alpha \to \infty  $. These two  results are well supported by simulations:
 for elongated rectangles,  Vigil and Ziff\cite{VZ89,Z91}  obtained
an algebraic law  $\theta_\infty(\alpha) \sim 1/\alpha^{0.20-0.22}$, whereas  the  value of the
exact exponent  of Eq.~(\ref{eq:35}) is  $-0.2612\ldots$.  The increase in
crossover  time has  also been  reported  but not   quantitatively
estimated\cite{VZ90}.

A structural analysis  of the configurations of anisotropic  particles
generated   by    RSA has been    performed   by considering  the pair
correlation   function\cite{RTTV94}.  For  anisotropic hard-core  particles  in two
dimensions, the pair  correlation   depends on three  variables:   the
distance  between the two particles $r$  and the orientation $\theta_i$ of
each    particle   with  respect  the    interparticle   vector.   The
center-to-center pair  correlation function $g_{0,0}(r)$,  which is the
average  over   the   orientations of  two   particles,   i.e.  has no
orientational dependence,  already provides useful information  on the
structure.  In particular, the highest  peak in $g_{0,0}(r)$ is for  a
distance $r$  close to $1/2(1+\alpha)$  when $\alpha  \leq 5$,  whereas the peak
shifts towards $r=1$ with a secondary weak peak emerging for distances
close to $\alpha$  when $\alpha \geq 5$ (all  distances are measured in units of
the short  axis). These results show that  the most probable relative
orientation of two particles  evolves  from perpendicular to  parallel
when   the   anisotropy parameter     becomes  larger.   Although   the
center-to-center  pair  correlation  function   is always   finite  at
contact,     one   observes a      logarithmic    divergence  of   the
surface-to-surface      pair     correlation       function         at
saturation\cite{RTTV94}.  This is reminiscent of the behavior observed
in the RSA  of spherical particles. As already stressed, nematic order 
is never observed, regardless of the aspect ratio.

\subsection{RSA of mixtures}

For a general mixture of $n$ components of arbitrarily hard particles
adsorbing irreversibly on a planar surface, one would like to be
able to predict the time dependent and saturation coverages. Even
in one-dimension, this is a challenging problem. However, 
one can deduce the behavior in some limiting cases. 

The RSA of a binary mixture of hard disks is characterized by two
parameters:
$\kappa=k_A/k_B$, where $k_A$ and $k_B$ are the adsorption rates
of the components on an empty surface, and the diameter ratio 
$\lambda=\sigma_A/\sigma_B$.
Talbot and Schaaf \cite{TS90} 
analyzed the case of
a binary mixture of hard disks of greatly differing diameters, 
$\lambda<<1$.
The time-dependent
coverages can be obtained from the numerical solution of two coupled
first-order differential equations,
\begin{equation}\label{eq:37}
\frac{\partial \theta_A}{\partial t}=\kappa\lambda^2\phi(\theta^{\rm eff}_A)
(1-x_B\theta_B)
\end{equation}
\begin{equation}\label{eq:38}
\frac{\partial \theta_B}{\partial t}=\phi_{BB}(\theta_A,\theta_B)\exp\biggl[
\frac{-x_A}{1-x_B\theta_B}\biggr]
\end{equation}
where $x_A=(1+\lambda^{-1})^2, x_B=(1+\lambda)^2$
and $\theta^{\rm eff}_A=\theta_A/[1-x_B\theta_B]$ is the
effective coverage of the smaller disks. As long as $\lambda$
is sufficiently small, $\phi_{BB}$ depends only on $\theta_B$
and the large disks approach
their saturation coverage exponentially. The small disks then behave
essentially as a one-component system in a reduced area, 
approaching their jamming limit coverage according to the
usual algebraic power law, $t^{-1/2}$\cite{TS90}. The final combined
saturation coverage is accurately given by
\begin{equation}\label{eq:39}
\theta_{A+B}(\infty)=\theta_B(\infty)+(1-x_B\theta_B(\infty))\theta_{\infty}
\end{equation}
where $\theta_{\infty}=0.547$.

Tarjus and  Talbot  examined the  asymptotic approach   to the jamming
limit of  a    continuous polydisperse   mixture  \cite{TT91}.     The
adsorption rate of  a  particle of radius $\sigma$  on  an empty  surface,
$K(\sigma)$,  is assumed to be   continuous for $\sigma_1\leq\sigma\leq\sigma_2$ and  zero
otherwise.          As     in        the    monocomponent       system
(section~\ref{sec:Approximation-scheme}),  the asymptotic kinetics are
determined by the  filling of isolated target  areas. If $K(\sigma_1)$  is
different  from  zero then  it   follows that   the jamming  limit  is
approached as
\begin{equation}\label{eq:40}
 \rho_{\infty}-\rho(t)\sim t^{-1/3}
\end{equation} 
This is consistent with  the idea that the  exponent is the inverse of
the  number  of  degrees  of freedom  of   the adsorbing species:  two
translational, plus one corresponding  to the  continuous distribution
of  particle diameters  (see also  section~\ref{sec:RSA-anis-part} for
anisotropic particles). Note that if $K(\sigma_1)=0$  then the exponent is
determined by the  order of the first  non-vanishing derivative of $K$
at $\sigma_1$ \cite{TT91}.

Meakin   and Jullien \cite{MJ92a,MJ92b}  conducted computer simulation
studies of a binary mixture of large and small disks (and spheres), as
well  as   continuous      mixtures   with   uniform   and    gaussian
distributions. Their results confirmed the theoretical analyses.

\section{Bulk transport issues}\label{sec:Bulk-transport-issu}
In the simple RSA model the position of the trial particle is selected
from  a  uniform,   random  distribution.   While  this  simple choice
facilitates  both simulation and  analytic solutions in come cases, it
is not clear that it is always consistent with the transport mechanism
of the  particle from the   bulk to  the  vicinity  of the  adsorption
surface.

An adsorbing particle is in general subject to Brownian, gravitational
and  hydrodynamic  forces, as well as  specific  interactions with the
adsorbing   surface and the pre-adsorbed   particles.  Examples of the
latter include van  der     Waals,  electrostatic and    short   range
repulsions.

A substantial  body of work has addressed  the issue  of the effect of
the transport mechanism of  the adsorbing particles on  the adsorption
kinetics  and  the   resulting structure  of  the deposited   particle
configurations. In  particular,   two limiting cases of   interest are
Diffusion   Random  Sequential Adsorption  (DRSA)  and   the Ballistic
Deposition (BD). In the former  the transport mechanism is pure
diffusion, while in the latter gravitational forces are dominant. For
spherical particles, one can define a dimensionless gravity number
(proportional to the P{\'e}clet number)

\begin{equation}\label{eq:41}
N_G = \frac{4\pi \sigma^4 \Delta\rho g}{3k_BT}
\end{equation}
where  $\Delta\rho$ is  the  difference  between the particle   and solution
mass density, $g$ the acceleration due to the gravity, $\sigma $ is the
particle diameter, $k_B$ is the Boltzmann constant, and T is the temperature.
The limits $N_G\to\infty$  and  $N_G\to 0$  correspond to  the  BD and  DRSA
models, respectively.

\subsection{Bulk diffusion }\label{sec:Bulk-diff}

Schaaf {\it et al.} analyzed the effect of diffusion on the asymptotic
kinetics
\cite{SJT91}. They found that the presence of bulk diffusion modifies the
asymptotic kinetics,
\begin{equation}\label{eq:42}
\rho_{\infty}-\rho(t)\sim t^{-2/3}.
\end{equation}
The saturation coverage is approached more rapidly  than in simple RSA
due to a  funneling effect of the  pre-adsorbed particles  forming the
target.     Yet,     this  power  law     has    never   been observed
experimentally.  This may   result from  the   presence  of hydrodynamic
interactions between the diffusing  particles and the particles on the
surface,  interactions that are neglected in the diffusion
RSA  model and that   may modify the  kinetics  and  the structure  of  the
adsorbed layer\cite{BSVS93,PR94,WA98}.

Senger {\it et  al}  \cite{SVSJST91,SSVJST92}  examined the  kinetics  and saturation
coverage of the DRSA  process  by simulating   the  random walk  of  a
spherical particle of radius  $R$ on  a  cubic lattice with  a lattice
parameter $a$. Each trajectory starts from a randomly selected lattice
site in a plane at height $z=3R$.  A particle is adsorbed, and remains
permanently fixed,  once its  center reaches the  plane  $z=R$. If the
particle reaches the plane  $z=5R$ it is considered  lost to the  bulk
and  a new  trajectory  is  initiated.    At each  Brownian step,  the
particle  moves  to    one  of six   neighboring     sites with  equal
probability. If the selected displacement results in an overlap with a
preadsorbed particle, the particle is returned to its initial position
and a new direction is chosen.

The  principal  result of  the  simulations  was  that  to within  the
statistical error, the coverage and structure of the adsorbed particle
configurations generated by a  DRSA process are  identical to those of
the  simple RSA model.  In order  to gain insight into this unexpected
result,  various one-dimensional models  were studied.  In particular,
the generalized parking  process  proposed  by two  of   us\cite{TV92}
proved useful in understanding the effect of the transport mechanism.

In  $(1+1)$ dimensions, the  kinetic equation describing the generalized
parking process is
\begin{equation}\label{eq:43}
\frac{\partial G(h,t)}{\partial t}=-k_a(h)(h-\sigma )G(h,t)+2\int_{h+\sigma }^{\infty}
k_a(h')G(h',t)dh'
\end{equation}
Where the rate of adsorption  per unit length in a  gap of size $h$ is
denoted by $k_a(h)$. In simple RSA,  $k_a(h)=k_aH(h-\sigma )$ where $H(x)$ is the
Heaviside unit step function.

In the  generalized  parking process,  the   rate of deposition  of  a
particle  in a gap depends on   the width of the   gap, but is uniform
within the  gap.    It has been shown   that  all  generalized parking
processes have the same jamming limit coverage.  Here we reproduce the
argument given  by Bafaluy et   al.  \cite{BCST94}. The  problem is to
determine  the  number, $N_{\infty}(h)$ of   particles adsorbed  on a line
segment  of length   $h$ after an   infinite time.   Insertion of  one
particle into this gap   leads to two  new  gaps  of length  $h'$  and
$h-h'-\sigma $ so one has the following recursion formula:
\begin{equation}\label{eq:44}
N_{\infty}(h)=1+2\int_0^{h-\sigma }N_{\infty}(h')P(h,h')dh',
\end{equation}
where  $P(h,h')$ is the probability that  insertion of a disk into the
gap of length $h$ produces gaps of length  $h'$ and $h-h'-\sigma $. In all
generalized parking processes, including simple RSA,
\begin{equation}\label{eq:45}
P(h,h')=(h-1)^{-1}
\end{equation}
since,  by  definition, the trial   particle arrives randomly and {\it
uniformly}  in the gap. Thus,  the final state of  the system does not
depend  on  $k(h)$ and  we   conclude  that  all  generalized  parking
processes have  the same jamming  limit. The kinetics, of  course, can
vary greatly depending on the form of $k(h)$.

The story is not  yet  complete, however,  since a careful  simulation
study revealed that the saturation coverage  of 1D DRSA \cite{STSSV93}
at $0.7529$ is slightly  larger than that  of 1D RSA ($0.7476\ldots$).
This  difference is a result of  the non-uniform flux  of particles at
the   surface.   Eq.~(\ref{eq:43}) can   be  modified  to  allow for  a
non-uniform distribution of incoming particles:

\begin{equation}\label{eq:46}
\frac{\partial G(h,t)}{\partial t}=-k(h)G(h,t)+2\int_{h+\sigma }^{\infty}
k(h',h)G(h',t)dh'
\end{equation}
where $k(h)$ is the total rate that gaps of length $h$ are destroyed
by  addition of a  new particle and  $k(h',h)$ is  the probability per
unit length per unit time that deposition of a disk in a gap of length
$h'$ produces gaps  of length $h$  and $h'-h-\sigma$.  One can also obtain
information about  the jammed state without  solving Eq.~(\ref{eq:46})
directly with an  approach similar to that  described earlier for  the
generalized (uniform)   parking process. It is  possible  to obtain an
analytic expression  for the   flux of  particles   on a line  segment
bounded by two preadsorbed disks \cite{BCST94} and the result obtained
for the  jamming coverage is  equal to the  simulated value within the
confidence interval.   More recent work  has examined situations where
both gravity and and diffusion play a role\cite{CT98}.

\subsection{Ballistic transport and generalizations}\label{sec:Ball-transport-gener}

\begin{figure}
\begin{center}
\resizebox{14cm}{!}{\includegraphics{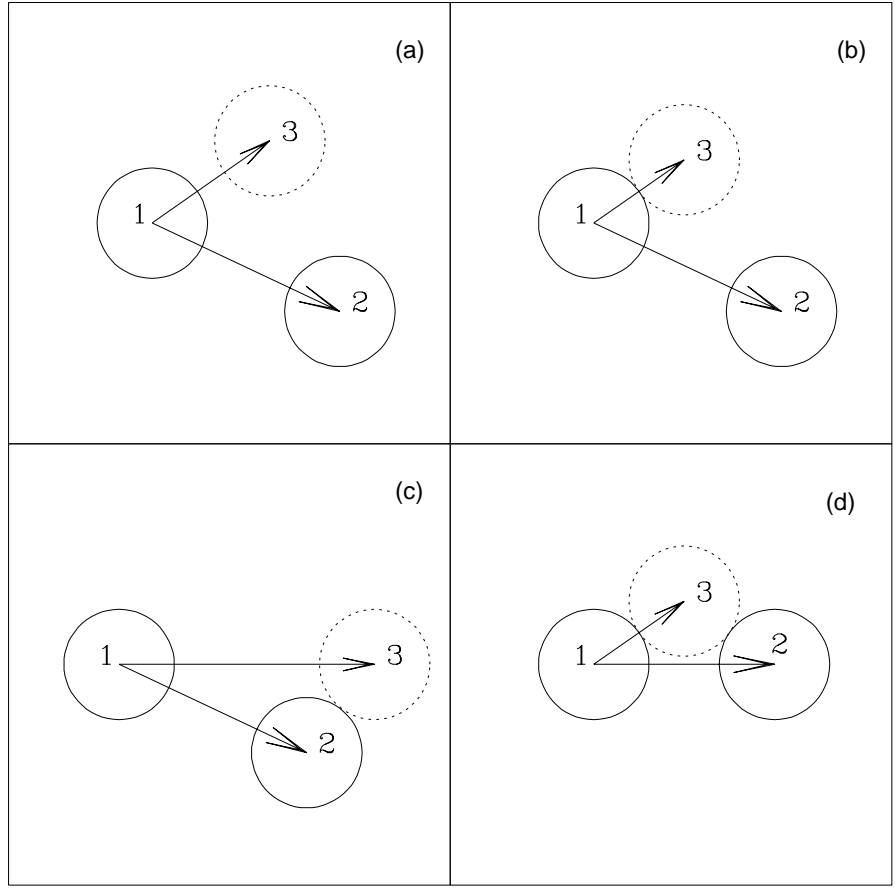}}

\caption[99]{Generalized ballistic deposition:  schematic illustration
of the  four possibilities for inserting  particle $3$ in the vicinity
of two previously adsorbed particles $1$ and $2$: (a) corresponds to a
direct deposition, (b) and   (c)  correspond  to a deposition    after
rolling on  one  preadsorbed sphere,  and (d)   to a deposition  after
rolling on two spheres. This process can be  described by a  CSA of disks with
 effective rates of adsorption (see Eq.~(\ref{eq:87})). }
\label{fig:6}
\end{center}
\end{figure}

The ballistic  deposition  model  describes situations  in  which  the
transport is  dominated by   gravitational effects, corresponding   to
large values of   $N_G$   (Eq.~(\ref{eq:41})).  Adsorbing  particles  follow   linear
trajectories from the bulk to  the adsorption surface. If the particle
arrives at  an unoccupied  region of  the surface, it  is  immediately
accepted and remains permanently fixed in place. If  it should land on
top of a pre-adsorbed particle, it is not immediately rejected as with
simple  RSA.  Instead, it follows a  path of steepest descent over the
pre-adsorbed particles.  If it reaches   the surface, it is  accepted,
while if   it remains in   an elevated position  it is  rejected.  The
possible trajectories for 2+1D BD are shown in Fig.~\ref{fig:6}.

\subsubsection{$1+1$-dimensional models}
Like simple RSA, the model is exactly soluble  in one-dimension with a
gap-density approach \cite{TR92}.  This  a consequence of the  already
mentioned             shielding            property               (see
section~\ref{sec:Usefulness-1-D}).  Specifically, the kinetic equation
is
\begin{equation}\label{eq:47}
{\partial G(h,t)  \over \partial (k_at)}  = -H(h-\sigma)(h+\sigma )G(h,t)  + 2G(h+\sigma ,t) +
2\int_{h+\sigma }^{\infty}G(h',t)dh'\;\;,
\end{equation}
where $k_a$ is the rate of adsorption of disks of size $\sigma$.  A gap of
length $h$ will be destroyed if  the center of  an incoming disk falls
anywhere on an interval  $h+\sigma $  centered on the  gap of  length $h$.
Conversely, a gap of length $h$ may be created by the impact of a disk
on either of the adsorbed  disks bounding a gap of  length $h+\sigma $, or
by direct deposition  in  a gap of  length $h'>h+\sigma  $.  To solve  the
kinetic equations, one  sets $G(h,t) = e^{-k_a(h+\sigma )t}F(k_a\sigma t)/\sigma^2
$ for $h\geq \sigma $,  which leads to  a solvable differential equation for
$F(t)$. $G(h,t)$ for  $h   < \sigma $ can be found using Eq.~(\ref{eq:47}). By
employing Eq.~(\ref{eq:4}),    the  solution for   the  time-dependent
density is then obtained as
\begin{equation}\label{eq:48}
\rho(t)=\frac{1}{\sigma }\int_{0}^{k_a\sigma t}du(1+2u)e^{2(1-u-e^{-u})}\frac{F(u)}{u^2}
\end{equation}
where $F(t)$ is given by Eq.~(\ref{eq:7}).  As expected, the saturation
coverage, $\rho(\infty)\sigma =0.80865..$ is greater than  that of the RSA process.
Moreover, the saturation coverage is approached exponentially:
\begin{equation}\label{eq:49}
\rho(\infty)-\rho(t)\sim\left( \frac{2 e^{2(1-\gamma)}}{\sigma}\right) \frac{e^{-2k_a\sigma t}}{(k_a\sigma t)}
\end{equation}
where $\gamma$ is    the Euler  constant.  The saturation   state  is then
obtained   more rapidly than  the   corresponding  RSA processes  (see
section~\ref{sec:car-parking-problem}). Another distinct feature of BD
is the presence of connected clusters of  particles, which result from
the rolling mechanism.

Viot  et al. introduced a  model that generalizes both RSA
and BD \cite{VTT93}. A disk  that arrives on  an unoccupied surface is
accepted  with   probability $p$.   Otherwise, if    it alights  on  a
pre-adsorbed  particle, it follows  the  path of steepest  descent. If
this disk reaches the surface,  it is accepted with probability $1-p$,
otherwise  it is rejected. RSA  is recovered when $a=p/(1-p)=0$, while
$a=1$ corresponds  to the BD   model.  As $a\to\infty$  only deposition via
rolling   is   permitted.  This  limit    corresponds to  an Eden-type
off-lattice ballistic aggregation model.

In $(1+1)$ dimensions the kinetic equations describing the process are
\begin{equation}\label{eq:50}
\frac{\partial{G(h,t)}}{\partial (k_at)}= -H(h-\sigma)(h-\sigma +2a\sigma) G(h,t)+ 2aG(h+\sigma ,t)
+ 2 \int_{h+\sigma }^{\infty} dh' G(h',t),
\end{equation}
and they   may be solved analytically. The
number density is
\begin{equation}\label{eq:51}
\rho(t)=\frac{1}{\sigma }\int_{0}^{k_a\sigma t}du(1+2au)e^{2a(1-u-e^{-u})}\frac{F(u)}{u^2}
\end{equation}
For $k_a\sigma t>>1/a$, the saturated density is approached as
\begin{equation}\label{eq:52}
\rho(\infty )-\rho(t)\sim 2 e^{2(a-\gamma)}\frac{e^{-2k_a\sigma a t}}{k_a\sigma^2t},
\end{equation}
showing that saturation  is  approached exponentially for  $a>0$ while
the  usual power law is  recovered  for $a=0$.  For  small, but finite
values of $a$,  an intermediate critical regime for  $ 1< <k_a\sigma t< <a$
is present in which the number density increases like $1/t$; the final
exponential regime  has a  contribution that  vanishes  when  $a\to 0$.
(This absence of      discontinuity, when the  tuning   parameter  $a$
decreases to zero, is similar to  the situation encountered in the RSA
of anisotropic  particles when the anisotropy  parameter goes to zero:
see section~\ref{sec:RSA-anis-part}.)

\begin{figure}
\begin{center}

\resizebox{16cm}{!}{\includegraphics{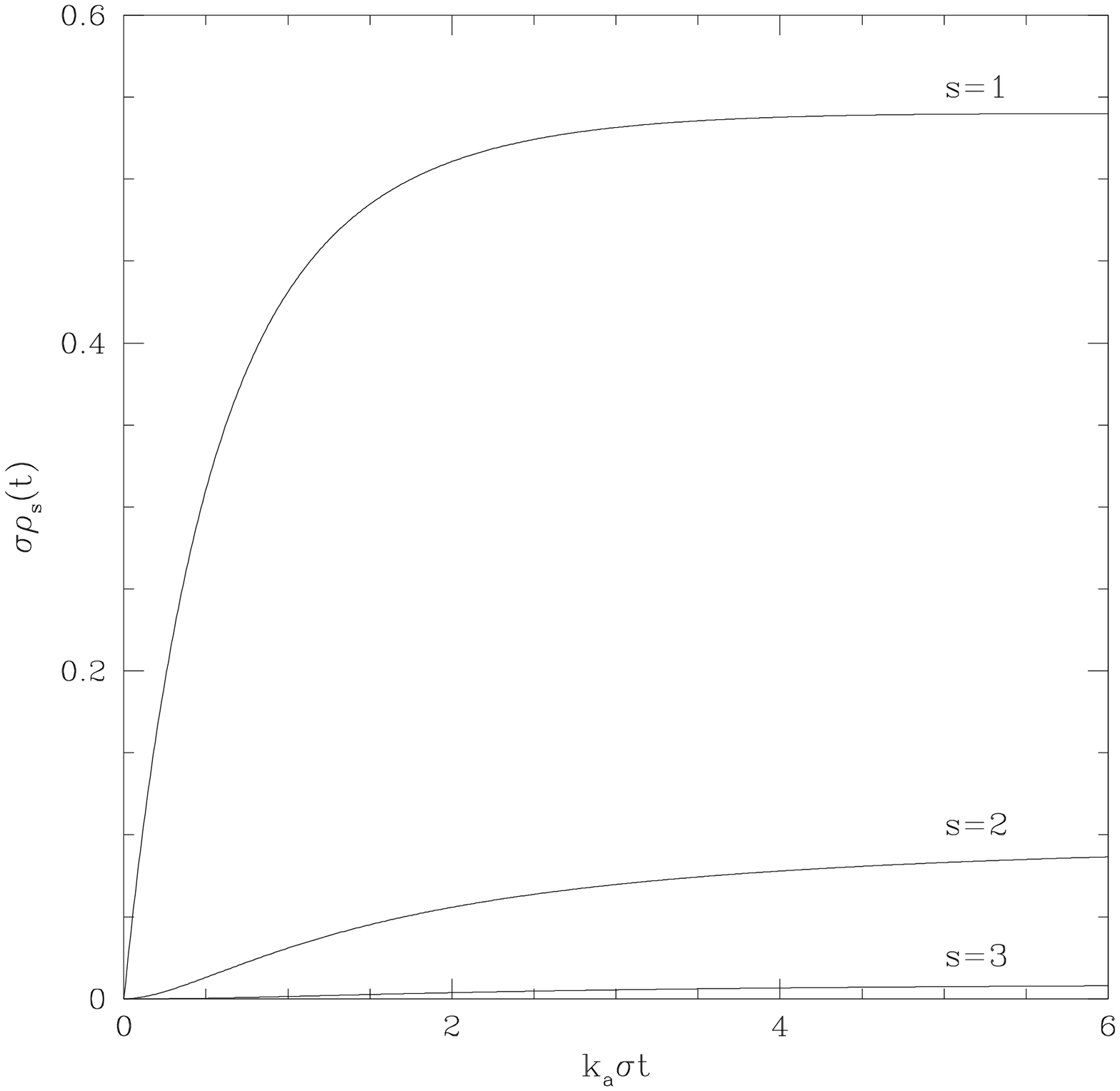}
\includegraphics{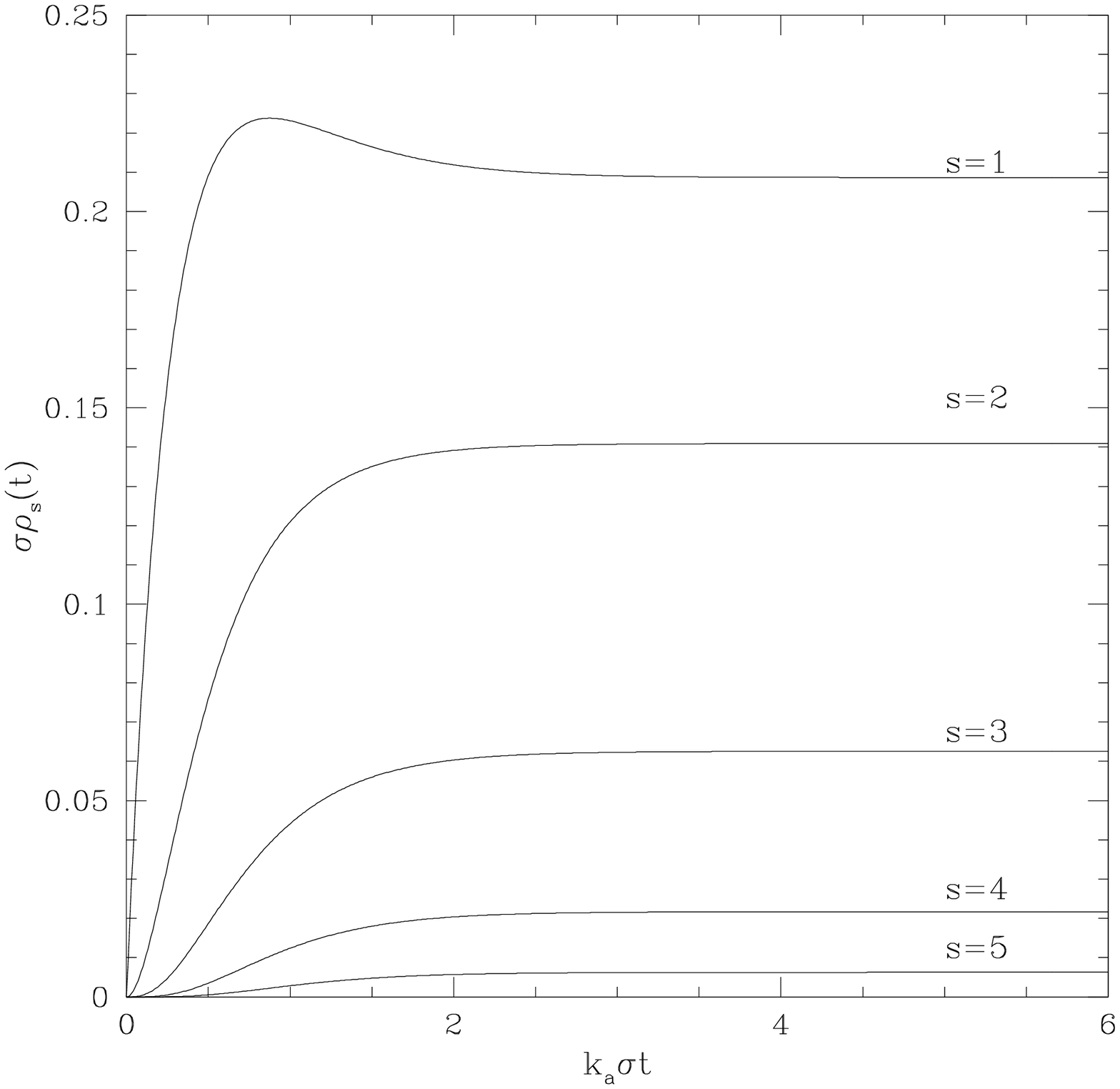}}
\resizebox{10cm}{!}{
\includegraphics{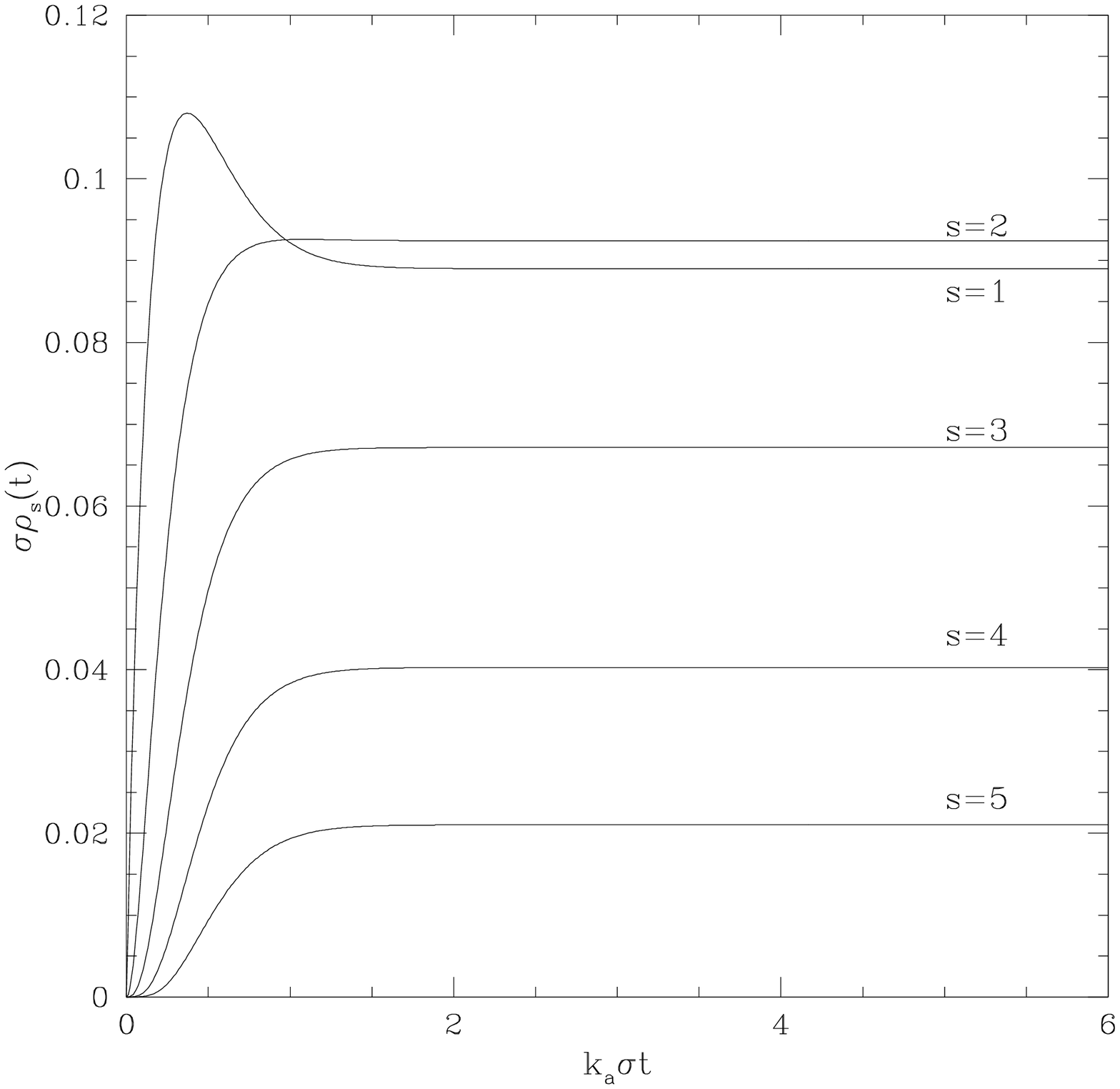}
}

\caption[99]{Density of clusters formed by $s$ particles at contact
$\rho_s(t)$
 as a function of time
for different  values of $a>0$  and  $s$.  (a) $a=0.1$,  $s=1,2,3$ (b)
$a=1$, $s=1,2,3,4,5$, (c) $a=3$,  $s=1,2,3,4,5$. Note that the density
of monomers $\rho_1(t)$  has a maximum   at a finite  value of  time for
$a=1$ and  $3$ but not for  $a=0.1$. For $a=3$,  the efficiency of the
rolling mechanism leads to a density of dimers larger than the density
of monomers at the jamming limit.}\label{fig:7}
\end{center}

\end{figure}

The rolling mechanism leads to the  formation of connected clusters of
particles of different  sizes, the   distribution  of which is    very
sensitive to the value of the tuning  parameter $a$. Some examples are
shown in Fig.~\ref{fig:7}. The presence of clusters has a signature in
the pair  correlation function which  has an infinity of singularities
for each integral multiple  of $\sigma$\cite{BTV96}. The amplitude of
these    singularities  decreases      super-exponentially   with   the
distance. Nevertheless, no  percolation transition is expected  in one
dimension for finite values of $a$.

\subsubsection{$2+1$-dimensional models}
The behavior of the model in (2+1) dimensions is qualitatively similar
to the (1+1)D version\cite{JM92,CTTV93}: a
saturation coverage  of 0.611  (higher  than  the   RSA value)  and   an
exponential approach to saturation.
As discussed in section
\ref{sec:Usefulness-1-D}, the irreversible nature of the process
allows us to  determine he asymptotic  behavior, to leading order, by
investigating the filling of    targets. Characterizing  the
targets  by the the surplus area relative  to the  minimum value $s=S-S_m$,
where $S_m$ is  the smallest  target defined   by  three preadsorbed spheres,  the
density $n(s,t)$  of  such targets, when $s\to0$, evolves in the  asymptotic regime,
$t\geq t_c$, according to
\begin{equation}\label{eq:53}
\frac{\partial }{\partial t}n(s,t)=-ak_a\frac{4}{\pi}(S_m+s)n(s,t),
\end{equation}
where one  neglects the (much less efficient) filling  by direct deposition.  One finds that,
as    in  the $(1+1)-$dimensional model,   the   coverage is approached
exponentially\cite{CTTV93}:
\begin{equation}\label{eq:54}
\theta(a,\infty)-\theta(a,t)\sim \frac{e^{-ak_a\sigma^22\frac{\sqrt{3}}{\pi} t}}{(ak_a\sigma t)^2}
\end{equation}

Using the methodology outlined in section~\ref{sec:Stat-mech-sequ} and
\ref{sec:Approximation-scheme}, one  can   also  perform  density  (or
coverage) expansions.   The generalized ballistic deposition model can
indeed  be   shown to  be  equivalent  to  a   purely  two-dimensional
cooperative sequential   adsorption model of hard    disks in which an
incoming disk interacts with   preadsorbed disks through  effective
potentials   involving  up to $5-$body  irreducible interactions.  For
instance,  the  $2-$  and $3-$body normalized    adsorption rates (see
Eq.~(\ref{eq:19})) are expressed as
\begin{eqnarray}
e({\bf r}_1|{\bf r}_2)&=&H(r_{12}-\sigma)+a\delta(r_{12}-\sigma)\\\label{eq:87}
 e({\bf  r}_1|\{ {\bf r}_2,  {\bf  r}_3\}  )&=&e({\bf  r}_1|{\bf  r}_2)e({\bf r}_1|{\bf
r}_3)\nonumber\\
&+&\frac{a}{2}\delta(r_{12}-\sigma)\delta(r_{13}-\sigma)\left(\sin(\alpha_{1,23})-\frac{a}{2}\right)
\end{eqnarray}
where $H(r_{12}-\sigma)$ is  the Heaviside step-function that describes  a
hard-disk  interaction, $ \delta(r_{12}-\sigma)$ is   a  Dirac function  that
describes  a sticky-disk interaction\cite{B68}  and $\alpha_{1,23}$ is the
angle between  the   vectors ${\bf r}_{12}$   and ${\bf  r}_{13}$ (see
Fig.~\ref{fig:6}).   Using   then     the  Kirkwood-Salsburg-like hierarchy
described in  Eqs.~(\ref{eq:22})-~(\ref{eq:27}), expanding in density,
and calculating analytically the   various geometric factors,  one can
obtain the rate   of adsorption $ \Phi(a,\theta)$   to the  third  order in
coverage:
\begin{eqnarray}\label{eq:55}
\Phi(a,\theta)&=&1+4(a-1)\theta+\left[\frac{6\sqrt{3}}{\pi}+
a\left(\frac{16}{3}-\frac{10\sqrt{3}}{\pi}\right)
+a^2\left(-\frac{16}{3}+4\frac{\sqrt{3}}{\pi}\right)\right]\theta^2
\nonumber\\
&+&\left[\left(\frac{40\sqrt{3}}{3\pi}-\frac{176}{3\pi^2}\right)
+a\left(\frac{376}{27}-\frac{488\sqrt{3}}{9\pi}+\frac{1834}{9\pi^2}\right)
+a^2\left(-\frac{920}{27}+\frac{532\sqrt{3}}{9\pi}-\frac{228}{\pi^2}\right)
+\right .
\nonumber\\ & + & \left.
a^3\left(\frac{496}{27}-\frac{200\sqrt{3}}{9\pi}+\frac{24}{\pi^2}\right)
\right]\theta^3 +\cdots
\end{eqnarray}

Eq.~(\ref{eq:55}) reduces  to  that of RSA and  of  BD  for $a=0$  and
$a=1$,\cite{TG92} respectively.    In  the  latter  case,  the  the    first- and
second-order  terms of the expansion  vanish, expressing the inability
of   one  or  two pre-adsorbed  particles   to prevent an  incoming
particle  from reaching the   surface. 

Note finally that  although configurations  of particles generated  by
the simple   ballistic deposition model   ($a=1$) do  not  contain any
spanning clusters, above a threshold value of approximately $a_c=3.05$
there is a percolation transition  as  the surface coverage  increases
and one   can construct  a   percolation phase  diagram  of  the model
\cite{CTTV95}.  We also showed that  the critical exponents $\beta,\gamma,\nu$
are consistent with those of ordinary 2D lattice percolation.

The ballistic  deposition model and  its generalization  are useful to
describe the adsorption (or deposition)  of particles that are  denser
than the   solvent   and  are   therefore subject  to    gravitational
forces.  This is often the  case  for colloidal particles. Wojtaszczyk
{\it et al.}\cite{WSSZV93}    in  their  experimental study  of    the
deposition  of   melamine particles on   mica  surface showed that the
saturation  coverage  $\theta_\infty $ is   close to  $0.61$,  and that  it is
reached with  a fast kinetics. These  features are describable  by the
two-dimensional ballistic deposition model $(a=1)$, and the connection
can even  be made more  precise   by comparing the  experimental  pair
correlation function  in the deposited layer   with the predicted pair
correlation function of the  ballistic deposition model\cite{W95} (see
Fig~\ref{fig:8}).  For particles not as   dense as melamine particles,
saturation coverages intermediate  between  the RSA ($\theta_\infty  \simeq 0.55$)
and  BD ($\theta_\infty \simeq 0.61$) values are obtained\cite{W95}.  Recently,
Cs{\'u}cs and  Ramsden\cite{CR98} have  shown  that the  adsorption of bee
venom phospholiphase $A_2$ on a  planar metal surface can be described
by this model with $a=0.47$.

\section{Surface events: conformational and orientational changes }\label{sec:Surf-events:conf}

In the  standard RSA model,  adsorbing particles are assumed to remain
indefinitely in  their  initial adsorbed  state.   This  feature is in
contrast with  many experimental  observations of non-spherical and/or
flexible   proteins and other    biomolecules whose adsorbed state may
change    over  time.   For  example,   the   highly elongated protein
fibrinogen (aspect  ratio  of $7.5$)  is known to  adsorb  initially in an
end-on orientation and, following some time,  convert to a more stable
side-on orientation \cite{SDS87}.  Other proteins, upon adsorbing to a
surface, exhibit changes in  secondary  or tertiary  conformation.   A
number of early and more recent experimental studies have investigated
these post-adsorption transitions for  various proteins.  In  previous
articles,       we     review     these      findings     in      some
detail\cite{BTTVV94,VVTT94,VTTV96,VTV97,VGRTVT98}.    Here we   simply
summarize the emerging picture of the transition process:

\begin{enumerate}
\item Post-adsorption transitions in conformation and orientation do occur 
frequently in protein adsorption systems.

\item  The transition usually leads to a larger contact region, thus decreasing 
the probability that incoming proteins land on unoccupied surface.

\item  The transition usually leads to stronger surface-binding, thus decreasing the rate 
of desorption.
\item  The transition tends to be disfavored when the surface is crowded, 
probably due to steric blocking by neighboring proteins.
\end{enumerate}

It is clear that the  presence of a post-adsorption transition affects
the  kinetics of  the  adsorption  process  and vice-versa.   A  truly
universal adsorption  model  should incorporate the possibility   of a
post-adsorption transition,  should account for the above experimental
observations, and should predict, in  addition to the total  adsorbate
surface density, the  fraction of molecules in  an altered  state.  In
the   remainder of  this  section, we    review  recent work aimed  at
developing   RSA-like   models that  account    for a  post-adsorption
transition.  The  basic idea  is  that particles first  adsorb to  the
surface  as   in RSA  and then  may  change  size   subject to certain
geometric exclusion rules.  The change in  size represents a change in
conformation   or  orientation.  We   also   highlight  the  phenomena
displayed  by these new models and the link to  experimental  results.

\subsection{One-dimensional models}\label{sec:One-dimensional-mode}

\begin{figure}
\resizebox{14cm}{!}{\includegraphics{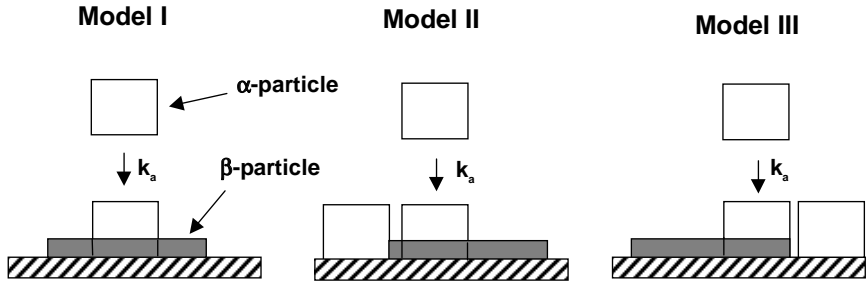}}
\caption[99]{Illustration of a two-state transition on the
substrate. The particle is a hard rod of length $\sigma_\alpha $ when it reaches the 
surface. (a) Model I: if the neighbors are sufficiently far, the
particle grows symmetrically up to a length $\sigma_\beta  $, (b)  Model 
II: growth can  also occur close to a neighboring particle, but the
spreading is asymmetric (c) Model III: the particle adsorbs side on
and tilts to the left or the right if space is available.
}\label{fig:9}
 
\end{figure}

Recently, three  one-dimensional models of  irreversible adsorption with
subsequent transition have been introduced \cite{BTTVV94}.  In each of
these,  particles  are modeled  as  line  segments of  initial  length
$\sigma_\alpha$ that deposit randomly and  sequentially onto an infinite  line
at a rate $k_a$.  Once placed, the particles may spread immediately to
a larger   size   $\sigma_\beta$.  One of   these (Model   I) incorporates  a
symmetric  spread to the larger particle   length.  Another (Model II)
accounts  for an asymmetric spread due  to contacting another particle
during the spread.   The third (Model III)  accounts for a tilting  of
the   particle  to    either   side    following  adsorption.     (See
Fig.~\ref{fig:9}).  In each case, the transition  occurs only if space
allows    (i.e.  the presence   of  other   particles   may block  the
transition).  If  the transition is  assumed  to occur instantaneously
following  adsorption, these   models   become  exactly  solvable   by
introducing the gap density function $G(h,t)$, defined so that $G(h,t) dh$
is the density of empty line segments of length between $h$ and $h+dh$ at
time $t$.  A
kinetic equation can be written  for the time  evolution of $G(h,t)$ for
each of these models as follows.

\begin{equation}\label{eq:56}
\frac{\partial{G(h,t)}}{\partial (k_at)} = -(h-\sigma_\alpha)G(h,t)+
2\int_{h+\sigma_\alpha}^{h+(\sigma_\alpha+\sigma_\beta)/2}dh'G(h',t)+
2\int_{h+\sigma_\beta}^{+\infty}dh'G(h',t)
\hspace*{0.5cm}\rm Model \hspace*{.2cm}I
\end{equation}

\begin{equation}\label{eq:57}
\frac{\partial{G(h,t)}}{\partial (k_at)} = -(h-\sigma_\alpha)G(h,t)+
2\int_{h+\sigma_\beta }^{+\infty}dh'G(h',t)+ (\sigma_\beta-\sigma_\alpha)G(h+\sigma_\beta,t)
\hspace*{0.5cm}\rm Model\hspace*{.2cm}II
\end{equation}

\begin{equation}\label{eq:58}
\frac{\partial{G(h,t)}}{\partial (k_at)} = -(h-\sigma_\alpha)G(h,t)+
3\int_{h+\sigma_\beta }^{h+2\sigma_\beta-\sigma_\alpha}dh'G(h',t)+
2\int_{h+2\sigma_\beta-\sigma_\alpha}^{+\infty}dh'G(h',t)
\hspace*{0.5cm} \rm Model\hspace*{.2cm}III
\end{equation}

In each  of   these, the  first  term on  the right  accounts  for the
destruction  of gaps due to incoming   particles landing on empty line
segments and subsequent terms account for the  creation of gaps formed
by a previously  adsorbed  particle and one landing  on  the line (the
factors    of  2  account     for the  left-right   symmetry  of   the
line)\footnote{This property is lost  for a class  of models in  which
the  incoming     direction of  particles   is  not   vertical (shadow
models)\cite{PBR95}}.   One can also write  similar equations for gaps
of $h<\sigma_\alpha $, but they are not needed to  determine the surface densities.
To  arrive  at a solution,  one takes $G(h,t)=F(t)\exp[-k_a(h-\sigma_\alpha)t]$
and solves for F(t) (note that F(t) is different for each model).

The  kinetic  equations for    the  time  evolution of  the   particle
densities,  $\rho_\alpha$ and $\rho_\beta$, are  directly calculable from the gap
density distribution via

\begin{equation}\label{eq:59}
\frac{\partial\rho_\alpha}{\partial (k_at)} = 
\int_{\sigma_\alpha}^{+\infty}dh({\rm Min}(\sigma_\beta,h)-\sigma_\alpha )G(h,t)
\end{equation}

\begin{equation}\label{eq:60}
\frac{\partial\rho_\beta}{\partial (k_at)} =
 \int_{\sigma_\beta}^{+\infty}dh(h-\sigma_\beta)G(h,t)
\end{equation}
These are  integrated  to obtain an  analytical expression for
the densities as functions of time.
 
In each model, the line fills initially with $\beta$-particles.  At short
times, one obtains  $\rho_\beta  \sim t$  and $\rho_\alpha \sim   t^2$ (Model I)  and
$\rho_\alpha\sim t^3$  (Models II and III).   The exponents reflect the number
of particles on the line needed to block the  transition (one in Model
I and two in Models II and III).  Line filling later in the process is
dominated  by  $\alpha$-particles;   these  approach their   saturation as
$t^{-1}$.    $\beta$-particles,    conversely,     approach    saturation
exponentially.

\begin{figure}
\resizebox{15cm}{!}{\includegraphics{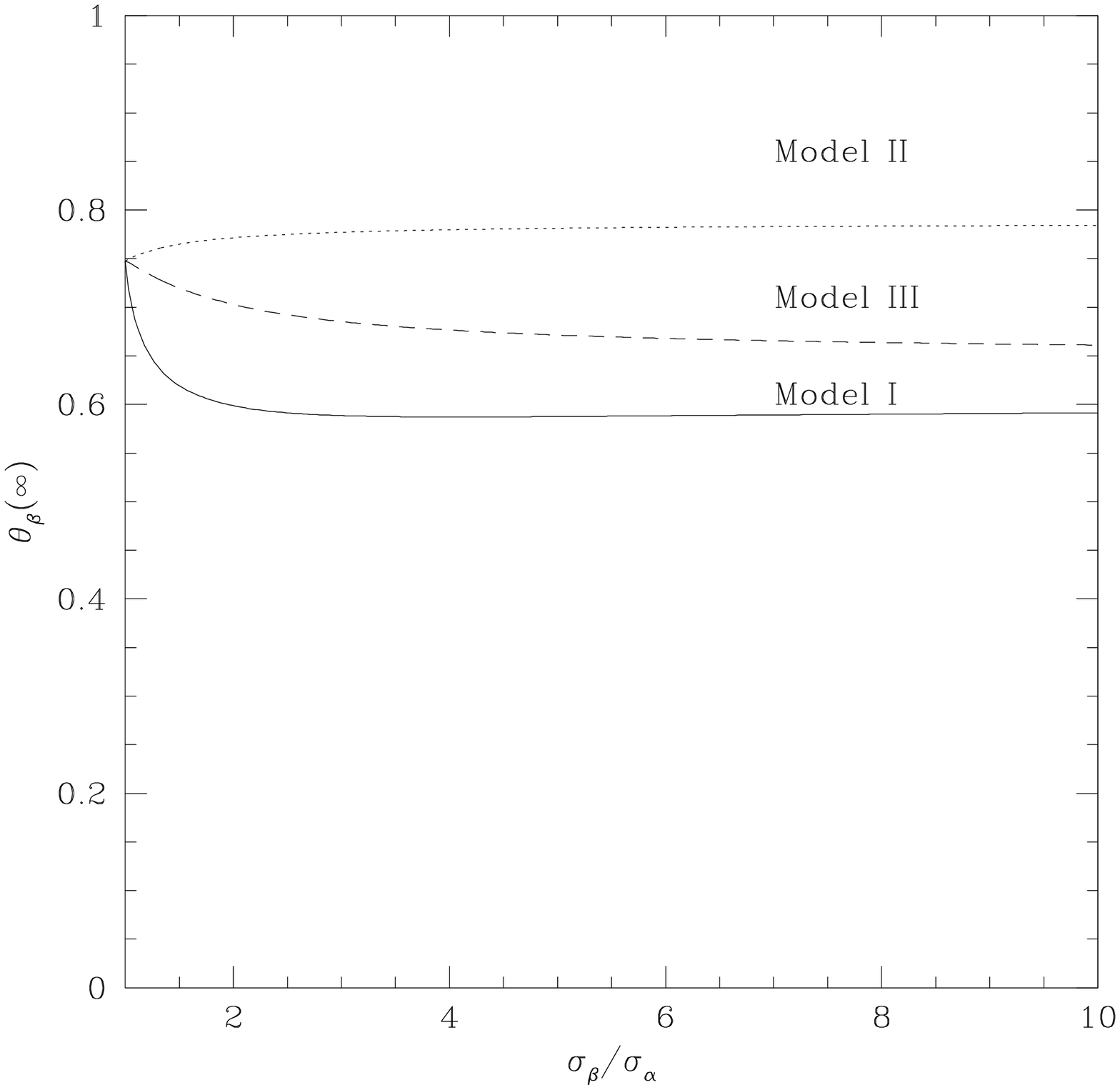}
\includegraphics{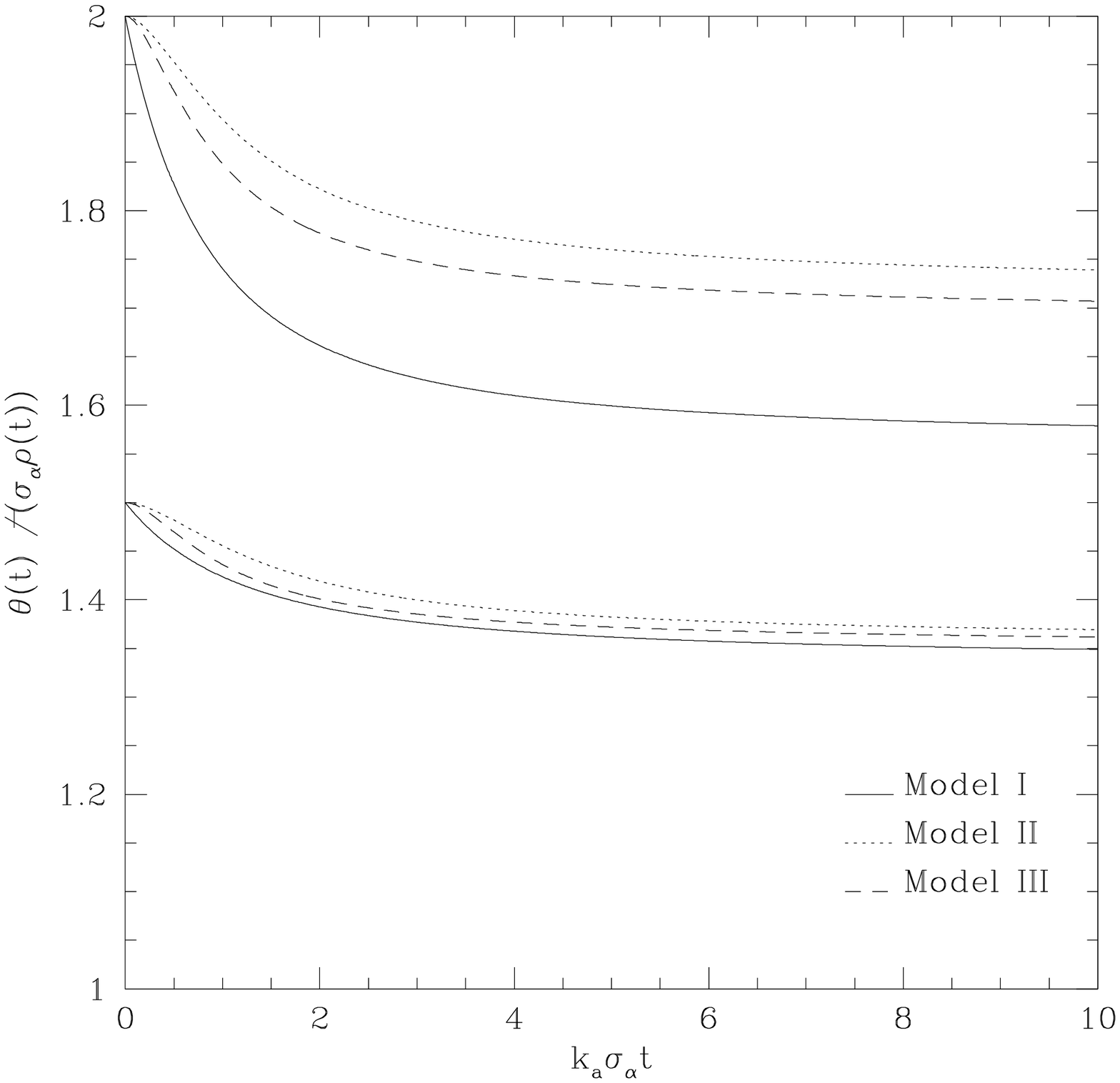}
}
\caption[99]{(a) Dependence
on $\sigma_\beta/ \sigma_\alpha$ of the saturation coverage of the largest particles,
$\theta_\beta(\infty  )=\rho_\beta (\infty)\sigma_\beta$,    for the  three     models.  (b) Time
evolution  of the mean  size of the adsorbed  particles  for the three
models and for $\sigma_\beta/ \sigma_\alpha=1.5$, $2.0$.

}\label{fig:10}
\end{figure}

In  all of these models,  the  saturation values of the  $\alpha$-particle
density  and  total coverage ($\theta=\rho_\alpha\sigma_\alpha+\rho_\beta\sigma_\beta$) increase and
those  of the $\beta$-particle  density  and total density  decrease with
increasing  $\sigma_\beta/\sigma_\alpha$.   Interestingly, the $\beta$-particle coverage
increases with $\sigma_\beta/\sigma_\alpha$ in Model II and decreases in Models I and
III (see Fig.~\ref{fig:10}a).  This quantity is important experimentally
since  often the  unaltered   fraction may  be removed  by  surfactant
elution.   The observed behavior is due   to the greater efficiency of
the  spreading mechanism in Model  II.   This spreading efficiency  is
also evident  in   the  average particle  diameter,    $\theta/\rho$.   (see
Fig.~\ref{fig:10}b).

\subsection{Two-dimensional models}\label{sec:Two-dimensional-mode}
\begin{figure}
\resizebox{12cm}{!}{\includegraphics{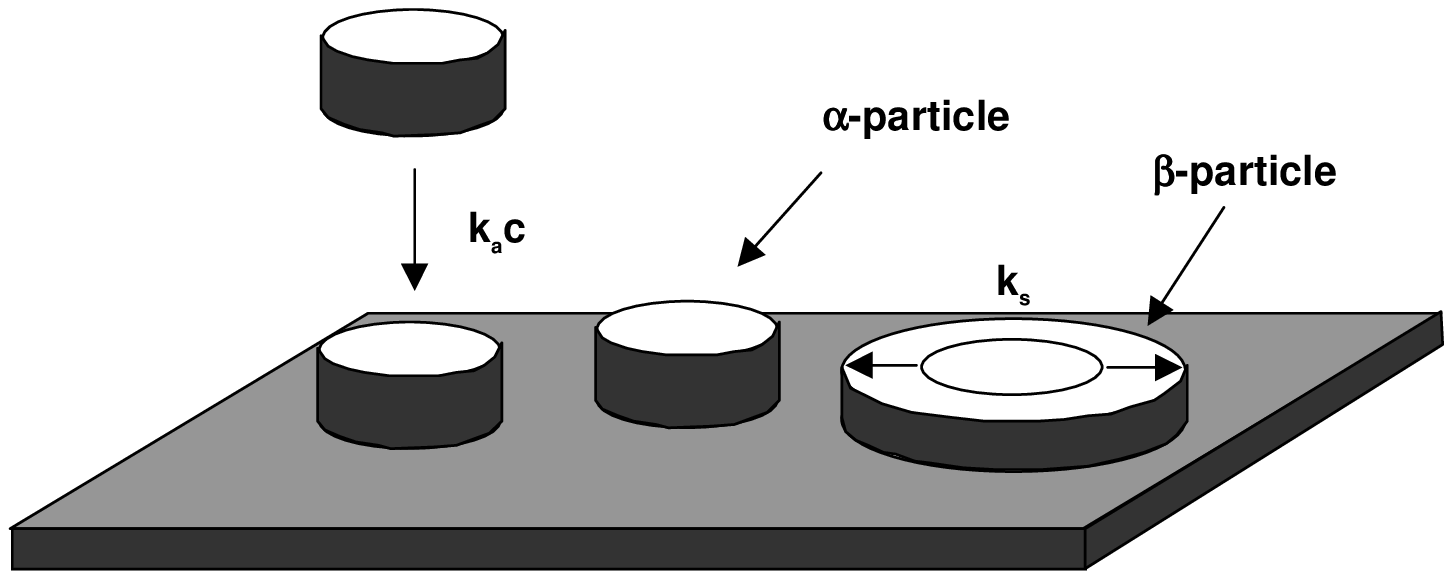}}
\caption[99]{Protein adsorption with surface-induced configurational
change.    Illustration of  the  two-dimensional  symmetric  spreading
model. Once adsorbed, an  $\alpha-$particle  spreads symmetrically at a
rate $k_s$ on   the
substrate if neighboring   preadsorbed particles do not  prevent  the
transition. }\label{fig:11}
\end{figure}

The  experimental situation of  interest is usually  adsorption onto a
two-dimensional surface.  With this  in mind, recent work has focused
on  the  development  of  two-dimensional  models of    irreversible
adsorption with a post-adsorption transition
\cite{VVTT94,VTTV96,VTV97,VTVT97}.   Adsorbing particle are modeled as
disks of size  $\sigma_\alpha$ that  adsorb randomly  and sequentially onto  a
plane at a rate $k_a=\hat{k}_ac$ (c  is the bulk  solution concentration).  Once
adsorbed, if space is available,   the disk will expand  symmetrically
and instantaneously to a larger diameter $\sigma_\beta$  at a rate $k_s$.  If
space is  not available, the disk  remains permanently of size $\sigma_\alpha$
(see Fig.~\ref{fig:11}).  The kinetic equations  for the time evolution
of $\alpha$- and $\beta$-particles are

\begin{equation}\label{eq:61}
\frac{\partial\rho_\alpha}{\partial t} = \hat{k}_ac \Phi_\alpha(\rho,\Sigma,K_s)-k_s\rho_\alpha\Psi_{\alpha\beta}(\rho,\Sigma,K_s)
\end{equation}

\begin{equation}\label{eq:62}
\frac{\partial\rho_\beta}{\partial t} = k_s\rho_\alpha\Psi_{\alpha_\beta}(\rho,\Sigma,K_s)
\end{equation}

where  $\Phi_\alpha$ is the  probability  that  an  $\alpha$-particle lands   on
unoccupied surface and  $\Psi_{\alpha\beta}$ is the probability that sufficient
space exists  for the transition to occur.   These functions depend on
the  overall  surface   density,  $\rho$,    the  ratio of  particle
diameters, $\Sigma   =\sigma_\beta/\sigma_\alpha$, and the  ratio  of the  spreading to
adsorption  rates, $K_s=k_s/\hat{k}_aca_\alpha$  ($a_\alpha$   is    the  area of     an
$\alpha$-particle).   In the special  case  where $k_s$=0, these equations
reduce to those  of the standard RSA  problem.   When $k_s$ approaches
infinity, which is just the two-dimensional analog of Model I above, one
has kinetic equations

\begin{equation}\label{eq:63}
\frac{\partial\rho_\alpha}{\partial t} = \hat{k}_ac 
(\Phi_\alpha(\rho,\Sigma)-\Phi_\beta(\rho,\Sigma))
\end{equation}

\begin{equation}\label{eq:64}
\frac{\partial\rho_\beta}{\partial t} = \hat{k}_ac \Phi_\beta(\rho,\Sigma)
\end{equation}
where $\Phi_\beta$  is the probability that  an incoming particle will have
sufficient space to immediately spread to a diameter $\sigma_\beta$.

Analytical  solutions    to Eqs.~(\ref{eq:63}-\ref{eq:64})    are not
available.  However, $\Phi_\alpha , \Phi_\beta$, and $\Psi_{\alpha\beta}$ may be expressed
as power series in terms of  the total surface  density.  This is done
by    first    generalizing  the   Kirkwood-Salsburg-like   hierarchy   (see
Eq.~(\ref{eq:14})   )  to  express   sets  of   mixed  particle-cavity
distribution  functions in  terms of an   infinite series of integrals
over Mayer functions and multiplet density distribution functions (see
section~\ref{sec:Stat-mech-sequ}).

\begin{eqnarray}\label{eq:65}
\lefteqn{\Phi_{\lambda_1\lambda_2...\lambda_{n+m}}({\bf r}_1,{\bf
r}_2,...,{\bf r}_n,{\bf r}^*_{(n+1)},...,{\bf r }^*_{(n+m)}) = } 
\nonumber \\
 &    &    \prod_{n'=1}^n   \prod_{m'=1}^m     [1+f_{n'm'}^{\lambda_{n'}\lambda_{m'}}]
 \sum_{s_\alpha=1}^{\infty} \sum_{s_\beta=1}^{\infty}
\frac{1}{s_\alpha!s_\beta!} \int d{\bf r}_{(n+m+1)}...d{\bf r}_{(n+m+s_\alpha+s_\beta)} \nonumber \\
 &                               &                         \prod_{m'=1}^m
 f_{ (n+m')(n+m+1)}^{\lambda_{n+m'}\lambda_{n+m+1}}...f_{ (n+m')(n+m+s_\alpha+s_\beta)}^{\lambda_{n+m'}\lambda_{n+m+s_\alpha+s_\beta}}
 \rho_{\lambda_1\lambda_2...\lambda_{n+s_\alpha+s_\beta}}^{(n+s_\alpha+s_\beta)}({\bf r}_{1},{\bf r}_{ 2},...,{\bf r}_{ n+m+s_\alpha+s_\beta})
\end{eqnarray}

\begin{eqnarray}\label{eq:66}
\lefteqn{\rho_{\lambda_1 \lambda_2 ... \alpha} ^{(n)}({\bf r}_
 { 1},{\bf r}_{ 2},...,{\bf r}_
 { n})\Psi_{\lambda_1\lambda_2...\lambda_{n-1}\alpha\beta}({\bf r}_
 { 1},{\bf r}_{ 2},...,{\bf r}_{ n-1},{\bf r}_{n}^*) = } \nonumber \\
 &      &   \prod_{n'=1}^{n}      [1+f_{n'n}^{\lambda_{n'}\beta}]\sum_{s_\alpha=0}^{\infty}
 \sum_{s_\beta=1}^{\infty}\frac{1}{s_\alpha!s_\beta!}\int     
   d{\bf r}_{n+1}...d{\bf r}_{ n+s_\alpha+s_\beta}
 \nonumber  \\     &                &       f_{n(n+1)}^{\beta\lambda_{n+1}}...
 f_{n(n+s_\alpha+s_\beta)}^{\beta\lambda_{n+s_\alpha+s_\beta}}\rho_{\lambda_1
 \lambda_2...\lambda_{n-1}\alpha\alpha...\beta...}^{(n+s_\alpha+s_\beta)}
 ({\bf r}_{ 1},{\bf r}_{2},...,{\bf r}_{n+s_\alpha+s_\beta})
\end{eqnarray}

$\Phi_{\lambda_1\lambda_2...\lambda_{n+m}}({\bf r}_{ 1},{\bf r}_{ 2},...,{\bf r}_{ n},{\bf r}_{n+1}^*,...,{\bf  r}_{n+m}^*)$     is       the
probability density of finding a $\lambda_1$-particle at position
${\bf r}_{1}$, ...  a
$\lambda_n$-particle  at position  ${\bf r}_n$, sufficient space   (a  cavity) for a
$\lambda_{n+1}$-particle at position ${\bf r}_{n+1}$, ..., and  sufficient space for a
$\lambda_{n+m}$-particle    at   position  ${\bf r}_{n+m}$.
    $\rho_{\lambda_1  \lambda_2 ...\alpha}
^{(n)}({\bf r}_1,{\bf r}_2,...,{\bf r}_n)\Psi_{\lambda_1\lambda_2...\lambda_{n-1}\alpha\beta}({\bf r}_1,{\bf r}_2,...,{\bf r}_{n-1},{\bf r}_n^*)$ is the
probability density of finding a $\lambda_1$-particle at position ${\bf r}_1$, ..., a
$\lambda_{n-1}$-particle at position  ${\bf r}_{n-1}$, and an $\alpha$-particle at position
${\bf r}_n$  that has sufficient space to spread and become a $\beta$-particle.  The
terms in  these summations represent the  blockage  of cavities on the
surface by   $s_\alpha$ $ \alpha$-   and   $s_\beta$  $ \beta$-particles.   All   of  the
distribution functions (the $\Phi$'s, $\Psi$'s, and $\rho$'s) may be written
as  series  expansions  in  the total  singlet    density $\rho$.  Using
Eqs.~(\ref{eq:65})-~(\ref{eq:66})     and    kinetic        equations,
Eqs.~(\ref{eq:61})-~(\ref{eq:62})  for the   time   evolution  of  the
multiplet  density  distribution functions,  the   coefficients of the
density expansion may  be determined term  by term \cite{VTTV96}.   In
the  case when  $K_s$  is finite,  a  renormalization  is performed to
change variables from ($\rho,K_s$) to ($\rho,\zeta$) where $\zeta=\rho K_s$.  This
variable change is reflected in the  form of the kinetic equations for
all density functions and results in the series expansion coefficients
having a dependence on the new variable $\zeta$.  In the case where $K_s$
is  infinite,  no renormalization  is needed and  the series expansion
coefficients depend only on $\Sigma$.

Power series are accurate only at low  to moderate densities.  As the
system approaches saturation, one must use  other tools to analyze the
kinetics.  As is  the case in standard  RSA, it is  useful to consider
the geometry of  small, isolated targets on  the crowded surface  that
are available for addition of $\alpha$- and $\beta$-particles.  By expressing
the  rate   of adsorption in  terms  of  the density and  size  of the
targets, one finds  that  $\alpha$-particles approach their saturation  as
$t^{-1/2}$  and    that   $\beta$-particles  approach   their  saturation
exponentially in time.  An exception is when $\Sigma$=1;  in this case the
$\beta$-particles approach  their saturation  as $t^{-3/2}$   since their
formation is  not inhibited  by   surface blockage.  Consideration  of
surface targets also  allows one to show that  the  $\Sigma$ derivative of
the saturation $\alpha$- and  $\beta$-particle  (but not the  total)  density
diverges as $\Sigma$ approaches 1.  This result suggests  that even a very
small transition  could result in  quite a large fraction of particles
remaining in the unaltered state.   Finally, an asymptotic analysis of
the  structure of  the adsorbed   layer shows  that the $\beta-\beta$   pair
correlation function, $g_{\beta\beta}(r)$, diverges at contact (as does g(r)
in standard RSA) while $g_{\alpha\alpha}(r)$ and $g_{\alpha\beta}(r)$ remain finite.

The  asymptotic  regime  scaling   laws can  be   combined   with  the
short/moderate time  regime density series expansion via interpolation
yielding  expressions that   are   accurate over the entire    filling
process.   When $K_s$ is infinite,  methods similar to those developed
for standard RSA may  be applied.  One obtains  equations for the time
evolution of the total density and total fractional surface coverage,

\begin{equation}\label{eq:67}
\frac{\partial \rho}{\partial t} = \frac{\hat{k}_ac(1-x)^3}{1+b_1x+b_2x^2}
\end{equation}

\begin{equation}\label{eq:68}
\frac{\partial \theta}{\partial t} = \frac{\hat{k}_ac a_\beta (1-x)^3}{1+c_1x+c_2x^2+c_3x^3}
\end{equation}
where $x=\rho/\rho_\infty$ and the coefficients $b_1, b_2, c_1, c_2, c_3$, and
$\rho_\infty$ are determined  by matching the density  expansion of the Pad{\'e}
ratios with the coefficients determined above.  These equations may be
integrated to  give  $\rho_\alpha$  and $\rho_\beta$ as  functions   of time that
compare quite favorably with computer simulation \cite{VTTV96}.

When $K_s$ is  finite, it is  more accurate to introduce  an effective
area, $a_{eff}(\rho,\zeta)$, that   serves as   a  mapping
variable of  the RSA+spreading model onto  the standard RSA model.  It
is defined  as the  particle area in  the standard  RSA of  disks that
would yield   the    same  fractional available surface    as   in the
RSA+spreading  problem at the same  overall surface  density, that is,
$\Phi(\theta=\rho a_{eff})=\Phi_\alpha(\rho)$  where  $\Phi$  is  the available  surface
function for  the RSA of disks.   Also introducing the average area of
particles landing on the surface  of density $\rho$, $ \mbox{{\~a}}(\rho,\zeta)$,
kinetic equations for the density and coverage may be written:

\begin{equation}\label{eq:69}
\frac{\partial \rho}{\partial t} = \hat{k}_ac \Phi(\rho a_{eff}(\rho,\zeta))
\end{equation}

\begin{equation}\label{eq:70}
\frac{\partial \theta}{\partial t} = \hat{k}_ac a_\beta \Phi(\rho \mbox{{\~a}}(\rho,\zeta))
\end{equation}
$a_{eff}$ and $\mbox{{\~a}}$ are  themselves written as  Pad{\'e} approximants
whose   coefficients are  determined     by matching the   low density
expansion coefficients
\cite{VTTV96}.  
This  method yields   extremely accurate total   surface densities and
reasonably  accurate partial  densities   when  compared to   computer
simulation.

\begin{figure}
\begin{center}
\resizebox{12cm}{!}{\includegraphics{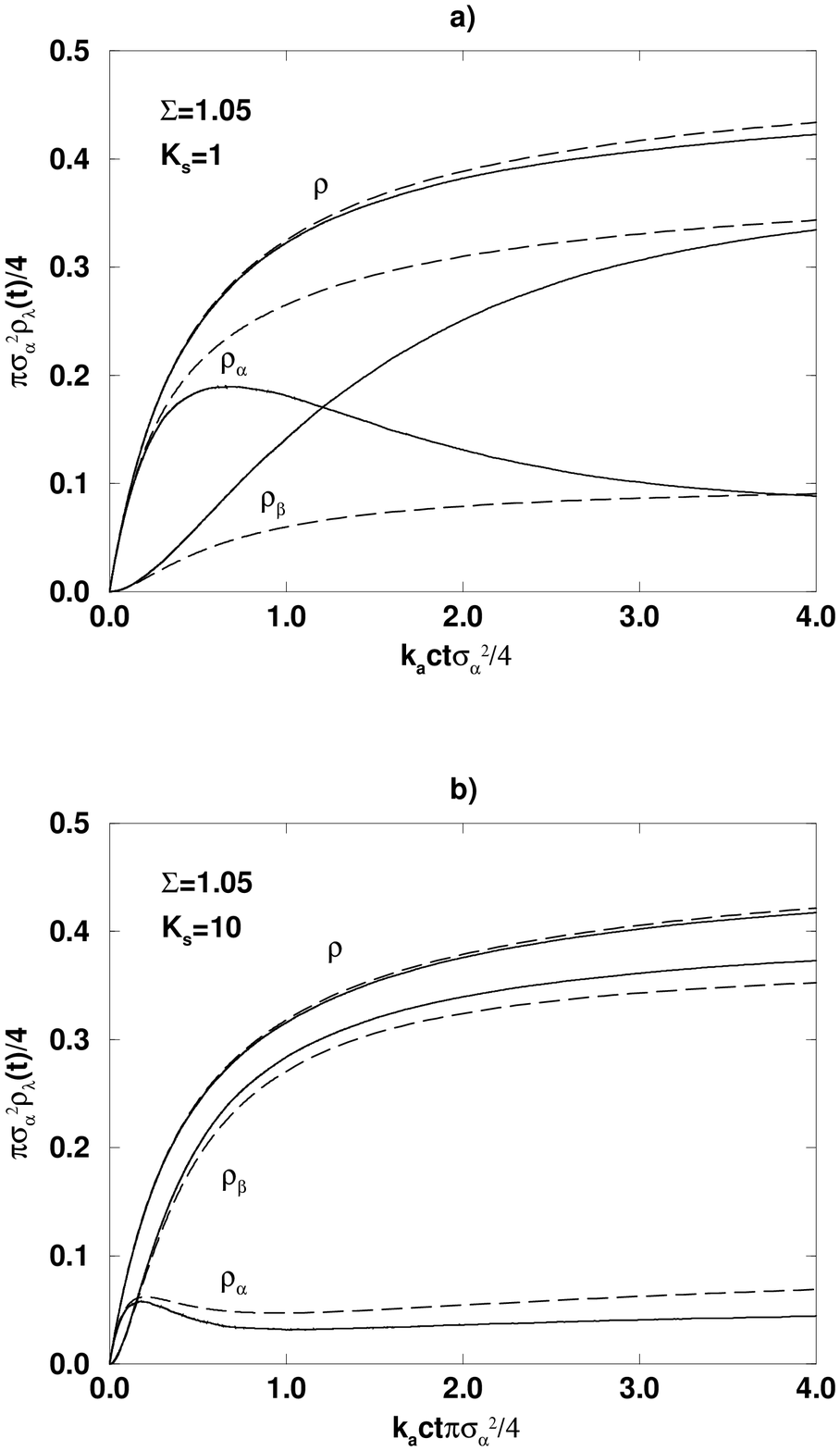}}
\caption[99]{Kinetics of RSA+spreading model. Effective area
interpolation calculation (dashed lines) of the reduced total and
partial densities as function of reduced time for  $\Sigma=1.05$ and (a)
$K_s=1$ and (b) $K_s=10$. Also shown are simulation results (solid lines).
}\label{fig:12}
\end{center}

\end{figure}

When  $K_s$  is  infinite,   results  of computer simulation  of   the
two-dimensional   problem are qualitatively  similar   to those of the
one-dimensional   problem.  One difference  is   that  the density  of
$\alpha$-particles ($\beta$-particles) tends to be higher  (lower) in 2-D due
to the greater probability of landing near  to a preadsorbed particle.
When $K_s$ is finite,  some important differences emerge.  First,  the
$\alpha$-particle density  may become  non-monotonic in   time due to  the
delayed spreading  (see Fig.~\ref{fig:12}).  Second, the $\beta$-particle
coverage, $\rho_\beta a_\beta$,  approaches zero   for  large $\Sigma$.  This   is
because for any  finite rate of spreading,  one can choose a spreading
magnitude larger   than  the  characteristic  separation  of  adsorbed
particles.   Increasing $K_s$ favors the spreading  event and causes a
decrease in the $\rho_\alpha$ and $\rho$ and an increase in $\rho_\beta$ and $\theta$.

\begin{figure}
\begin{center}
\resizebox{12cm}{!}{\includegraphics{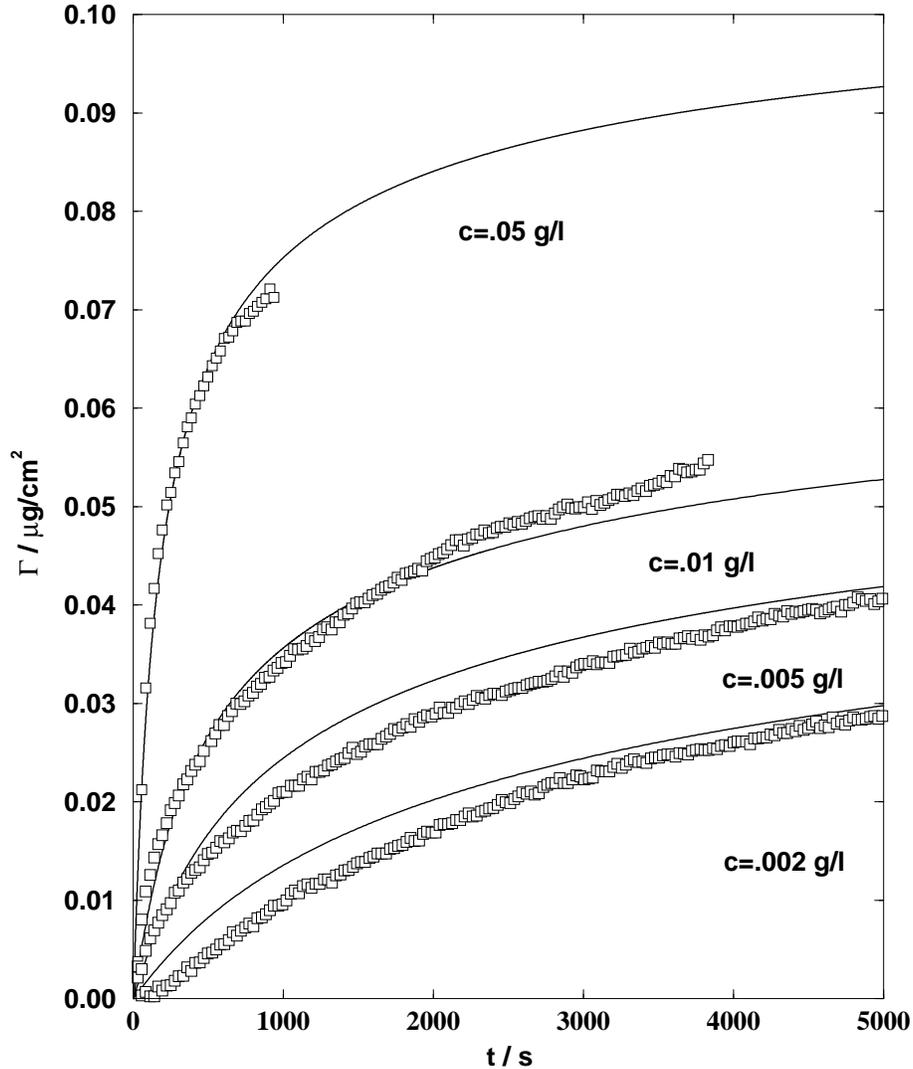}}
\caption[99]{Surface density of adsorbed fibronectin $\Gamma $ as a
function of time. Curves corresponding to bulk protein concentration
of 0.05, 0.01, 0.005, 0.002 g/l are shown. Also displayed are the curves
predicted from Eq.~(\ref{eq:69}) with  $\hat{k}_a=1\times 10^{-5}cm/s$ and
$\sigma_\alpha =20nm$. More details are given in Ref.\cite{VGRTVT98}.
}\label{fig:13}
\end{center}

\end{figure}

The    need to  employ  a  concentration-dependent  particle size when
fitting  experimental  data   to the   standard  RSA  model\cite{R93b}
provided  an   incentive  to  develop irreversible   adsorption models
incorporating a  post-adsorption transition.  With the RSA+spreading
model,  several  kinetic isotherms,    each  differing only    in bulk
concentration, may be accurately predicted with a single particle area
that is close in  value to the  known  size of the  adsorbing particle
(see Fig.~\ref{fig:13})
\cite{VGRTVT98}.

\section{Influence of desorption processes}\label{sec:Desorption-processes}
Up to this  point, we  have  considered adsorption processes  that are
strictly irreversible,   that  is, ones  in which    proteins or other
macromolecules remain forever on the surface following adsorption.  In
many  experimental situations, proteins  do indeed  adsorb tenaciously
and little   or no desorption  occurs.   However, other examples exist
where desorption may occur spontaneously or in response to a change in
experimental conditions  (pH, ionic strength, surfactant  or detergent
concentration).   It  is clear   that,   in order  to  be  of  general
applicability, a protein adsorption model must  be able to account for
the possibility of desorption.  In this  section, we introduce several
models  that  extend the irreversible adsorption  approaches presented
above to include desorption.

\subsection{Multistep adsorption / Depletion model}\label{sec:Mult-adsorpt}

Protein adsorption   need not  always be conducted    as a single step
process.  Adsorption  may be  interrupted  by a   buffer rinse or   by
exposure to a surfactant  or detergent solution.  This usually results
in removal of   a  fraction of the   adsorbed  proteins.  Due  to  the
non equilibrium   nature of protein adsorption,   we  suspect that when
adsorption  is resumed, the behavior   may differ nontrivially from an
uninterrupted process.

We have investigated  the effect  of a desorption (or
depletion) step, separating  two irreversible adsorption steps, on the
structure, kinetics, and saturation density of adsorbed disks
\cite{VTVT97,VVTRT99}.  A starting point is  to consider the evolution
of  the density distribution functions during  a random depletion from
an initial density $\rho_1$ as follows:

\begin{equation}\label{eq:71}
\frac{\partial \rho^{(n)}({\bf r}_1,{\bf r}_2,...,{\bf r}_n;\rho_1,\rho)}{\partial t} = 
- k_d n \rho^{(n)}({\bf r}_1,{\bf r}_2,...,{\bf r}_n;\rho_1,\rho)
\end{equation}

It is straightforward to show  that this  leads  to a conservation  of
distribution functions,  that is $g^{(n)}({\bf r}_1,{\bf r}_2,...,{\bf
r}_n;\rho_1,\rho)=g^{(n)}({\bf   r}_1,{\bf r}_2,...,{\bf  r}_n;\rho_1,\rho_1)$
for all $\rho_2\leq  \rho\leq \rho_1$ where  $\rho_1$ and $\rho_2$ are the densities
before   and after depletion ($g^{(n)}=\rho^{(n)}/\rho^n$)  \cite{VTVT97}.
This  result    holds for  depletion  of   any immobile  collection of
particles, including those at equilibrium.  An interesting implication
is that  one could create a  low-density configuration that would have
the structure of  one taken at a  much higher density. This result can
be used to  analyze the kinetics and saturation   density of a  second
adsorption step beginning  with the depleted configuration.  To  third
order,

\begin{equation}\label{eq:72}
\Phi(\rho_1,\rho_2,\rho)=\Phi^{RSA}(\rho)+A_{32}\rho^2_2(\rho_1-\rho_2)
\end{equation}

where $A_{32}$ is the  contribution  of the  third order coefficient  of
$\Phi^{RSA}$   from the pair   density.\cite{VVTRT99} Since   $A_{32}>0$
\cite{ST89a}, the available surface function for the multistep process
is greater than that for the simple  RSA process and an enhancement in
the saturation density is expected.

\begin{figure}\begin{center}
\resizebox{13cm}{!}{\includegraphics{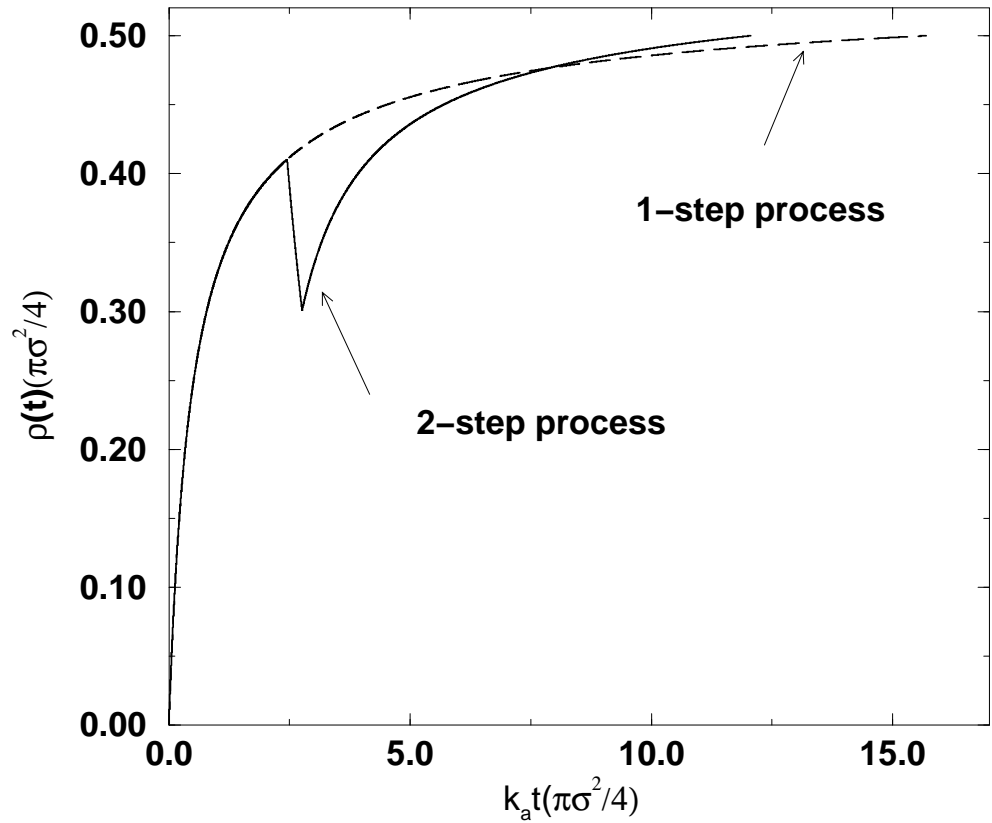}}
\caption[99]{ Influence of a depletion step on the adsorption
kinetics. The dashed line corresponds to a single-step RSA  and 
the full line to two RSA steps separated by a depletion step (with
$k_d=k_a$). In both cases, the process is stopped as soon as a
coverage of $0.5$ is reached.}   \label{fig:14} \end{center}

\end{figure}

Using    an interpolation scheme, one can calculate the
saturation density and show that it is always enhanced compared to the
uninterrupted   adsorption process and   that  the maximum enhancement
occurs when   $\rho_2=\frac{2}{3}\rho_1$  \cite{VVTRT99}.  An  interesting
aspect  of these results is  that in order to  reach a certain surface
density as  rapidly as possible,  it is  sometimes most  efficient  to
incorporate a desorption step.  This is shown in Fig.~\ref{fig:14}.

\subsection{Partial reversibility}\label{sec:Part-reversibility}

A  number  of  experimental  results suggest  that  in  some instances
protein adsorption is partially reversible, that is, a fraction of the
adsorbed molecules may be removed by a buffer rinse while others cling
to the surface    irreversibly \cite{VTV97}.  These observations   are
consistent with the two-state   adsorption picture first put forth  by
Soderquist and Walton where proteins  adsorb initially in a reversible
manner and   then   later   become irreversibly
adsorbed by changing conformation  and/or orientation \cite{SW80}.  We
proposed  a  model   of   partially    reversible protein   adsorption
\cite{VTV97} incorporating   Morrisey's
observation    that neighboring  proteins  can   block this transition
\cite{M77},

In  this model,  proteins are modeled  as disks  of diameter $\sigma_{\alpha}$
that adsorb   onto the surface sequentially and   without overlap at a
rate $k_a=\hat{k}_ac$.  Once   adsorbed, the proteins desorb  at  rate $k_d$  or
spread to  a larger diameter $\sigma_{\beta}$ at  a rate $k_s$ (the latter is
subject to  no overlap with  other adsorbed proteins).  The  rate laws
are as follows:

\begin{equation}\label{eq:73}
\frac{\partial\rho_\alpha}{\partial t} = 
\hat{k}_ac \Phi_\alpha(\rho,\Sigma,K_s,K_d)- k_s\rho_\alpha\Psi_{\alpha\beta}(\rho,\Sigma,K_s,K_d)-k_d\rho_\alpha
\end{equation}

\begin{equation}\label{eq:74}
\frac{\partial\rho_\beta}{\partial t} = 
k_s\rho_\alpha\Psi_{\alpha_\beta}(\rho,\Sigma,K_s,K_d)
\end{equation}
where  $\Phi_{\alpha}$ and $\Psi_{\alpha\beta}$  are, respectively, the probabilities
of adsorbing and  spreading without overlap, $\Sigma=\frac{\sigma_\beta}{\sigma_\alpha}$,
$K_s=\frac{k_s}{\hat{k}_aca_\alpha }$, and $K_d=\frac{k_d}{\hat{k}_aca_\alpha }$ where
$a_\alpha =\pi\sigma_\alpha^2/4$.

Because the adsorption  is not  strictly reversible  in this model,  a
density         expansion      approach     as       introduced     in
sections~\ref{sec:Two-dimensional-sequ} and
\ref{sec:Approximation-scheme} is not likely to succeed.  However, the
asymptotic regime   may be  analyzed by again considering
the time evolution  of isolated targets on  the surface for $\alpha$-  and
$\beta$-particles. Using  this   approach, one  can show that   both
$\rho_\alpha$ and  $\rho_\beta$ approach saturation  as $t^{-1/2}$  \cite{VTV97}.
The  change from a faster exponential  approach to  a slower algebraic
approach  for the $\beta$-particles     when $K_d$ becomes non-zero    is
because, in this  case, $\alpha$-particles may  spread at long times  when
cavities  form due to desorption  of  neighboring $\alpha$-particles.  (In
the  special  case of  $\Sigma=1$,  no  steric  hindrance to spreading  is
present and an even faster $t^{-3/2}$ approach is found.)

To evaluate the above  kinetic   equations over all times,    computer
simulation is required  \cite{VTV97}.  An  interesting result is  that
for certain   parameter  values, both  the    total density  and   the
$\alpha$-particle density  may become non-monotonic  in time.  This is due
to the gradual replacement of  initially adsorbed $\alpha$-particles by  a
smaller  number of larger $\beta$-particles.  This  behavior is similar to
what is observed in experimental ``overshoots''  where the total density
decreases at  long times \cite{WAL95}.   Another interesting result is
that   saturation densities  tend   to  be   higher  than  in   purely
irreversible adsorption.  This is  due to a more liquid-like  particle
structure,  as  evidenced   by   the secondary  peaks   in  the radial
distribution  function, which leads  to more  efficient packing on the
surface.

\begin{figure}
\resizebox{13cm}{!}{\includegraphics{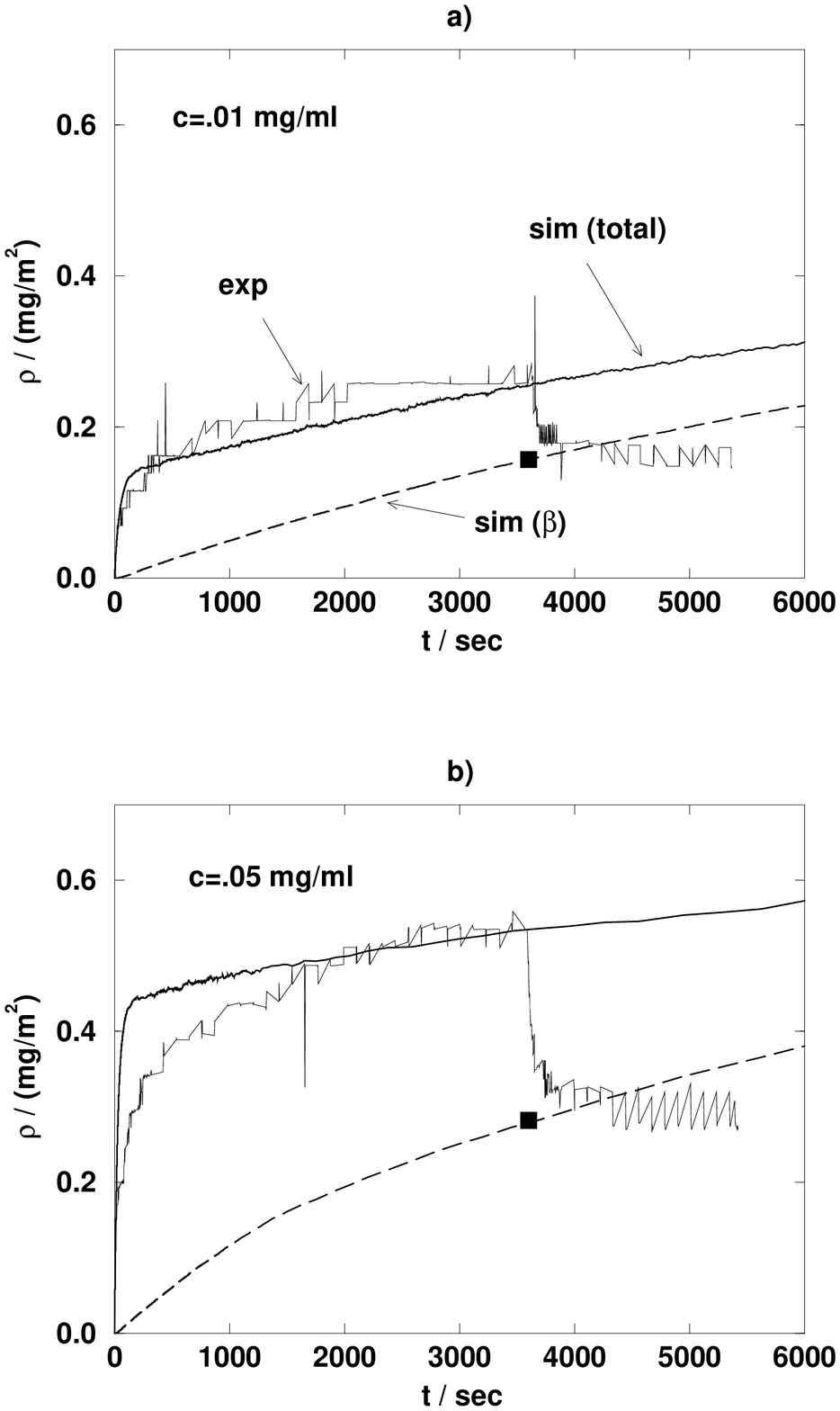}}
\caption[99]{A comparison between the model of partially reversible
adsorption and the lysozyme adsorption experiments  of Walgren {\it et
al}\cite{WAL95}. Simulated total (solid line) and $\beta-$particle (dashed
line) densities are shown along with the experimental total amount for
two bulk   protein concentrations.  (For   the  determination  of  the
parameters, see Ref.\cite{VTV97}). We   note  good agreement in the    total
amount  adsorbed  before rinsing    (at 3600s).   The   amount
remaining after rinsing   is well approximated by  the $\beta-$particle
density at the time (indicated by squares).  }\label{fig:15}
\end{figure}

The chief utility  of this model  is its applicability to experimental
situations in  which desorption  and post-adsorption transitions occur.
Good  quantitative predictions of  the   adsorption kinetics and   the
reversible fraction  of lysozyme on   silica are  obtained  (see
Fig.~\ref{fig:15}) \cite{VTV97,WAL95}.

\subsection{Parking lot model}\label{sec:Parking-lot-model}
In the  adsorption-desorption  model,   particles  are  placed  in   a
$D$-dimensional space according  to the RSA  rules at  a constant rate
$k_{a}$ and in which    an additive uniform desorption process   takes
place such  that all  objects in  the  system  are subject  to removal
(desorption) at random  with a constant rate  $k_{d}$.  In the parking
lot model, the  substrate is  a line and   the objects are hard  rods.
This   one-dimensional model has been solved   in some limiting cases.
When $k_{d}=0$, the adsorption is totally irreversible and the process
corresponds to the Car Parking Problem.  When $k_{a}=0$, starting with
any  configuration of   particles,   the process   corresponds to  the
desorption model studied in section \ref{sec:Mult-adsorpt}.  The limit
$k_{d}\to   0+$   allows a   slow   but eventual rearrangement of
particles on the  line leading to  a final coverage  of unity.  (It is
worth noting the finite discontinuity of  the final saturation density
between the case  $k_d \to 0+$ and  $k_d=0$, i.e., RSA).  Moreover, the
final density is  independent   of the initial configuration   of
particles on the line; it depends only on the ratio of desorption and
adsorption rates.  For  $ k_d\to0$, accurate descriptions  have  been
obtained\cite{JTT94,KN94}.   In this case, the process cleanly divides into
two  sub-processes.  The  initial phase  consists  of an  irreversible
adsorption    and is     followed   by  an    infinite sequence     of
desorption-adsorption events in which a  rod detaches from the surface
and the gap that is  created is immediately filled by  one or two  new
rods.  The latter possibility causes the system to evolve continuously
to the close-packed state with $\rho  \sigma =1$, as \cite{JTT94,KN94} $1-\sigma
\rho(t) \simeq  1/\ln(t) $ where $t$ now  represents a rescaled time.  For
the general case, where both $k_a$ and $k_d$ are  non zero, a complete
solution is not yet available.

\begin{figure}
\begin{center}
\resizebox{10cm}{!}{\includegraphics{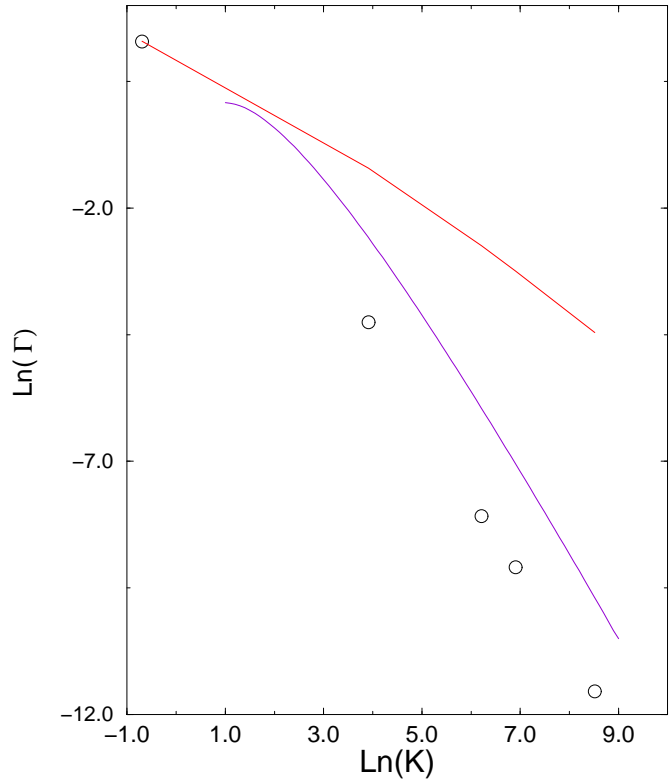}}
\end{center}

\caption[99]{Relaxation rate $\Gamma$ for the final exponential approach to
equilibrium in the Parking Lot model. Upper  curve:  mean-field  approximation,
Eq.~(\ref{eq:79}).  Intermediate curve: Eq.~(\ref{eq:80})
Open circles: simulation results.
}\label{fig:16}
\end{figure}

The properties of the parking lot model depend  only on the ratio $K =
k_a/k_d$.  Large   values of  $K$     correspond to small
desorption rates. With an  appropriate rescaled time, the  kinetics is
given by
\begin{equation}\label{eq:75}
\frac{\partial(\rho\sigma)}{\partial t} = \Phi(t)-\frac{\rho\sigma}{K},
\end{equation}
where $\Phi(t)$,  the insertion  probability  at  time $t$ (or   density
$\rho$), is the   fraction of the   substrate that is available for  the
insertion of a new particle.  The presence  of a relaxation mechanism,
even infinitesimally small, implies that the system eventually reaches
a steady state   that corresponds to  an equilibrium  configuration of
hard particles with $\rho_{\rm eq}\sigma  = K\Phi_{\rm eq}(\rho_{\rm eq}\sigma)$,  where
$\rho_{\rm eq}$  denotes the equilibrium  density.  At  equilibrium, the
insertion probability is given exactly by
\begin{equation}\label{eq:76}
\Phi_{\rm eq}(\rho)= (1-\rho\sigma)\exp(-\rho\sigma/(1-\rho\sigma)).
\end{equation}
Inserting   Eq.~(\ref{eq:76})   in Eq.~(\ref{eq:75})    leads  to  the
following expression for the equilibrium density:
\begin{equation}\label{eq:77}
\rho_{\rm eq}\sigma = \frac{L_w(K)}{1+L_w(K)}
\label{isotherm}
\end{equation}
where $L_W(x)$ (the  Lambert-W function) is  the solution of $x=ye^y$.
In the limit of  small  $K$, the  isotherm   takes the Langmuir   form
($\rho_{eq}\sigma \sim K/(1+K) $) while for large $K$, $\rho_{eq}\sigma\sim 1-1/\ln(K)$.  At
small values of  $K$,  equilibrium is  rapidly obtained, but  at large
values   the  densification displays   a  dramatic  slowing  down.  An
adiabatic (mean-field)   treatment consists of  assuming that,  at any
density $\rho(t)$, the structure of the adsorbate follows an equilibrium
form.  This  means   that $\Phi$  is  similar to  Eq.~(\ref{eq:76})  with
$\rho(t)$ in place of $\rho_{eq}$.  Denoting $\delta \rho(t) = \rho(t) - \rho_{\infty}$,
with $\rho_{\infty}=  \rho_{eq}(K)$, one   performs   a first  order   density
expansion of Eq.~(\ref{eq:75})
\begin{equation}\label{eq:78}
\frac{\partial }{\partial t}\delta\rho = -\Gamma_{MF}(K)\delta\rho + O(\delta\rho^2)
\end{equation}
with
\begin{equation}\label{eq:79}
\Gamma_{MF}(K) \simeq  \ln(K)^2/K
\,\,\mbox{\rm when K is large}
\end{equation}
equivalent    to   a relaxation time     of    $K/\ln(K)^2$ for  large
$K$\cite{BKNJN98}. However,   a   careful  analysis of   the   kinetic
equations for the gap  distribution function has been done\cite{TTV99}
and it has shown that the relaxation rate $\Gamma $ is actually given by
\begin{equation}\label{eq:80}
\Gamma\simeq  2\frac{(\ln K)^3}{K^2} + O(\frac{(\ln K)^2}{K^2})
\end{equation}
and   thus   much    smaller   than   the      mean-field  prediction,
Eq. (\ref{eq:79}); simulation results  are compared with both the mean
field result, and with     the prediction from the  gap   distribution
approach, Eq.~(\ref{eq:80}), in Fig.~\ref{fig:16}. (It is worth noting
that  the difference  between a mean-field  treatment\cite{KN94} and a
more detailed  description leads to  a difference which only occurs in
subdominant terms  of the kinetics).  In addition to
the anomalous relaxation behavior of this model, the power spectrum of
fluctuations exhibits two  frequency scales\cite{KNT99} which could be
a general feature  of models with very  slow relaxation. Similar power
spectra  have been observed  in   densification of monodisperse  glass
beads during a long series of vibrations\cite{NKBJN98}.

\subsection{Scaled Particle Theory}\label{sec:Scale-Part-Theory}

The Parking Lot Model described in the previous  section shows both that
when the desorption rate  is  very small,  the relaxation kinetics  is
very slow and the  whole process is not  well described by a mean-field
approximation.  On the other hand, if desorption  is not too small, an
adiabatic approximation provides a   fair description of  the process.
The final state is  an equilibrium one and  the difference between the
transient states and the  corresponding equilibrium states at the same
coverage is  small.   This property  is  useful  when  one considers the
adsorption of  complex systems, such as  mixtures, because it allows a
drastic simplification of the analysis by using an approximation which
essentially  requires the   knowledge of equilibrium  properties.  The
Scaled Particle Theory (SPT) introduced by Reiss et al.
\cite{RFL59}   is particularly   convenient  since  it yields   simple
analytic expressions  for  the   available  surface functions for
systems at equilibrium.   

The kinetic equation generalized to mixtures is
\begin{equation}\label{eq:81}
\frac{\partial \theta_i}{\partial t} = k_{a,i}c\Phi_i(\{ \theta_j \})-k_{d,i}\theta_i
\end{equation}
where $\Phi_{i}(\{\theta_j\})$ is the  available surface function of species
$i$. To apply  SPT one first  approximates $\Phi$ by  $\Phi_{eq}$ and then
the connection to the equation of state of  the equilibrium mixture is
provided by the relation \cite{W66}.
\begin{equation}\label{eq:82}
\ln\Phi_i^{eq} = -\mu_i^R/kT
\end{equation}
where $\mu_i^R$ is the residual chemical potential of component $i$.

For  a two-component  mixture of  hard   disks of diameter  $\sigma_1$ and
$\sigma_2$ the explicit equations are:
\begin{equation}\label{eq:83}
\Phi_1^{eq}=(1-\theta_1-\theta_2)\exp\biggl[-{\frac{3\theta_1+\gamma^{-1}
(\gamma^{-1}+2)\theta_2}{1-\theta_1-\theta_2}}-
{\frac{(\theta_1+\gamma^{-1}\theta_2)^{2}}{(1-\theta_1-\theta_2)^2}}\biggr],
\end{equation}
where $\gamma=\sigma_2/\sigma_1$ is the size ratio.

\begin{equation}\label{eq:84}
\Phi_2^{eq}=(1-\theta_1-\theta_2)\exp\biggl[-{\frac{3\theta_2+\gamma
(\gamma+2)\theta_1}{1-\theta_1-\theta_2}}-
{\frac{(\theta_2+\gamma\theta_1)^{2}}{(1-\theta_1-\theta_2)^2}}\biggr].
\end{equation}

The time-dependent coverages obtained by integrating Eq.~(\ref{eq:81})
with  Eqs.~(\ref{eq:83})    and (\ref{eq:84}) agree    very  well with
numerical  simulations   of   the  adsorption-process.  For    certain
parameter  values,  the model  exhibits  an ``overshoot''  in the time
dependent   coverage    -  a   phenomenon   that   has   been observed
experimentally.

Adsorbed configurations of a single species  that can adsorb in two or
more orientations with  respect to the  surface can be considered as a
two-dimensional mixture. Kinetic and isotherms  based on SPT have been
developed to describe this situation
\cite{T97,M99,JMTW99}. The equations presented above can be readily
generalized to continuous multicomponent mixtures and some interesting
kinetic phenomena  have been  reported \cite{OT99}.   Finally, kinetic
equations based  on  SPT have  also  been developed for  the spreading
model\cite{BV99}.

\section{Multilayer formation}\label{sec:Multilayer-formation}
The rules of the RSA model and its extensions discussed thus far allow
only for a single adsorbed layer to be  placed on a surface.  In fact,
a number of experimental   situations exist where adsorption does  not
stop at one layer but instead continues  with later arriving particles
adhering to previously adsorbed ones.    The fabrication of thin  film
materials via solution deposition of macromolecular building blocks is
an important example of multilayer adsorption and is currently used to
make sensors, superconducting materials, and  cathode ray tubes.   The
physical,  electrical, and  optical    properties of these thin   film
materials depends strongly on their structure and density.  Models
are thus needed to predict thin film structure and density in terms of
the  particle properties and  deposition conditions.  In this section,
we discuss multilayer adsorption models developed for this purpose.

A  number of lattice   and continuum studies of  multilayer deposition
have appeared  recently.\cite{BP91,K92,BE94}. Our  work focuses on the
first layers  (typically less than  $20-30$)  and that it incorporates
all three  events that can  occur when  a particle contacts  a growing
interface: it may stick to the interface (i.e. adsorb), it may diffuse
away  from the interface (i.e.  desorb),  or it  may remain in contact
while descending further toward the surface (i.e.  roll)
\cite{VV97,YVV98}.

Particles are  modeled   as $(D+1)$-dimensional   spheres that   descend
vertically onto a $D-$dimensional substrate.  If a particle lands on the
surface,  it remains there  permanently.  If it  lands on a previously
placed  particle,  an overhang amount  $\delta$  is calculated either with
respect   to  the surface   or     to the  contacting particle    (see
Fig.~\ref{fig:17}    ).   The  overhang  extent  is   compared to  two
parameters, an  adsorption parameter  $\delta_1$  and a  rolling parameter
$\delta_2$.    If $\delta<\delta_1$, then   the  particle  will adsorb  and remain
permanently in  its place.  If  $\delta_1<\delta<\delta_2$, then the particle will
desorb.  If $\delta>\delta_2$, then the particle will  roll toward the surface
and, if  its path  is unblocked, will  adsorb to  the surface.  If the
path   is  blocked by  another  particle,   then  the rolling particle
desorbs.    The rationale   behind  these deposition    rules is  that
particles landing  directly  on top of  other particles   will tend to
adhere more strongly due to a greater contact  area.  By adjusting the
values  of $\delta_1$ and $\delta_2$,  one can perhaps mimic most experimental
situations.

In  one dimension, an analytical solution  is available for the lowest
(surface contacting) layer when overhangs  are calculated with respect
to the  surface.  This is  of particular  importance since  in certain
experimental situations, higher  layer   particles can be  removed  by
rinsing  and changing  the   solution conditions.  To  arrive  at  the
solution,   we introduce the  gap   distribution  function,  $G(h,t)$,
defined so that $G(h,t) dh$  is the density  of gaps of length between
$h$ and $h+dh$ at time $t$.  The time evolution of this function obeys
the following:

\begin{eqnarray}
\frac{\partial G(h,t)}{\partial (k_at)} &=& - (h+\sigma+2(\delta_1-\delta_2))G(h,t)
+2   \int_h^{h+\delta_1}G(h',t) dh'\nonumber\\&&    + 2  \int_{h+\sigma}^\infty
G(h',t) +2 k_a (\sigma-\delta_2) G(h+\sigma,t)\label{eq:85}
\end{eqnarray}
where $k_a$  is the  rate  of  deposition and  $\sigma$  is  the  particle
diameter.     By   taking $G(h,t)=e^{-k_a(h+\sigma+2(\delta_1-\delta_2))t}F(k_a\sigma
t)/\sigma^2$, the  integrodifferential  equation reduces   to an  ordinary
differential equation  in $t$.  Its solution  is then used to find the
surface density via

\begin{equation}\label{eq:86}
\frac{\partial \rho}{\partial (k_at)} = \int_\sigma^\infty(h+\sigma-2\delta_1)G(h,t) dh
\end{equation}

For  higher layer particles  and for particle overhang rules, computer
simulation  is required.  The   simulation procedure follows the above
deposition rules and involves  a line  of length  256 $\sigma$ (1-D)  or a
plane of size $256 \sigma$ x $256 \sigma$ (2-D).

We find that the saturation density of the lowest layer increases with
decreasing $\delta_1$ or   $\delta_2$.  In the  former case,  screening of the
surface by higher layer particles is reduced.  In the latter case, the
rolling  mechanism   (a very efficient way   to  fill the  surface) is
enhanced due to the screening effect. An interesting result is that the
radial distribution  function of the  lowest  layer shows only  a weak
correlation   between   particles  and  decays  to     zero at $r=2\sigma$
\cite{VV97}.  

\begin{figure}\begin{center}
\resizebox{12cm}{!}{\includegraphics{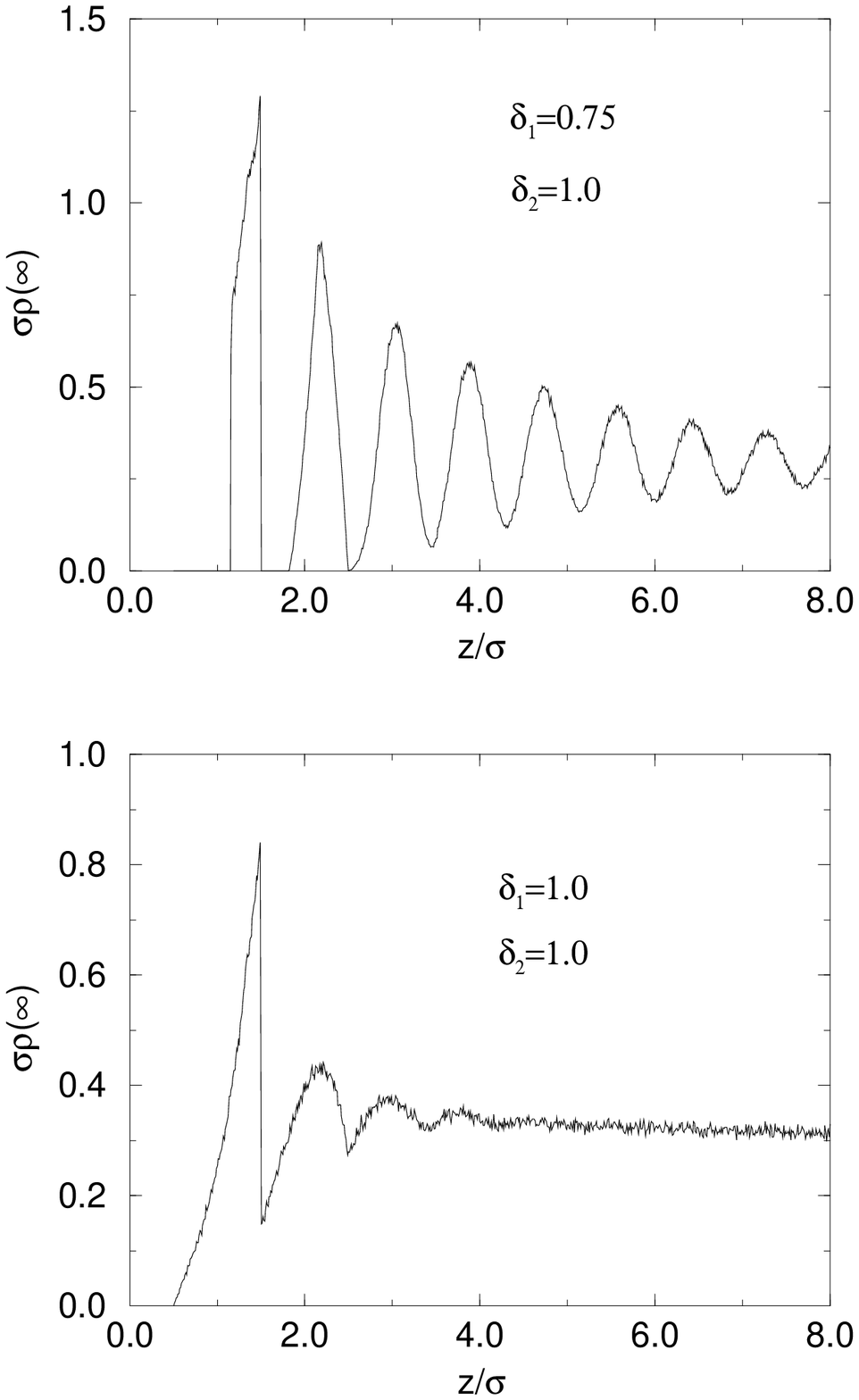}}
\caption[99]{Density of higher-layer particles as a function of height 
above the surface in two dimensions for different values of $\delta_1$ and 
$\delta_2$.}\label{fig:18}
 \end{center}

\end{figure}

When either $\delta_1$ or $\delta_2$ is unity, the  use of surface or particle
overhang rules   leads  to  the same   result\cite{LPR93}.  Otherwise,
particle overhang  rules always result in a  more dense  lowest layer.
1-D and 2-D results are qualitatively similar.  The density profile of
higher layer particles  exhibits  density oscillations whose  form may
vary markedly  depending on the  values of the  parameters  used.  For
example, when $\delta_1=\delta_2=1$,  the oscillations decay within 4 particle
diameters   of the   surface\cite{LPR93} whereas    for $\delta_1=.75$,  $
\delta_2=1$, the oscillations  continue  past  a  height  of more  than  8
particle diameters (Fig.~\ref{fig:18}).

\section*{Acknowledgements} The Laboratoire de Physique Th{\'e}orique des
Liquides is Unit{\'e} Mixte de Recherche No  7600 au CNRS. The Laboratoire
de  Physique  Th{\'e}orique  is Unit{\'e}   Mixte   de Recherche  No 8627   au
CNRS.  P.R.V.T wishes to acknowledge   the National Science Foundation
for support via CAREER  award  CTS-9733310. J.T. thanks  the  National
Science Foundation for financial support.

\begin{table} \caption[99]{Saturation coverage for RSA processes. Estimates from approximants and    simulation results.}

  \begin{center}  \begin{tabular}{|c||c|c|} \hline 
$\theta_\infty $ & Estimate  &     Computer  Simulation    \\       \hline 
  $D=2$   hard   disks  &0.548-0.553\cite{E89,DWJ91} &0.547\cite{HFJ86}   \\
   $D=3$   hard spheres  & 0.365\cite{TST91b}   & 0.382\cite{TST91b,C88}
 \\   $D=2$ parallel hard  squares &  0.5625\cite{DWJ91,BHM93}& 0.5620\cite{BZV91} \\  $D=3$ parallel
  hard   cubes   &  0.40-0.46 \cite{BHM93}  &0.42-0.43 \cite{JT80,C88} \\   \hline  \end{tabular}
  \end{center}
\end{table}

\begin{figure}
%\resizebox{8cm}{!}{\includegraphics{fig1a.ps}}\resizebox{8cm}{!}{\includegraphics{fig1b.ps}}
\caption[99]{Typical RSA (a) and equilibrium (b) configurations of disco-rectangles with an aspect ratio
$\alpha=15$ at the RSA  saturation coverage $\theta=0.45$.  Note that the  orientational
order is purely local   for RSA   whereas  a  nematic  phase  is  present  at
equilibrium.}\label{fig:1}
\end{figure}


\begin{figure}\begin{center}
%\resizebox{8cm}{!}{\includegraphics{fig8a.ps}}\resizebox{8cm}{!}{\includegraphics{fig8b.ps}}
%\resizebox{8cm}{!}{\includegraphics{fig8c.ps}}\resizebox{8cm}{!}{\includegraphics{fig8d.ps}}
	      \end{center}

\caption[99]{Pair correlation function for  deposited colloid layer at  four different
coverages: (a) $\theta  =0.075$, (b) $\theta  =0.30$, (c) $\theta =0.46$, (d)  $\theta
=0.53$. Diamonds  correspond   to the  experimental results  (melamine
latex spheres on mica) and the lines are determined from simulation of
the ballistic deposition model.}\label{fig:8}
 
\end{figure}



\begin{figure}\begin{center}
\resizebox{14cm}{!}{\includegraphics{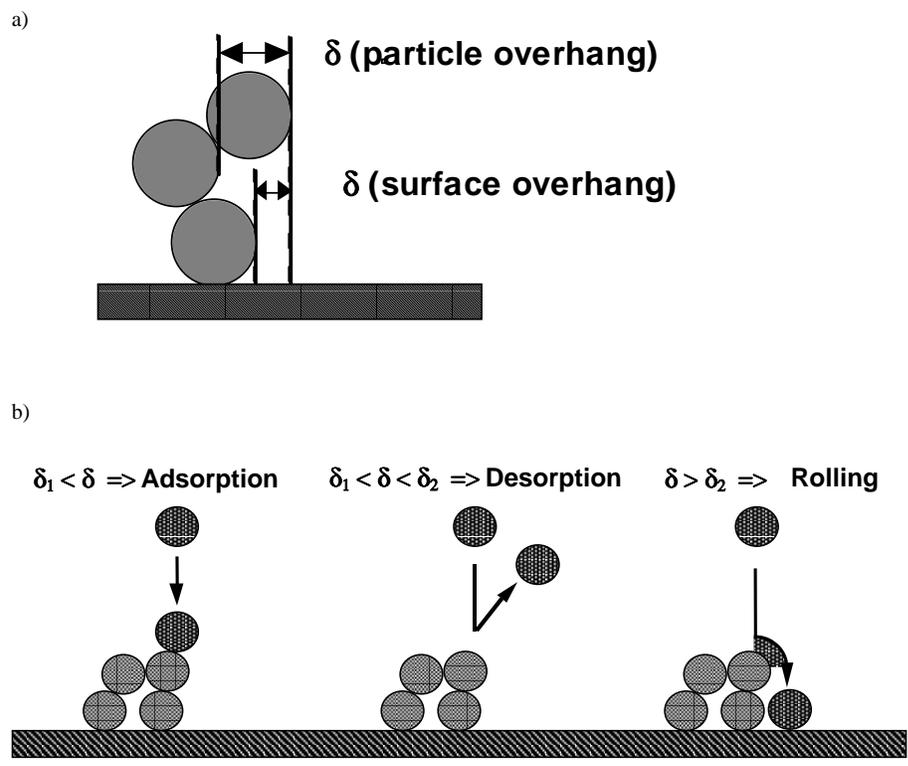}}
\caption[99]{(a) Depiction of the surface and particle overhangs. (b)
Schematic of the three possible events that can occur when an
incoming particle contacts a previously placed particle.}
			     \end{center}

\end{figure}\label{fig:17}

\begin{thebibliography}{99}


\bibitem{F80b} J. Feder and I. Giaever, J. Colloid Interface Sci. {\bf
78}, 144, (1980).
\bibitem{OL86} G.Y. Onoda and E.G. Liniger, Phys. Rev. Lett. {\bf 33},
715 (1986).
\bibitem{AH86} J.D. Andrade and V.L. Hlady, Adv. Polymer. Sci. {\bf
79}, 1 (1986).
\bibitem{N86} W. Norde, Adv. Colloid and Interface Sci. {\bf 25}, 267 (1986).
\bibitem{R93} J. J. Ramsden, Quart. Rev. of Biophysics {\bf 27}, 1 (1993).
\bibitem{ASZB94} Z. Adamczyk, B. Siwer, M. Zembala, and P.  Belouschek,
Adv. in  Colloid and Interface Sci. {\bf 48}, 151 (1994).
\bibitem{BP91}  M.C. Bartelt and V. Privman, Int. J. Mod. Phys. B {\bf 5}, 2883 (1991).  
\bibitem{E93}   J.W. Evans, Rev. Mod. Phys. {\bf 65}, 1281 (1993).
\bibitem{R63} A. R{\'e}nyi, Sel. Trans. Math. Stat. Prob. {\bf 4}, 205
(1963).
\bibitem{GHH74} J.J. Gonzales, P.C. Hemmer and J.S. Hoye,
Chem. Phys. {\bf 3}, 228 (1974).
\bibitem{BBV94} B. Bonnier, D. Boyer and P. Viot, J. Phys. A {\bf 27},
3671 (1994).
\bibitem{ST89a} P. Schaaf and J. Talbot, Phys. Rev. Lett. {\bf 62},
175 (1989).
\bibitem{ST89b} P. Schaaf and J. Talbot, J. Chem. Phys. {\bf 91}, 4401
(1989).
\bibitem{TST91a}G. Tarjus, P. Schaaf and J. Talbot, J. Stat.
Phys. {\bf 63}, 167 (1991).
\bibitem{F80a} J. Feder, J. Theor. Biol. {\bf 87}, 237, (1980).  
\bibitem{HFJ86} E.L. Hinrichsen, J. Feder, and T. J{\o}ssang, J. Stat. Phys. {\bf 44}, 793 (1986).
\bibitem{R93b}J. J. Ramsden, Phys. Rev. Lett., {\bf 71}, 295 (1993).
\bibitem{AZSW90} Z. Adamczyk, M. Zembala, B. Siwek, and P.
Warsz{\'y}nski, J. Colloid Interface Sci. {\bf 140}, 123 (1990); 
\bibitem{AB91} Z. Adamczyk and P. Belouschek J. Colloid Interface Sci.,
{\bf 146}, 123  (1991)
\bibitem{WSSZV93} P. Wojtaszczyk, P. Schaaf, B. Senger, M. Zembala,
J.C. Voegel  J. Chem. Phys. {\bf 99}, 7198 (1993).
\bibitem{W66} B. Widom, J. Chem. Phys. {\bf 44}, 3888 (1966).
\bibitem{MPV87} M. M{\'e}zard, G. Parisi, and M.A. Virasoro, Spin Glass
Theory and Beyond (World Scientific, Singapore, 1987).
\bibitem{G92} J.A. Given, Phys. Rev. A{\bf 45}, 816 (1992).
\bibitem{P80} Y. Pomeau, J. Phys.  {\bf A13},  L193 (1980).
\bibitem{S81} R.H. Swendsen, Phys. Rev. A {\bf 24}, 504 (1981).
\bibitem{TT92} G. Tarjus and J. Talbot Phys. Rev A {\bf 45}, 4162 (1992).
\bibitem{TTS89} J. Talbot, G.Tarjus, and P. Schaaf, Phys. Rev. A {\bf 40}, 4808 (1989).
\bibitem{VT90}	P. Viot and G. Tarjus, Europhys. Lett. {\bf 13}, 295
(1990). 
\bibitem{TT91} G. Tarjus and J. Talbot J. Phys. A{\bf 24}, L913 (1991).
\bibitem{HM86} J.P. Hansen and I.R. Mc Donald, {\it Theory of Simple
Liquids}, Academic Press (1986).
\bibitem{AW96}  Z. Adamczyk and P. Warszy{\'n}ski,  Adv. Colloid and Interface Sci. {\bf 63}, 41 (1996).
\bibitem{E89} J. W. Evans, Phys. Rev. Lett. {\bf 62}, 2642 (1989).
\bibitem{DWJ91} R. Dickman, J.-S. Wang, and I. Jensen,
J. Chem. Phys. {\bf 94}, 8252 (1991).
\bibitem{BHM93} B. Bonnier, M. Hontebeyrie, and C. Meyers, Physica
A{\bf 198} 1 (1993).
\bibitem{BTVT95} D. Boyer, G. Tarjus, P. Viot, and J. Talbot,
J. Chem. Phys. {\bf 95}, 1607 (1995).
\bibitem{P60}  I. Palasti, Publ. Math. Inst. Hung. Acad. Sci. {\bf 5},
353 (1960).
\bibitem{VTV97} P. R. Van Tassel, G. Tarjus, and P. Viot, J. Chem. Phys. {\bf  106}, 761 (1997).
\bibitem{RTV94} S.M. Ricci, J. Talbot, and P. Viot, Mol. Simul. {\bf  13}, 1 (1994).
\bibitem{W98} J.S. Wang, Physica {\bf 254}, 179 (1998) and references therein.
\bibitem{SDJS92} P. Schaaf, P. D{\'e}jardin, A. Johner, and A. Schmitt,
Langmuir {\bf 8}, 514 (1992).
\bibitem{MVBBS91} M.J. Mura-Galelli, J.C. Voegel, S. Behr, E.F. Bres, and
P. Schaaf, Proc. Natl. Acad. Sci., {\bf 88}, 5557, (1991).   
\bibitem{VZ89}R. D. Vigil and R. M. Ziff, J. Chem. Phys. {\bf 91}, 2599 (1989).
\bibitem{VZ90}R. D. Vigil and R. M. Ziff, J. Chem. Phys. {\bf 93}, 8270 (1990).
\bibitem{TVRT91} G. Tarjus, P. Viot, S. Ricci and J. Talbot, Mol. Phys. {\bf 73}, 773 (1991).
\bibitem{TV91} G. Tarjus and P. Viot, Phys. Rev. Lett. {\bf 67}, 1875 (1991).
\bibitem{RTTV92} S. Ricci, J. Talbot, G. Tarjus, and P. Viot, J. Chem. Phys. {\bf 97}, 5219 (1992).
\bibitem{VTRT92} P. Viot, G. Tarjus, S. Ricci, and J. Talbot, Physica A {\bf 191}, 248 (1992).
\bibitem{RTTV94} S.M. Ricci, J. Talbot, G. Tarjus, and P. Viot, J. Chem. Phys. {\bf 101}, 9164 (1994).
\bibitem{B75} T. Boubl\'{\i}k, Mol. Phys., {\bf 29}, 421 (1975).   
\bibitem{BTV95}  D. Boyer, G. Tarjus, and P. Viot,  Phys. Rev. E, {\bf
51}, 1043  (1995).  
\bibitem{ZV91}   R.M. Ziff and R.D. Vigill, J. Phys. A{\bf 23} 5103 (1990). 
\bibitem{Z91} R.M. Ziff (unpublished).
\bibitem{TS90} J. Talbot and P.Schaaf, Phys. Rev. A {\bf 40}, 422 1990.
\bibitem{MJ92a} P. Meakin and R. Jullien, Phys. Rev. A, {\bf 46}, 2029
(1992).
\bibitem{MJ92b} P. Meakin and R. Jullien, Physica. A, {\bf 187}, 475 (1992).
(1992).
\bibitem{SJT91} P. Schaaf, A. Johner, and J. Talbot, Phys. Rev. Lett. {\bf 66} 1603 (1991).
\bibitem{BSVS93} F.J. Bafaluy, B. Senger, J.C. Voegel, and P. Schaaf, Phys. Rev. Lett {\bf 70}, 623 (1993).
\bibitem{PR94} I. Pagonabarraga and J.M. Rubi, Phys. Rev. Lett. {\bf 73}, 114 (1994).
\bibitem{WA98} P. Wojtaszczyk and J.B. Avalos, Phys. Rev. Lett. {\bf
80}, 754 (1998).
\bibitem{SVSJST91} B. Senger, J.-C Voegel, P. Schaaf, A. Johner, A.
Schmitt, and J. Talbot, Phys. Rev. A {\bf 44}, 6926 (1991).
\bibitem{SSVJST92} B. Senger, P. Schaaf, J.-C Voegel, A. Johner, A. Schmitt,and J. Talbot, J. Chem. Phys {\bf97}, 3813 (1992).
\bibitem{TV92} G. Tarjus and P. Viot, Phys. Rev. Lett. {\bf 68}, 2354 (1992).
\bibitem{BCST94}F. J. Bafaluy, H. S. Choi, B. Senger and J. Talbot, Phys. Rev. E, {\bf 51}, 5985 (1995). 
\bibitem{STSSV93}B. Senger, J. Talbot, P. Schaaf, A. Schmitt and J.-C. Voegel, Europhys. Lett. {\bf 21}, 135 (1993). 
\bibitem{CT98} H.S.Choi and J.Talbot J. Stat. Phys. {\bf 92}, 891 (1998).
\bibitem{TR92} J. Talbot and S. Ricci,   Phys. Rev. Lett. {\bf 68}, 958 (1992).
\bibitem{VTT93} P. Viot, G. Tarjus, and J. Talbot, Phys. Rev. E {\bf 48}, 480 (1993).
\bibitem{BTV96} D. Boyer, G. Tarjus and P. Viot, J. Phys. A {\bf 29}, 2309 (1996).
\bibitem{JM92} R. Jullien and P. Meakin, J. Phys.A  {\bf 25}, L189 (1992).
\bibitem{CTTV93} H.S. Choi, J. Talbot, G. Tarjus and P. Viot, J. Chem. Phys. {\bf 97}, 4256 (1993).
\bibitem{TVCT94} G. Tarjus, P. Viot, H.S. Choi, and J. Talbot, Phys. Rev. E {\bf 49}, 3239 (1994).
\bibitem{B68} R.J. Baxter, J. Chem. Phys. {\bf 49}, 2770 (1968).     
\bibitem{TG92} A. P. Thompson and E.D. Glandt, Phys. Rev. A {\bf 46}, 4639 (1992).
\bibitem{CTTV95}H.S. Choi, J. Talbot, G. Tarjus and P. Viot, Phys. Rev. E {\bf  51}, 1353 (1995).
\bibitem{W95}  P. Wojtaszczyk, PhD thesis (Strasbourg University, 1995).50 
\bibitem{CR98}G. Cs{\'u}cs and J. J. Ramsden, J. Chem. Phys., {\bf 109},  779 (1998).
\bibitem{SDS87} P. Schaaf, P. Dejardin, and A. Schmidt, Langmuir {\bf
3}, 1131 (1987).
\bibitem{BTTVV94} D. Boyer, G. Tarjus, J. Talbot, P. Van Tassel, and P. Viot, Phys. Rev. E {\bf 49}, 5525 (1994).
\bibitem{VVTT94} P. R. Van Tassel, P. Viot, G. Tarjus, and J. Talbot,
J. Chem. Phys. {\bf 101}, 7064 (1994).
\bibitem{VTTV96} P. R. Van Tassel, J. Talbot, G. Tarjus, and P. Viot, Phys. Rev. E {\bf 53}, 785 (1996).
\bibitem{VGRTVT98} P.R. Van Tassel, L. Guemouri, J.J. Ramsden,
G. Tarjus, P. Viot, and J. Talbot, J. Colloid Interface Sci. {\bf 207}, 317 (1998).
\bibitem{PBR95} I. Pagonabarraga, J. Bafaluy, and J. M. Rubi, Phys, Rev. Lett.
{\bf 75}, 461 (1995).   
\bibitem{VTVT97} P.R. Van Tassel, J. Talbot, P. Viot, and G. Tarjus,
Phys. Rev. E {\bf 56}, R1299, (1997).
\bibitem{VVTRT99} P.R. Van Tassel, P. Viot, G. Tarjus, J.J. Ramsden, and
J. Talbot, J. Chem. Phys., submitted (1999).   
\bibitem{SW80} M.E. Soderquist and A.G. Walton, J. Colloid Interface Sci.
{\bf75}, 386 (1980).
\bibitem{M77} B.M. Morrisey, Ann. N.Y. Acad. Sci. {\bf288}, 50 (1977).
\bibitem{WAL95} M. Wahlgren, T. Arnebrant, and I. Lundstrom, J. Colloid
Interface Sci. {\bf175}, 506 (1995). 
\bibitem{JTT94} X. Jin, G. Tarjus and J. Talbot, J. Phys. A. {\bf 27},
L195 (1994).
\bibitem{KN94}P. L. Krapivsky and  E.  Ben-Naim, J. Chem. Phys.
{\bf 100}, 6778, 1994.
\bibitem{BKNJN98}   E.   Ben-Naim, J.  B.     Knight, E.  R.  Nowak,
H. M. Jaeger, and S. R. Nagel, Physica D {\bf 123}, 380 (1998).
\bibitem {KNT99} Amy J. Kolan, Edmund R. Nowak, Alexei V. Tkachenko, 
cond-mat/9809434.
\bibitem{TTV99} J. Talbot, G. Tarjus and P. Viot,  J. Phys.A {\bf 32}
2997 (1999).
\bibitem{NKBJN98} E. R. Nowak, J. B. Knight, E. Ben-Naim, H. M. Jaeger
and S. R. Nagel, Phys. Rev. E {\bf 57}, 1971 (1998).
\bibitem{RFL59}H. Reiss, H. L. Frisch and J. L. Lebowitz, J. Chem. Phys.,
{\bf 31}, 369 (1959).    
\bibitem{T97}J. Talbot, J. Chem. Phys., {\bf 106}, 4696 (1997).
\bibitem{M99}A. P. Minton, Biophysical Journal, {\bf 76}, 176 (1999).
\bibitem{JMTW99}X. Jin, Z. Ma, J. Talbot and N.-H. L. Wang, Langmuir, 
in press (1999).
\bibitem{OT99} C.B. Olson and J. Talbot, to be submitted.
\bibitem{BV99} M.A. Brusatori and P.R. Van Tassel, submitted to
J. Colloid Int. Sci.
\bibitem{K92} P.L.  Krapivsky, J. Chem. Phys. {\bf 97}, 2134 (1992).
\bibitem{BE94}  M.C. Bartelt and J.W. Evans, J. Stat. Phys. {\bf 76}, 867
(1994).          
\bibitem{VV97} P. R. Van Tassel, and P. Viot, Europhys. Lett. {\bf 40}, 293 (1997).
\bibitem{YVV98} S. Yang, P.Viot and P.R. Van Tassel, Phys. Rev. E {\bf
58}, 3324 (1998).
\bibitem{LPR93} B.D. Lubachevsky, V. Privman, and S.C. Roy, Phys. Rev. E 
{\bf 48}, 48 (1993).
\bibitem{TST91b} J. Talbot, P. Schaaf, and G. Tarjus,  Mol. Phys. {\bf 72}  1397 (1991).
\bibitem{C88} D.W. Cooper, Phys. Rev. A{\bf 38} 522 (1988).
\bibitem{BZV91} B.J. Brosilow, R.M. Ziff, and R.D. Vigil,
 Phys. Rev. A {\bf 43}, 631 (1991).
\bibitem{JT80} W.S. Jodrey and E.M. Tory, J. Stat. Comput. Simul. {\bf
10}, 87 (1980).
\bibitem{C88} D.W. Cooper, J. Appl. Prob. {\bf 26}, 664 (1988).      
\end{thebibliography}
\end{document}